\documentclass[11pt]{article} 

\usepackage[utf8]{inputenc} 
\usepackage{geometry} 
\usepackage{amsmath,amssymb, amsfonts}
\usepackage{bm,cite,color,epsf,epsfig,graphicx,multirow,paralist,times,subfigure}
\usepackage{algorithm,algorithmic,bbm}

\newcommand{\EE}{\mathbb{E}}
\newcommand{\NN}{\mathbb{N}}
\newcommand{\PP}{\mathbb{P}}

\newcommand{\indic}{\mathbbm{1} }

\newcommand{\bp}{\noindent{\bf Proof.}\ }
\newcommand{\ep}{\hfill $\Box$}

\newcommand{\al}[1]{ \begin{align} #1  \end{align}}
\newcommand{\eq}[1]{ \begin{equation} #1  \end{equation}}
\newcommand{\als}[1]{ \begin{align*} #1  \end{align*}}
\newcommand{\eqs}[1]{ \begin{equation*} #1  \end{equation*}}

\newcommand{\sk}{\nonumber\\} 
\newcommand{\Lp}{\left(}
\newcommand{\Rp}{\right)}
\newcommand{\Lb}{\left[}
\newcommand{\Rb}{\right]}

\newcommand{\el}{\end{flushleft}}
\newcommand{\bl}{\begin{flushleft}}
\newcommand{\floor}[1]{  \lfloor #1 \rfloor} 
\newcommand{\ceil}[1]{  \lceil #1 \rceil}

\newcommand{\separator}{
  \begin{center}
    \rule{\columnwidth}{0.3mm}
  \end{center}
}
\newenvironment{separation}{ \vspace{-0.3cm}  \separator  \vspace{-0.2cm}}
{  \vspace{-0.4cm}  \separator  \vspace{-0.1cm}}

\newtheorem{proposition}{Proposition}
\newtheorem{theorem}{Theorem}[section]
\newtheorem{lemma}[theorem]{Lemma}
\newtheorem{corollary}[theorem]{Corollary}

\newtheorem{assumption}{Assumption}

\DeclareGraphicsExtensions{.pdf,.eps}

\graphicspath{{.//}}

\title{Optimal Rate Sampling in 802.11 Systems}
\author{Richard Combes$^\dag$,  Alexandre Proutiere$^{\dag,\times}$, Donggyu Yun$^\ddag$, Jungseul Ok$^\ddag$, Yung Yi$^\ddag$\thanks{$\dag$: KTH Royal Institute of Technology, Sweden. $\times$: INRIA/Microsoft joint research centre, Palaiseau, France. $\ddag$: Department of Electrical Engineering, KAIST, South Korea.}}

\begin{document}
\maketitle

\begin{abstract}
Rate Adaptation (RA) is a fundamental mechanism in 802.11 systems. It allows transmitters to adapt the coding and modulation scheme as well as the MIMO transmission mode to the radio channel conditions, and in turn, to learn and track the (mode, rate) pair providing the highest throughput. So far, the design of RA mechanisms has been mainly driven by heuristics. In contrast, in this paper, we rigorously formulate such design as an online stochastic optimisation problem. We solve this problem and present ORS (Optimal Rate Sampling), a family of (mode, rate) pair adaptation algorithms that provably learn as fast as it is possible the best pair for transmission. We study the performance of ORS algorithms in both stationary radio environments where the successful packet transmission probabilities at the various (mode, rate) pairs do not vary over time, and in non-stationary environments where these probabilities evolve. We show that under ORS algorithms, the throughput loss due to the need to explore sub-optimal (mode, rate) pairs does not depend on the number of available pairs, which is a crucial advantage as evolving 802.11 standards offer an increasingly large number of (mode, rate) pairs. We illustrate the efficiency of ORS algorithms (compared to the state-of-the-art algorithms) using simulations and traces extracted from 802.11 test-beds.    
\end{abstract}

\section{Introduction}  

In wireless communication systems, Rate Adaptation (RA) is a fundamental mechanism allowing transmitters to adapt the coding and modulation scheme to the radio channel conditions. In 802.11 systems, the transmitter may choose from a finite set of rates with the objective of identifying as fast as possible the rate providing maximum throughput, i.e., maximising the product of the rate and the successful packet transmission probability. The challenge stems from the facts that these probabilities are not known a priori at the transmitter, and that they may evolve over time. The transmitter has to learn and track the best transmission rate, based on the measurements and observations made on the successive packet transmissions. 

Over the last decade, a large array of RA mechanisms for 802.11 systems has been proposed. We may categorise these mechanisms depending on the feedback and measurements from past transmissions available at the transmitter, and actually used to sequentially select rates for packet transmissions. Traditionally in 802.11 systems, RA mechanisms are based on rate sampling approaches, i.e., the rate selection solely depends on the number of successes and failures of previous packet transmissions at the various available rates. Examples of such mechanisms include ARF (Auto Rate Fall back) \cite{kamerman1997wavelan} and SampleRate \cite{bicket2005bit}. As 802.11 standards evolve, the number of available rates increases. In 802.11n systems, for a given packet transmission, a MIMO mode (e.g. a diversity oriented single-stream (SS) mode or a spatial multiplexing driven double-stream (DS) mode) and a rate have to be jointly selected.  The number of possible decisions can then become quite large, making the use of sampling approaches questionable. 

A natural alternative to RA sampling approaches consists in using channel measurements. So far, such measurements have not been explicitly used in practice. The most accessible measurement, the receiver signal strength indication (RSSI), is known to lead to poor predictions of the packet error rate (PER) at the various rates (see e.g. \cite{aguayo2004, reis2006, camp2008,zhang2008,halperin2010,deek2013}). These poor predictions are for example due to the fact that RSSI does not reflect frequency-selective fading. Note that 802.11n NICs actually measure and report the channel quality at the OFDM subcarrier level, also known as channel state information (CSI), which provides better information than the simple RSSI. CSI feedback could be used to improve PER prediction accuracy \cite{halperin2010}. However it is difficult and costly to get and store this complete information \cite{crepaldi2012}, and CSI feedback is supported by very few 802.11n devices. A promising solution could then consist in storing and using only parts of this information, as proposed for example in \cite{deek2013}. 

As of now, it seems difficult to predict whether measurement-based RA mechanisms will be widely adopted in the future, or whether rate sampling approaches will continue to prevail. In this paper, we investigate the fundamental performance limits of sampling-based RA mechanisms. Our objective is to design the {\it best} possible rate sampling algorithm, i.e., the algorithm that identifies as fast as possible the rate maximising throughput. Our approach departs from previous methods to design RA mechanisms: in existing mechanisms, the way sub-optimal rates are explored to learn and track the best rate for transmission is based on heuristics. In contrast, we look for the optimal way of exploring sub-optimal rates. 

We rigorously formulate the design of the best rate sampling algorithm as an online stochastic optimisation problem. In this problem, the objective is to maximise the number of packets successfully sent over a finite time horizon. We show that this problem reduces to a Multi-Armed Bandit (MAB) problem \cite{lai1985}. In MAB problems, a decision maker sequentially selects an action (or an arm), and observes the corresponding reward. Rewards of a given arm are random variables with unknow distribution. The objective is to design sequential action selection strategies that maximise the expected reward over a given time horizon. These strategies have to achieve an optimal trade-off between exploitation (actions that has provided high rewards so far have to be selected) and exploration (sub-optimal actions have to be chosen so as to learn their average rewards). For the rate adaptation problem, the various arms correspond to the decisions available at the transmitter to send packets, i.e., in 802.11a/b/g systems, an arm corresponds to a modulation and coding scheme or equivalently to a transmission rate, whereas in MIMO 802.11n systems, an arm corresponds to a (mode, rate) pair. When a rate is selected for a packet transmission, the reward is equal to 1 if the transmission is successful, and equal to 0 otherwise. The average successful packet transmission probabilities at the various rates are of course unknown, and have to be learnt.   

The sequential rate (or (mode, rate) in MIMO 802.11n systems) selection problem is referred to as a {\it structured MAB problem} in the following, as it differs from classical MAB problems. First, the rewards associated with the various rates are stochastically correlated, i.e., the outcomes of transmissions at different rates are not independent: for example, if a transmission at a high rate is successful, it would be also successful at lower rates. Then, the average throughputs achieved at various rates exhibit natural structural properties. For 802.11b/g systems, the throughput is an {\it unimodal} function of the selected rate. For MIMO 802.11n systems, the throughput remains unimodal in the rates within a single MIMO mode, and also satisfies some structural properties across modes. We model the throughput as a so-called {\it graphically unimodal} function of the (mode, rate) pair. As we demonstrate, correlations and graphical unimodality are instrumental in the design of RA mechanisms, and can be exploited to learn and track the best rate or (mode, rate) pair quickly and efficiently. Finally, most MAB problems consider stationary environments, which, for our problem, means that the successful packet transmission probabilities at different rates do not vary over time. In practice, the transmitter faces a non-stationary environment as these probabilities could evolve over time. We consider both stationary and non-stationary radio environements.

In the case of stationary environments, we derive an asymptotic upper bound of the expected reward achieved in structured MAB problems. This provides a fundamental performance limit satisfied by {\it any} rate adaptation algorithm. This limit quantifies the inevitable performance loss due to the need to explore sub-optimal rates. It also indicates the performance gains that can be achieved by devising rate adaptation schemes that optimally exploit the correlations and the structural properties of the MAB problem. As it turns out, the performance loss due to the need of exploration does not depend on the number of available rates (or (mode, rate) pairs), i.e., on the size of the decision space. This suggests that rate sampling methods can perform well even if the number of decisions available at the transmitter grows large. We present two rate sampling algorithms: ORS (Optimal Rate Sampling) and G-ORS (G stands for Graphical), an extension of ORS to MIMO systems. We show that their performance matches the upper bound derived previously, i.e., ORS and G-ORS are asymptotically optimal. We extend the results and algorithms to non-stationary radio environments: we propose SW-ORS and SW-G-ORS algorithms (SW stands for Sliding Window) and analyse their performance. We show that again, the latter does not depend on the size of the decision space, and that the best rate (or (mode, rate) pair) can be efficiently learnt and tracked even. Finally we compare the performance of the proposed algorithms to that of existing rate sampling algorithms using simulations and traces extracted from real 802.11 test-beds. Our algorithms outperform existing RA schemes. This should not be surprising: the design of most existing algorithms is based on heuristics, whereas ORS algorithms are by design optimal.

\medskip
\noindent
{\bf Contributions and paper organisation.} 
\begin{enumerate}
\item The next section is devoted to the related work. Existing RA algorithms (using either sampling approaches, or based on measurements) are discussed. A brief state-of-the-art about MAB problems is also presented.
\item In the next two sections (Sections 3 and 4), we formulate the design of rate sampling algorithms as an online stochastic optimization problem, and we show how the latter can be mapped to a structured MAB problem.
\item We derive a performance upper bound satisfied by any rate sampling algorithm in the case of stationary radio environements, and show that this bound does not depend on the size of the decision space (Section 5).
\item We present, in Section 6, ORS, a rate sampling algorithm whose performance matches the upper bound derived in Section 5.
\item Next in Section 7,  we present SW-ORS, an extension of ORS to non-stationary radio environments and provide guarantees on its performance. Again we show that its performance does not depend on the number of available rates.
\item The algorithms and performance results are extended to MIMO 802.11n systems in Section 8. The proposed algorithm, referred to as G-ORS (G stands for {\it Graphical}), optimally exploits the fact that the throughput is a graphically unimodal function of the (mode, rate) pair.
\item Finally in Section 9, the performance of our algorithms are illustrated using simulation results and traces extratced from real 802.11 test-beds. 
\end{enumerate}

\section{Related work}

\subsection{RA mechanisms in 802.11 systems}

In recent years, there has been a growing interest in the design of RA mechanisms for 802.11 systems, perhaps motivated by the new functionalities (e.g. MIMO, and channel width adaptation) offered by the evolving standards. 

\medskip
{\bf Sampling-based RA mechanisms.}  ARF \cite{kamerman1997wavelan}, one of the earliest rate adaptation algorithms, consists in changing the transmission rate based on packet loss history: a higher rate is probed after $n$ consecutive successful packet transmissions, and the next available lower rate is used after two consecutive packet losses. In case of stationary radio environments, ARF essentially probe higher rates too frequently (every 10 packets or so). To address this issue, AARF \cite{lacage2004ieee} adapts the threshold $n$ dynamically to the speed at which the radio environment evolves. Among other proposals, SampleRate \cite{bicket2005bit} sequentially selects transmission rates based on estimated throughputs over a sliding window, and has been shown to outperform ARF and its variants. The aforementioned algorithms were initially designed for 802.11 a/b/g systems, and they seem to perform poorly in MIMO 802.11n systems \cite{Pefkianakis:2010}. One of the reasons for this poor performance is the non-monotonic relation between loss and rate in 802.11n MIMO systems, when considering all rates options and ignoring modes. When modes are ignored, the loss probability does not necessarily increase with the rate. As a consequence, RA mechanisms that ignore modes may get stuck at low rates. To overcome this issue, the authors of \cite{Pefkianakis:2010} propose MiRA, a RA scheme that zigzags between MIMO modes to search for the best (mode, rate) pair. In the design of RAMAS \cite{nguyen2011}, the authors categorise the different types of modulations into modulation-groups, as well as the MIMO modes into what is referred to as {\it enhancement} groups; the combination of the modulation and enhancement group is mapped back to the set of the modulation and coding schemes. RAMAS then adapts these two groups concurrently. As a final remark, note that in 802.11 systems, packet losses are either due to a mismatch between the rate selection and the channel condition or due to collisions with transmissions of other transmitters. Algorithms such as LD-ARF \cite{pang2005rate}, CARA \cite{kim2006cara}, and RRAA \cite{wong2006robust} have been proposed to distinguish between losses and collisions. 

It is important to highlight the fact that in all the aforementioned RA algorithms, the way sub-optimal rates (or (mode, rate) pairs) are explored to identify the best rate is based on heuristics. This contrasts with the proposed algorithms, that are designed, using stochastic optimisation methods, to learn the best rate for transmission as fast as possible. The way sub-optimal rates are explored under our algorithms is optimal.

\medskip
{\bf Measurement-based methods.}
As mentioned in the introduction, measurement-based RA algorithms could outperform sampling approaches if the measurements (RSSI or CSI) used at the transmitter could be used to accurately predict the PER achieved at the various rates. However this is not always the case, and measurement-based approaches incur an additional overhead by requiring the receiver to send channel-state information back to the transmitter. In fact, sampling and measurement-based approaches have their own advantages and disadvantages. We report here a few measurement-based RA mechanisms. 

In RBAR \cite{holland2001rate} (developed for 802.11 a/b/g systems), RTS/CTS-like control packets are used to ``probe'' the channel. The receiver first computes the best rate based on the SNR measured over an RTS packet and then informs the transmitter about this rate using the next CTS packet. OAR \cite{sadeghi2002opportunistic} is similar to RBAR, but lets the transmitter send multiple back-to-back packets without repeating contention resolution procedure. CHARM \cite{judd2008} leverages the channel reciprocity to estimate the SNR value instead of exchanging RTS/CTS packets. In 802.11n with MIMO, ARAMIS \cite{deek2013} uses the so-called {\em diff}SNR as well as the SNR to predict the PER at each rate. The {\em diff}SNR corresponds to the difference between the maximum and minimum SNRs observed on the various antennas at the receiver. ARAMIS exploits the fact that environmental factors (e.g., scattering, positioning) are reflected in the {\em diff}SNR. Recently, hybrid approaches combining SNR measurements and sampling techniques have also been advocated, see \cite{haratcherev2004}. It is also worth mentioning cross-layer approaches, as in \cite{vutukuru2009}, where BER (Bit Error Rate) are estimated using information provided at the physical layer.

In some sense, measurement-based RA schemes in 802.11 systems try to mimic rate adaptation strategies used in cellular networks. However in these networks, more accurate information on channel condition is provided to base station \cite{3gpp}. Typically, the base station broadcasts a pilot signal, from which each receiver measures the channel conditions. The receiver sends this measurement, referred to as CQI  (Channel Quality Indicator), back to the base station. The transmission rate is then determined by selecting the highest CQI value which satisfies the given BLER (Block Error Rate) threshold, e.g., 10\% in 3G systems. More complex, but also more efficient rate selection mechanisms are proposed in \cite{kim2008,freudenthaler2007}. These schemes predict the throughput more accurately by jointly considering other mechanisms used at the physical layer, such as HARQ (Hybrid ARQ).

\subsection{Stochastic MAB problems}

Stochastic MAB formalise sequential decision problems where the decision maker has to strike an optimal trade-off between exploitation and exploration. MAB problems have been applied in many disciplines -- their first application was in the context of clinical trials \cite{thompson1933}. Please refer to \cite{bubeck2012} for a recent survey. Most existing theoretical results concern {\it unstructured} MAB problems \cite{robbins1952}, i.e., problems where the average reward associated with the various arms are not related. For this kind of problems, Lai and Robbins \cite{lai1985} derived an asymptotic lower bound on regret and also designed optimal decision algorithms. When the average rewards are structured, the design of optimal decision algorithms is more challenging, see e.g. \cite{bubeck2012}. Unimodal bandit problems have received little attention so far. In \cite{yu2011}, the authors propose various algorithms, but they do not prove their optimality (the . In this paper, we first derive asymptotic regret lower bounds for these problems, and then devise asymptotically optimal algorithms. We also study unimodal bandit problems in non-stationary environments, where the average rewards of the different arms evolve over time. Non-stationary environments have not been extensively studied in the bandit literature. For unstructured problems, the performance of algorithms based on UCB \cite{auer2002} has been analyzed in \cite{kocsis2006,yu2009,garivier2011alt} under the assumption that the average rewards are abruptely changing. Here we consider more realistic scenarios where the average rewards smoothly evolve over time. To our knowledge, such scenarios have only been considered in \cite{slivkins2008, slivkins2011} but using different assumptions. Finally, the authors of \cite{radunovic2011} formalize rate adaptation and channel selection issues as a MAB problem, but they do not solve it, and the proposed algorithms are heuristics.


\section{Models and Objectives} \label{sec:prelim}

We present here the models and objectives for 802.11 a/b/g systems (using a single MIMO mode). The extension to MIMO 802.11n is discussed in Section \ref{sec:mimo}. We consider a single link (a transmitter-receiver pair). At
time 0, the link becomes active and the transmitter has packets to send
to the receiver. To do so, the transmitter can sequentially pick a
coding rate from a finite set ${\cal R}=\{r_1,\ldots,r_K\}$. This set is ordered, i.e., $r_1<r_2<\ldots<r_K$. After a packet
is sent, the transmitter is informed on whether the transmission has been
successful. Based on the observed past transmission successes and
failures at the various rates, the transmitter has to select a rate for
the next packet transminssion. We denote by $\Pi$ the set of all possible sequential rate selection schemes. Packets are assumed to be of equal size, and without loss of generality the duration of a packet transmission at rate $r_k$ is $1/r_k$ for any $k$.

\subsection{Channel models}

For the $i$-th packet transmission at rate $r_k$, a binary random variable $X_k(i)$ represents the success ($X_k(i)=1$) or failure ($X_k(i)=0$) of the transmission. 

\medskip
{\it Stationary radio environments.} In such environments, the success transmission probabilities at different rates do not evolve over time. This arises when the system considered is static (in particular, the transmitter and receiver do not move) -- refer to Section VI for a detailed discussion. Formally, $X_k(i)$, $i=1,2,\ldots$, are independent and identically distributed, and we denote by $\theta_k$ the success transmission probability at rate $r_k$: $\theta_k=\mathbb{E}[X_k(i)]$. We denote by $k^{\star}$ the index of the optimal rate, $k^{\star} \in \arg\max_{k} r_k\theta_k$. To simplify the exposition and the notation, we assume that the optimal rate is unique, i.e., $r_{k^{\star}}\theta_{k^{\star}}>r_k\theta_k$, for all $k\neq k^{\star}$. 

\medskip
{\it Non-stationary radio environments.} In practice, channel conditions may be non-stationary, i.e., the success probabilities at various rates could evolve over time. In many situations, the evolution over time is rather slow -- refer to \cite{radunovic2011} and to Section V for test-bed measurements. These slow variations allow us to devise rate adaptation schemes that efficiently track the best rate for transmission. In the case of non-stationary environment, we denote by $\theta_k(t)$ the success transmission probability at rate $r_k$ and at time $t$, and by $k^\star(t)$ the index of the optimal rate at time $t$.

Unless otherwise specified, for clarity, we consider stationary radio environments. Non-stationary environements are treated in Section \ref{sec:nonstat}, where we will present the models and objectives, and extend our rate selection algorithm for these scenarios.

\subsection{Structural properties} 

Our problem is to identify as fast as possible the rate maximizing throughput. To this aim, we shall desin algorithms that exploit two crucial structural properties of the problem: (i) The successes and failures of transmissions at various rates are correlated, and (ii) in practice, we have observed that the throughput is an {\it unimodal} function of the transmission rate.

\medskip
{\it Correlations.} If a transmission is successful at a high rate, it has to be successful at a lower rate, and similarly, if a low-rate transmission fails, then a transmitting at a higher rate would also fail. Formally assume that at a given time, rate $r_k$ (resp. rate $r_l$) has already been selected $(i-1)$ (resp. $(j-1)$) times. Then:
\begin{equation}\label{eq:dep1}
\left( k<l \hbox{ and } X_k(i)=0\right) \Longrightarrow (X_l(j)=0),
\end{equation}
\begin{equation}\label{eq:dep2}
\left( k>l \hbox{ and } X_k(i)=1\right) \Longrightarrow (X_l(j)=1).
\end{equation}
Using simple coupling arguments, we can readily show that an equivalent way of expressing the correlations is to state that the following assumption holds.

\begin{assumption}
$\theta=(\theta_1,\ldots,\theta_K)\in {\cal T}$, where ${\cal T}=\{\eta\in [0,1]^K: \eta_1\ge \ldots\ge \eta_K\}$.
\end{assumption}

{\it Unimodality.} In practice, we observe that the throughputs achieved by transmitting at various rates are unimodal. To formalize this observation, we make the following assumption, that will be extensively discussed and verified in Section \ref{sec:num}.

\begin{assumption}\label{as:uni}
$\theta\in {\cal U}$, where ${\cal U}=\{\eta\in [0,1]^K:\exists k^\star, r_1\eta_1<\ldots <r_{k^\star}\eta_{k^\star}, r_{k^\star}\eta_{k^\star}> r_{k^\star+1}\eta_{k^\star+1}>\ldots>r_k\eta_K\}$.
\end{assumption}

\subsection{Objective}

We now formulate the design of rate adaptation schemes as an online stochastic optimization problem. An optimal scheme maximizes the expected throughput up to a certain finite time $T$. The choice of $T$ is not really important as long as during time interval $T$, a large number of packets can be sent -- so that infering the success transmission probabilities efficiently is possible. 

Consider a rate adaption scheme $\pi\in \Pi$ that selects rate $r_{k^{\pi}(t)}$ for the $t$-th transmission. At time $T$, the number of packets $\gamma^\pi(T)$ that have been successfully sent under algorithm $\pi$ is: $\gamma^\pi(T) = \sum_k \sum_{i=1}^{s_k^\pi(T)}X_k(i)$, where $s_k^\pi(T)$ is the number of transmission attempts at rate
$r_k$ before time $T$. The $s_k(T)$'s are random variables (since the
rates selected under $\pi$ depend on the past random successes and
failures), and satisfy the following constraint:
$$
\sum_ks_k^\pi(T)\times {1\over r_k} \le T.
$$
Wald's lemma implies that the expected number of packets
successfully sent up to time $T$ is: $\mathbb{E}[\gamma^\pi(T)]=\sum_k
\mathbb{E}[s_k^{\pi}(T)] \theta_k.$ Thus, our objective is to design
an algorithm solving the following online stochastic optimization
problem:
\begin{align}\label{pb:1}
\max_{\pi\in \Pi} & \sum_k \mathbb{E}[s_k^{\pi}(T)] \theta_k,\\
\hbox{s.t. } & s_k^\pi(T)\in \mathbb{N}, \hbox{ and }\sum_k s_k^{\pi}(T)\times {1\over r_k} \le T,\quad\forall k.\nonumber
\end{align}

\section{An equivalent Structured Multi-Armed Bandit (MAB) problem}\label{section4}

In this section, we show that the online optimization problem (\ref{pb:1}) can be reduced to a {\it structured} Multi-Armed Bandit (MAB) problem.

\subsection{An alternative system}

\begin{figure}[htb]
\begin{center}
\includegraphics[width=6.2cm]{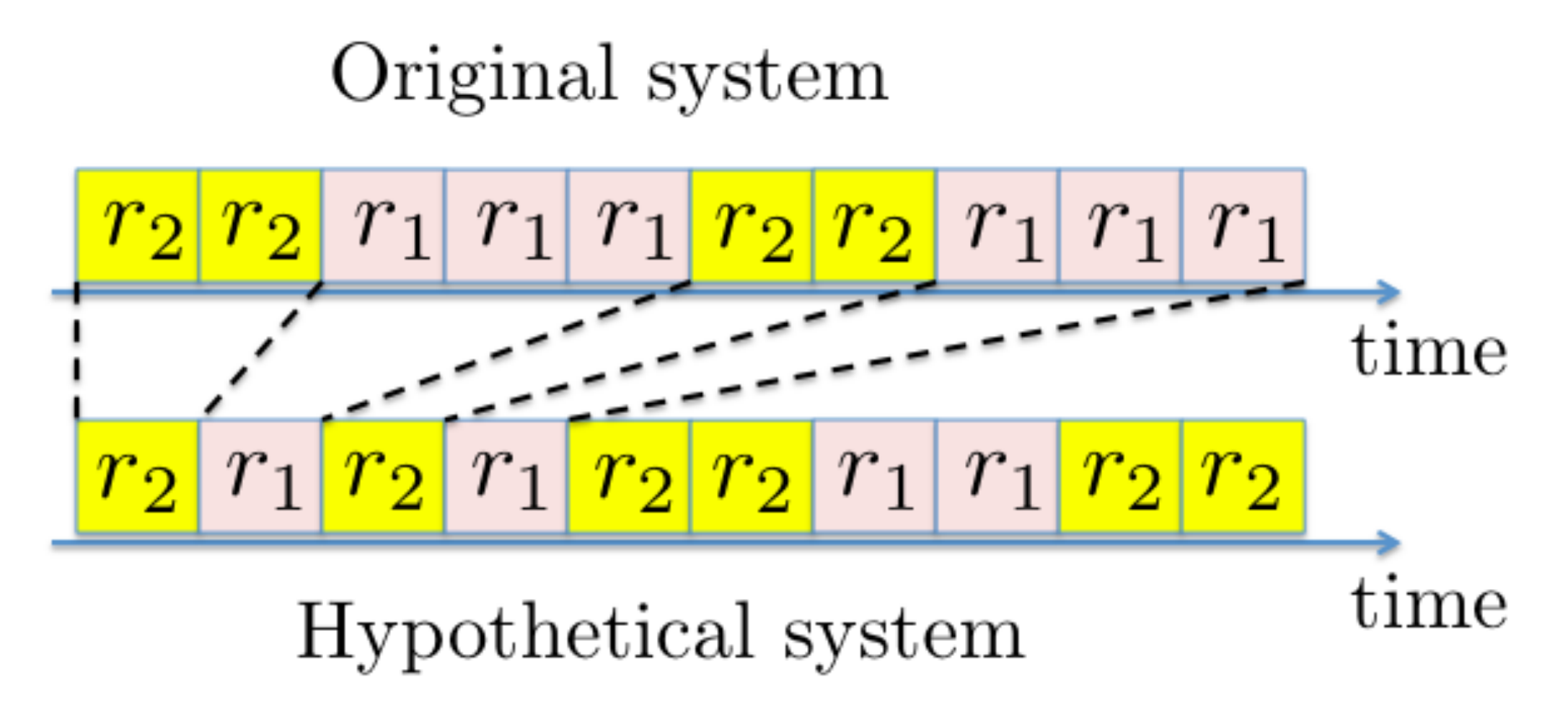} 
\end{center}
\vspace{-3mm}
\caption{Examples of rate selections made by the same algorithm $\pi\in \Pi$ in the original and alternative systems - $r_1=1/3$ and $r_2=1/2$.}
\label{fig1}

\end{figure}

Without loss of generality, we assume that time can be divided into slots whose durations are such that for any $k$, the time it takes to transmit one packet at rate $r_k$ corresponds to an integer number of slots. Under this convention, the optimization problem (\ref{pb:1}) can be written as:
\begin{align}\label{pb:1bis}
\max_{\pi\in \Pi} & \sum_k \mathbb{E}[t_k^{\pi}(T)] r_k\theta_k,\\
\hbox{s.t. } & \sum_k t_k^{\pi}(T)\le T, \hbox{ and }\nonumber\\
&t_k^\pi(T)\in {1\over r_k}\mathbb{N}:=\{ {u\over r_k}, u\in \mathbb{N}\}, \forall k,\nonumber
\end{align}
where $t_k^\pi(T)=s_k^\pi(T)/r_k$ represents the amount of time (in slots)
that the transmitter spends, before $T$, on sending packets at rate
$r_k$. The constraint $t_k(T)\in {1\over r_k}\mathbb{N}$ indicates that
when a rate is selected, this rate selection remains the same for the
next $1/r_k$ slots (to transmit an entire packet). By relaxing this constraint, we obtain an
optimization problem corresponding to a MAB problem. Indeed, consider
now an alternative system where rate selection is made {\it every}
slot. If at any given slot, rate $r_k$ is selected for the $i$-th times,
then if $X_k(i)=1$, the transmitter successfully sends $r_k$ bits in
this slot, and if $X_k(i)=0$, then no bit are received. A rate selection
algorithm then decides in each slot which rate to use. There is a
natural mapping between rate selection algorithms in the original system
and in the alternative system: let $\pi\in \Pi$, if for the $t$-th
packet transmission rate $r_k$ is selected under $\pi$ in the original
system, then $\pi$ selects the same rate $r_k$ in the $t$-th slot. This
mapping is illustrated in Figure \ref{fig1}.

For the alternative system, the objective is to design $\pi\in \Pi$ solving the following optimization problem, which can be interpreted as a relaxation of (\ref{pb:1bis}). 
\begin{align}\label{pb:2}
\max_{\pi\in \Pi} & \sum_k \mathbb{E}[t_k^{\pi}(T)] r_k\theta_k,\\
\hbox{s.t. } & \sum_k t_k^{\pi}(T)\le T,\hbox{ and }  t_k^{\pi}(T)\in \mathbb{N}, \forall k. \nonumber
\end{align}
The above optimization problem corresponds to a MAB problem, where in each slot a decision is taken (i.e., a rate is selected), and where when rate $r_k$ is chosen, the obtained reward is $r_k$ with probability $\theta_k$ and 0 with probability $1-\theta_k$. Note that in traditional MAB problems, the rewards obtained by taking various decisions are stochastically independent. This is not the case for our MAB problem.

\subsection{Regrets and Asymptotic Equivalence}

We may assess the performance of an algorithm $\pi\in\Pi$ in both original and alternative systems through the notion of {\it regret}. The regret up to slot $T$ compares the performance of $\pi$ to that achieved by an algorithm always selecting the best rate. If the parameter $\theta$ was known, then in both systems, it would be optimal to select rate $r_{k^\star}$. The regret of algorithm $\pi$ up to time slot $T$ in the original system is then defined by:
$$
R_1^{\pi}(T) = \theta_{k^\star}\lfloor r_{k^\star}T\rfloor -\sum_{k}\theta_k\mathbb{E}[s_k^{\pi}(T)],
$$
where $\lfloor x\rfloor$ denotes the largest integer smaller than $x$. 

The regret of algorithm $\pi$ up to time slot $T$ in the alternative system is similarly defined by:
$$
R^{\pi}(T) = \theta_{k^\star}r_{k^\star}T -\sum_{k}\theta_kr_k \mathbb{E}[t_k^{\pi}(T)].
$$

In the next section, we show that an asymptotic lower bound for the regret $R^\pi(T)$ is of the form $c(\theta)\log(T)$ where $c(\theta)$ is a strictly positive constant that we can explicitly characterize. It means that for all $\pi\in \Pi$, $\lim\inf_{T\to\infty}R^\pi(T)/\log(T)\ge c(\theta)$. It will be also shown that there exists an algorithm $\pi^\star\in \Pi$ that actually achieves this lower bound in the alternative system, in the sense that $\lim\sup_{T\to\infty}R^{\pi^{\star}}(T)/\log(T)\le c(\theta)$. In such a case, we say that $\pi^\star$ is asymptotically optimal. The following lemma states that actually, the same lower bound is valid in the original system, and that any asymptotically optimal algorithm in the alternative system is also asymptotically optimal in the original system. 

\begin{lemma}
Let $\pi\in \Pi$. For any $c>0$, we have:
$$
\left( \lim\inf_{T\to\infty}{R^{\pi}(T)\over \log(T)}\ge c\right)  \Longrightarrow \left( \lim\inf_{T\to\infty}{R_1^{\pi}(T)\over \log(T)}\ge c\right),
$$
and
$$
\left(\lim\sup_{T\to\infty}{R^{\pi}(T)\over \log(T)}\le c\right) \Longrightarrow \left(\lim\sup_{T\to\infty}{R_1^{\pi}(T)\over \log(T)}\le c\right).
$$
\end{lemma}

\medskip
\bp Let $T>0$. By time $T$, we know that there have been at least $\lfloor T r_1\rfloor$ transmissions, but no more than $\lceil Tr_K\rceil$. Also observe that both regrets $R^{\pi}$ and $R^{\pi}_1$ are increasing functions of time. We deduce that:
$$
R^{\pi}(\lfloor Tr_1\rfloor)\le R_1^{\pi}(T) \le  R^{\pi}(\lceil Tr_K\rceil).
$$
Now 
\begin{align*}
\lim\inf_{T\to\infty}{R_1^{\pi}(T)\over \log(T)} & \ge \lim\inf_{T\to\infty}{R^{\pi}(\lfloor Tr_1\rfloor)\over \log(T)}\\
& = \lim\inf_{T\to\infty}{R^{\pi}(\lfloor Tr_1\rfloor)\over \log(\lfloor Tr_1\rfloor)} \ge c.
\end{align*}
The second statement can be derived similarly.
\ep
\medskip

\subsection{Structured MAB problem}

Instead of trying to solve (\ref{pb:1}), we rather focus on analyzing the MAB problem (\ref{pb:2}). We know that optimal algorithms for (\ref{pb:2}) will also be optimal for the original problem. As already mentioned, the specificity of our MAB problem lies in its structure, i.e., in the correlations and unimodality of the rewards obtained using different rates. Let us summarize the problem:

\medskip\noindent
{\bf $(P_U)$ structured MAB.} We have a set $\{1,\ldots,K\}$ of possible decisions. If decision $k$ is taken for the $i$-th time, we received a reward $r_kX_k(i)$. $(X_k(i),i=1,2,...)$ are i.i.d. with Bernoulli distribution with mean $\theta_k$. The structure of rewards across decisions are expressed through Assumptions 1 and 2, or equivalently $\theta\in {\cal T}\cap {\cal U}$. The objective is to design a decision scheme minimizing the regret $R^\pi(T)$ over all possible algorithms $\pi\in \Pi$.

\section{Lower bound on regret}\label{sec:low}

In this section, we derive an asymptotic (as $T$ grows large) lower bound of the regret $R^{\pi}(T)$ satisfied by any algorithm $\pi\in \Pi$. This lower bound provides an insightful theoretical performance limit for any rate adaptation scheme. We also quantify the performance gains achieved by algorithms that smartly exploit the structural properties of ($P_U$), compared to algorithms that consider the rewards obtained at various rates as independent.

\subsection{Structured MAB}

To derive a lower bound on regret for MAB problem ($P_U$), we first introduce the notion of uniformly good algorithms. An algorithm $\pi$ is uniformly good, if for all parameters $\theta$, for any $\alpha >0$, we have\footnote{$f(T)=o(g(T))$ means that $\lim_{T\to\infty} f(T)/g(T)=0$.}: $\mathbb{E}[t_k^{\pi}(T)]=o(T^\alpha), \forall k\neq k^{\star}$, where $t_k^{\pi}(T)$ is the number of times rate $r_k$ has been chosen up to time slot $T$, and $k^{\star}$ denotes the index of the optimal rate ($k^{\star}$ depends on $\theta$). Uniformly good algorithms exist as we shall see later on. We further introduce for any $k=1,\ldots,K$, the set $N(k)$ defined by:
$$
N(k)=\{l\in \{k-1,k+1\}: r_k\theta_k\le r_l\}.
$$
Finally, we introduce the Kullback-Leibler (KL) divergence, a well-known measure for dissimilarity between two distributions. In the case we compare two Bernoulli distributions with respective parameters $p$ and $q$, the KL divergence is: $I(p,q) = p \log \frac{p}{q} + (1-p) \log \frac{1-p}{1-q}.$ The following theorem, proved in appendix, provides a regret lower bound.

\begin{theorem}\label{th:lower_dep}
Let $\pi\in \Pi$ be a uniformly good rate selection algorithm for MAB problem ($P_U$). We have: $\liminf_{T\to\infty} {R^\pi(T)\over \log(T)} \ge c({\theta}),$
where 
$$
c({\theta})=\sum_{k\in N(k^\star)}{r_{k^\star}\theta_{k^\star} - r_k\theta_k\over I(\theta_k, \frac{ r_{k^{\star}}\theta_{k^{\star}}}{r_k})}.
$$
\end{theorem}

\subsection{The value of exploiting structural properties}

It is worth quantifying the performance gains one may achieve when designing rate selection algorithms that exploit the structural properties of the MAB problem ($P_U$). To this aim, we can derive the regret lower bound that one would obtain in absence of correlations and unimodal structure, i.e., assuming that the rewards obtained at the various rates are independent. The corresponding MAB would be defined as follows:

\medskip\noindent
{\bf $(P_{I})$ MAB with independent arms.} We have a set $\{1,\ldots,K\}$ of possible decisions. If decision $k$ is taken for the $i$-th time, we received a reward $r_kX_k(i)$. $(X_k(i),i=1,2,...)$ are i.i.d. with Bernoulli distribution with mean $\theta_k$, and independent across decisions $k$. The objective is to design a decision scheme minimizing the regret $R^\pi(T)$ over all possible algorithms $\pi\in \Pi$.

The regret lower bound can be derived using the same direct technique initially used by Lai and Robbins \cite{lai1985}. Define: $k_0 = \min\{k \in \{1,\dots,k^{\star}\}: \frac{r_{k^{\star}}\theta_{k^\star}}{r_k} \leq 1 \}.$ Note that if $k<k_0$, then $r_{k^{\star}}\theta_{k^{\star}} > r_k$, which means that even if all transmissions at rate $r_k$ were successful, i.e., $\theta_k=1$, rate $r_k$ would be sub-optimal. Hence, there is no need to select rate $r_k$ to discover this fact, since by only selecting rate $r_{k^{\star}}$, we get to know whether $r_{k^{\star}}\theta_{k^{\star}} > r_k\ge r_k\theta_k$. 

\begin{theorem}\label{th:lower_indep}
Let $\pi\in \Pi$ be a uniformly good rate selection algorithm for MAB problem ($P_I$). We have: $\liminf_{T\to\infty} {R^\pi(T)\over \log(T)} \ge c'({\theta})$,
where 
$$
c'({\theta}) = \sum_{k=k_0: k\ne k^\star}^K {r_{k^\star}\theta_{k^\star} - r_k\theta_k\over I(\theta_k, \frac{ r_{k^{\star}}\theta_{k^{\star}}}{r_k})}.
$$
\end{theorem}

For completeness, the proof of the previous theorem is presented in Appendix. It should be observed that the constant $c(\theta)$ involved in the regret lower bound when exploiting structural properties (e.g. unimodality) is the sum of at most two terms, irrespective of the number of available rates $K$. This is an important property that indicates that it might be possible to devise rate selection algorithm whose performance (at least asymptotically) does not get worst when the number of rates increases -- which basically happens each time a new 802.11 standard comes out. In contrast, the constant $c'(\theta)$ may consist of the sum of $K-3$ terms, and linearly increases with the number of possible rates. When the number of rates is large, we expect that algorithms exploiting the structural properties of the MAB problem significantly outperform those that do not exploit the structure.

\section{Asymptotically Optimal Rate Selection algorithm} \label{sec:algos}

In this section, we present ORS (Optimal Rate Sampling), a rate selection algorithm whose regret matches the lower bound derived in Theorem \ref{th:lower_dep} -- in other words, ORS algorithm is asymptotically optimal. 

\subsection{ORS algorithm}

The algorithm we propose is an extension of the recently proposed KL-UCB (Kullback-Leibler Upper Confidence Bound) algorithm \cite{garivier2011}. The latter is a variant of the classical UCB algorithm initially proposed by Auer et al. in \cite{auer2002}, and it has been shown to be optimal in the case where rewards are Bernoulli and independent across arms. One of the main contributions of the paper is to show that ORS optimally exploits the reward structure of our MAB problem.

We need the following notations to describe ORS algorithm. Let $t_k(n)$ be the number of times rate $r_k$ has been selected before time $n$ and $\hat{\mu}_k(n)$ denotes the empirical average of the reward obtained by rate $r_k$ so far:
$$
\hat{\mu}_k(n)={ \frac{1}{ t_k(n)} }\sum_{i=1}^{t_k(n)} r_k X_k(i).
$$
The {\it leader} $L(n)$ at time $n$ is the index of the rate with maximum empirical throughput: $L(n) \in \arg\max_k  \hat{\mu}_k(n).$ We further define $l_k(n)$ as the number of times that rate $k$ has been the leader up to time $n$: $l_k(n) = \sum_{n^\prime=1}^{n-1} \indic\{ L(n^\prime) = k\},$ where $\indic\{\cdot\}$ is the indicator function. Finally, for any $k\in \{1,\ldots,K\}$, we define the set ${\cal N}(k)$ as follows. If $2\le k\le K-1$,
$$
{\cal N}(k)=\{k-1,k,k+1\},
$$
and ${\cal N}(1)=\{1,2\}$, ${\cal N}(K)=\{K-1,K\}$.

The sequential decisions under ORS algorithm are based on the indexes of the various rates and can be easily implemented. The index $b_k(n)$ of rate $r_k$ for the $n$-th packet transmission is:
\begin{align*}
b_k(n)= & \max \Big\{ q \in [0,r_k] : t_k(n)I\big(\frac{\hat{\mu}_k(n)}{r_k}, \frac{q}{r_k} \big) \\
&\leq \log (l_{L(n)}(n))+c\log(\log (l_{L(n)}(n))) \Big\},
\end{align*}
where $c$ is a positive constant. For the $n$-th transmission, KL-U-UCB selects the rate close to the leader $L(n)$ and with maximum index. More precisely, it selects the rate in ${\cal N}(L(n))$ with maximum index. Ties are broken arbitrarily.

\begin{separation}
\vspace{-0.1cm}
    {\bf Algorithm 1} Optimal Rate Sampling (ORS)
\vspace{-0.5cm}\separator
\vspace{-0.2cm}
For $n = 1,\dots,K$, select the rate with index $k(n) = n$.
\newline For $n = K+1,\dots$, select the rate with index $k(n)$ where:
\vspace{0.2cm}
 $$ k(n) = \begin{cases}  L(n)  & \text{if } (l_{L(n)}(n) - 1)/3 \in \NN, \\
	  \displaystyle \arg\max_{k\in {\cal N}(L(n))} b_k(n) & \text{otherwise.}
	  \end{cases} $$
\vspace{-0.5cm}
\end{separation}

The next theorem, proved in appendix, states that the regret achieved under ORS algorithm matches the lower bound derived in Theorem \ref{th:lower_dep}. 

\medskip
\begin{theorem}\label{th:KL-UCB-R2}
Fix $\theta\in {\cal T}\cap{\cal U}$. For all $\epsilon > 0$, under ORS algorithm, the regret at time $T$ satisfies:
\begin{equation}\label{eq:regretan}
R^{ORS}(T) \leq  (1 + \epsilon) c(\theta) \log (T) + O(\log(\log(T))).
\end{equation}
As a consequence:
$$
\lim\sup_{T\to\infty}{R^{ORS}(T)\over \log(T)}\le c(\theta).
$$
\end{theorem}

The regret achieved under ORS algorithm is at most a sum of two terms, each of them corresponding to neighbors of the optimal rate $r_{k^\star}$. In particular, the regret does not depend on the number of available rates. 

\subsection{KL-R-UCB algorithm}

We conclude this section by presenting KL-R-UCB (R stands for Rate), a simple extension of KL-UCB algorithm \cite{garivier 2011}. This algorithm does not exploit the structural properties of our MAB problem, and is asymptotically optimal for MAB problem $(P_I)$. The regret analysis of this algorithm will be useful when considering non-stationary radio environments.

Under this algorithm, each rate $r_k$ is associated with an index $q_k(n)$ at time $n$ defined by:
$$
q_k(n) = \max\{q\in [0,r_k]: t_k(n)I({\hat{\mu}_k(n)\over r_k},{q\over r_k})\le \log(n)+c\log\log(n)\}.
$$
The algorithm selects the rate with highest index:
\begin{separation}
\vspace{-0.1cm}
    {\bf Algorithm 2} KL-R-UCB
\vspace{-0.5cm}\separator
\vspace{-0.2cm}
For $n = 1,\dots,K$, select the rate with index $k(n) = n$.
\newline For $n = K+1,\dots$, select the rate with index $k(n)$ where:\vspace{0.2cm}
 $$ k(n) \in \arg\max_k q_k(n).$$
\vspace{-0.5cm}
\end{separation}

The regret analysis of KL-R-UCB can be conducted as that of KL-UCB, and we have:
\begin{theorem}\label{th:KL-R-UCB}
Fix $\theta\in {\cal T}$. For all $\epsilon > 0$, under $\pi=$KL-R-UCB algorithm, the regret at time $T$ satisfies:
\begin{equation}\label{eq:regretan}
R^{\pi}(T) \leq  (1 + \epsilon) c'(\theta) \log (T) + O(\log(\log(T))).
\end{equation}
As a consequence:
$$
\lim\sup_{T\to\infty}{R^{\pi}(T)\over \log(T)}\le c'(\theta).
$$
\end{theorem}

\section{Non-stationary environments}\label{sec:nonstat}

We extend previous algorithms and results to non-stationary radio environments, i.e., to the case where the transmission success probabilities $\theta(t)$ at various rates evolve over time. Based on the ORS algorithm, we design SW-ORS (SW stands for Sliding Window), an algorithm that efficiently tracks the best rate for transmission in non-stationary environments, provided that the speed at which the parameters $\theta(t)$ evolve remains upper bounded.

We design algorithms for non-stationary versions of the MAB problem ($P_U$). In particular, to simplify the exposition, we assume that time is slotted, and that at the beginning of each slot, a rate is selected for transmission -- in other words, we consider the alternative system as discussed in Section \ref{section4}.  

\medskip\noindent
{\bf $(NS-P_U)$ Non-stationary structured MAB.} We have a set $\{1,\ldots,K\}$ of possible decisions. If decision $k$ is taken at time $t$, we receive a reward $r_kX_k(t)$. $(X_k(t),t=1,2,...)$ are independent with Bernoulli distribution with evolving mean $\theta_k(t)=\mathbb{E}[X_k(t)]$. The structure and evolution of $\theta(t)$ will be made precise in \textsection \ref{sec:ntMAB}. The objective is to design a sequential decision scheme minimizing the regret $R^\pi(T)$ over all possible algorithms $\pi\in \Pi$, where
$$
R^\pi(T) = \sum_{t=1}^T\left( r_{k^\star}(t)\theta_{k^\star(t)}(t) - r_{k^\pi(t)} \theta_{k^\pi(t)}(t)\right),
$$
and $k^\star(t)$ (resp. $k^\pi(t)$) denotes the best transmission rate (resp. the rate selected under $\pi$) at time $t$. $k^\star(t)=\arg\max_k r_{k(t)} \theta_{k(t)}(t)$.

\medskip
The above definition of the regret is not standard: the regret is exactly equal to $0$ only if the algorithm is aware of the best transmission rate at any time. This notion of regret really quantifies the ability of the algorithm $\pi$ to track the best rate for transmission. In particular, as shown in \cite{garivier08}, under some mild assumptions on the way $\theta(t)$ varies, we cannot expect to obtain a regret that scales sublinearly with the time horizon $T$. The regret is linear, and our objective is to minimize the regret per unit time $R^\pi(T)/T$.

\subsection{The SW-ORS Algorithm}

MAB problems with non-stationary rewards have received little attention so far, but a natural and efficient way of tracking the changes of $\theta(t)$ over time is to select the rate at time $t$ based on observations made over a fixed time window preceding $t$, i.e., to account for transmissions that occurred between time $t-\tau$ and $t-1$, see e.g. \cite{garivier08}. The size $\tau$ of the time window is chosen empirically: it must be large enough (to be able to learn), but small enough so that the channel conditions do not vary significantly during a period of duration $\tau$.

The SW-ORS algorithm naturally extends ORS to non-stationary environments: it mimics the selections made under ORS, but based on the observations made over a sliding time window. We now provide a formal description of SW-ORS. Let $k(t)$ denote the index of the rate selected at time $t$. The empirical average reward of rate $r_k$ at time $n$ over a window of size $\tau+1$ is:
$$
\hat\mu_{k}^\tau(n) = \frac{1}{t_k^\tau(n)} \sum_{t = n-\tau }^{n-1}  r_k X_k(t) \indic \{k(t) = k \},
$$
where
$$
t_k^\tau(n) =  \sum_{t = n-\tau }^{n-1} \indic \{k(t) = k \}.
$$
By convention, $\hat\mu_k^\tau(n) = 0$ if $t_k^\tau(n) = 0$. Based on $\hat\mu_k^\tau(n)$, we can redefine as previously $L^\tau(n)$, the leader at time $n$, $l_k^\tau(n) = \sum_{t = n-\tau }^{n-1} \indic \{L^\tau(t) = k \}$, the number of times $k$ has been the leader over the window $\tau$ preceding $n$, and $b_k^\tau(n)$, the index of decision $k$ at time $n$. $b_k^\tau(n)$ is defined by:
\als{
b_k^\tau(n)= & \max \Big\{ q \in [0,r_k] : t_k^\tau(n)I\big(\frac{\hat{\mu}_k^\tau(n)}{r_k}, \frac{q}{r_k} \big) \\
&\leq \log (l_{L^\tau(n)}^\tau(n))+c\log(\log (l_{L^\tau(n)}^\tau(n))) \Big\}.
}
\begin{separation}
\vspace{-0.1cm}
    {\bf Algorithm 3} SW-ORS with window size $\tau$
\vspace{-0.5cm}\separator
\vspace{-0.2cm}
For $n = 1,\dots,K$, select the rate with index $k(n) = n$.
\newline For $n = K+1,\dots$, select the rate with index $k(n)$ where:
\vspace{0.2cm}
 $$ k(n) = \begin{cases}  L^\tau(n)  & \text{if } (l_{L^\tau(n)}^\tau(n) - 1)/3 \in \NN, \\
	  \displaystyle \arg\max_{k \in {\cal N}(L^\tau(n))} b_k^\tau(n) & \text{otherwise.}
	  \end{cases} $$
\vspace{-0.5cm}
\end{separation}

\subsection{Slowly varying environment}\label{sec:ntMAB}

We are interested in characterizing the regret per time unit achieved by SW-ORS in scenarios where the speed at which $\theta(t)$ evolves is controlled (this evolution has to be relatively slow to be able to design algorithms that track the best rate for transmission). Hence we make the following assumption, stating that $\theta(t)$ is a Lipschitz function.

\begin{assumption}\label{as:lip}
$t\mapsto \theta(t)$ is $\sigma$-Lipschitz, for some $\sigma>0$: for any $k=1,\ldots, K$, for any $t,t'$:
$$
|\theta_k(t')-\theta_k(t)| \le \sigma |t'-t|.
$$
\end{assumption}

Since the optimal rate changes over time, we cannot expect that the average rewards at various rates are always well separated and that strict unimodality always holds as stated in Assumption \ref{as:uni}. However, we still assume that the unimodal structure is preserved, and make assumptions about the periods of time where different rates yield very similar throughputs. More precisely, in the following analysis, we shall use one of the two following assumptions. Let $\overline{\cal U}$ be the smallest closed set containing ${\cal U}$: $\overline{\cal U}=\{\eta\in [0,1]^K:\exists k^\star, r_1\eta_1\le \ldots \le r_{k^\star}\eta_{k^\star}, r_{k^\star}\eta_{k^\star}\ge r_{k^\star+1}\eta_{k^\star+1}\ge \ldots\ge r_k\eta_K\}$.

\begin{assumption}\label{assum:uniclosed}
At any time $t$, $\theta(t)\in \overline{\cal U}$.
\end{assumption}

The performance of bandit algorithms in non-stationary environments will depend on the proportion of time where various decisions yield throughputs close to each other. Indeed, the closer they are, the harder it is to differentiate them with high probability. We define
$$
H(\Delta,T) =  \sum_{n=1}^{T} \sum_{k =1}^{K} \sum_{ k^{\prime} \in N(k)} \indic \{  | r_k \theta_k(n) - r_{k^\prime} \theta_{k^\prime}(n) |   \geq \Delta  \}.
$$
$H(\Delta,T)$ is the number of time instants at which there is a decision which is not well separated from one of its neighbors. $H$ depends on the evolution of the success transmission probabilities, i.e., on $n \mapsto ( \theta_k(n) )_{1 \leq k \leq K}$. We are interested in the performance of the algorithm in a \emph{slowly} changing environment, so that we will let $\sigma \to 0^+$. In this case, it is natural to choose a large window size $\tau \to +\infty$. We are interested in how the regret per unit of time scales with $\sigma$ in the regime $\sigma , \Delta \to 0^+$ , $\tau \to +\infty$. We consider the following assumption on $H$. 
	
\begin{assumption}\label{assum:separation2}
There exists a function $\Phi(K) \leq K$ such that for all $\Delta > 0$ (or for all $\Delta$ small enough):
$$
\lim \sup_{T \to +\infty} \frac{H(\Delta,T)}{T} \leq \frac{\Phi(K) \Delta}{r_1}.
$$
\end{assumption}
 
Assumption~\ref{assum:separation2} is satisfied for instance if for all $k$, $n \mapsto \theta_k(n)$ is a stationary ergodic process with uniform stationary distribution on $[0,1]$ (in this case, we can select $\Phi(K) = K$).

\subsection{The SW-KL-R-UCB algorithm and its regret}

Before analysing the regret achieved under SW-ORS, we introduce and study the SW-KL-R-UCB algorithm, a version of KL-R-UCB adapted to non-stationary environments using a sliding window. We derive an upper bound of the regret of SW-KL-R-UCB, and in the next subsection, we use this bound to analyse the regret of SW-ORS.

To define SW-KL-R-UCB, we use $\hat\mu_{k}^\tau(n)$ and $t_{k}^\tau(n)$ as defined earlier. Each rate $r_k$ is associated with an index $q_k^{\tau}(n)$ at time $n$ defined by:
$$
q_k^{\tau}(n) = \max\{q\in [0,r_k]: t_k^{\tau}(n)I({\hat{\mu}_k^{\tau}(n)\over r_k},{q\over r_k})\le \log(\tau)+c\log\log(\tau)\}.
$$
	
\begin{separation}
\vspace{-0.1cm}
    {\bf Algorithm 4} SW-KL-R-UCB with window size $\tau$
\vspace{-0.5cm}\separator
\vspace{-0.2cm}
For $n = 1,\dots,K$, select the rate with index $k(n) = n$.
\newline For $n = K+1,\dots$, select the rate with index $k(n)$ where:
\vspace{0.2cm}
  $$k(n) =  \arg\max_{1 \leq k \leq K} q_k^{\tau}(n).$$
\vspace{-0.5cm}
\end{separation}

Define
\begin{align*}		
G(T,I_{\min},\tau,\sigma) =  \sum_{n=1}^{T} \sum_{k \neq k^{\star}(n)} & \indic \{  r_k (\theta_k(n) + \tau \sigma) \geq  r_{k^\star(n)} (\theta_{k^\star}(n) - \tau \sigma)  \} \sk & \cup  \{    I(\theta_k(n) + \tau \sigma,  r_{k^\star(n)}(\theta_{k^\star}(n) - \tau \sigma)/r_k  \})  > I_{\min} \}.
\end{align*}
$G(T,I_{\min},\tau,\sigma)$ is the number of time instants up to time $T$ at which there exists at least a sub-optimal decision $k \neq k^\star(n)$ which cannot be distinguished from $k^\star(n)$ well enough. Namely, either the difference between the average reward of $k$ and $k^\star(n)$ is smaller than the error caused by the changing environment, or the KL-divergence number between them is smaller than a fixed threshold $I_{\min}$. Applying Pinkser's inequality (see Appendix), we have:
$$
G(T,I_{min},\tau,\sigma) \leq H(\sqrt{I_{min}/2} + 2 r_K \tau \sigma,T).
$$
The next theorem provides an upper bound of the regret of KL-R-UCB in a slowly changing environment. 

\begin{theorem}\label{th:KLUCBchanging}
Under Assumption \ref{as:lip}, the regret of $\pi=$SW-KL-R-UCB per unit of time satisfies: for all $I_{\min} > 0$ and $\epsilon > 0$,  
\als{	
\frac{R^{\pi}(T)}{T}  &\leq  r_K \Lp \sqrt{I_{\min}/2} + 2 \tau \sigma \Rp \frac{G(T,I_{\min},\tau,\sigma)}{T}  +  K r_K (1 + \epsilon) \frac{\log(\tau) + c \log(\log(\tau)) }{\tau I_{\min}} \sk
&+  \frac{C K}{(\tau \log(\tau)^c)^{g_0 \epsilon^2}},
}
where $C$ is constant, and
	\als{
	g_0 &=  \frac{1}{2} \min_{1 \leq n \leq T}  \min_{ \stackrel{k \neq k^{\star}(n),}{r_{k^\star(n)}( \theta_{k^\star}(n) - \sigma \tau) < r_k} }  \frac{r_{k^\star(n)}}{r_k}( \theta_{k^\star}(n) - \sigma \tau) \Lb 1 - \frac{r_{k^\star(n)}}{r_k}( \theta_{k^\star}(n) - \sigma \tau)  \Rb. 
	}
\end{theorem}

The following regret upper bound is obtained from the above theorem by choosing $I_{\min} = 2(\Delta - 2 r_K \tau \sigma)^2/r_K^2$ and $\epsilon = g_0^{-1/2}$, and using the fact that $G(T,I_{min},\tau,\sigma) \leq H(\sqrt{I_{min}/2} + 2 r_K \tau \sigma,T)$.

\begin{corollary}\label{cor:2}
Let $\Delta > 2 r_K \tau \sigma$. The regret per unit of time of $\pi=$SW-KL-R-UCB is upper bounded by: 
\als{	
\frac{R^{\pi}(T)}{T} &\leq  \frac{\Delta H(\Delta,T)}{T} +  K r_K^3 \Lp 1 + g_0^{-1/2} \Rp \frac{\log(\tau) + c \log(\log(\tau)) + C}{2 \tau (\Delta - 2 r_K \tau \sigma)^2},
}
where $C>0$ is a constant.
\end{corollary}
	
Finally, using Assumption~\ref{assum:separation2}, and Corollary \ref{cor:2}, we get a simple asymptotic upper bound for the regret of SW-KL-R-UCB:

\begin{corollary}\label{prop:klucbregret2}
Under Assumptions \ref{as:lip} and \ref{assum:separation2}, if $\tau = (K \sigma/\Phi(K))^{-4/5} / 4$, the regret per unit of time of $\pi=$SW-KL-R-UCB satisfies:	
\eqs{
\lim \sup_{T\to\infty} \frac{R^{\pi}(T)}{T}  \leq C \Phi(K) \Lp \frac{ K \sigma}{\Phi(K)} \Rp ^{\frac{2}{5}} \log(1 /\sigma),
}
where $C>0$ is a constant.
\end{corollary}

The previous result is proved using the bound derived in Corollary \ref{cor:2} with $\Delta = r_K (K \sigma/\Phi(K))^{1/5}$. 

\subsection{Upper bound of the regret under SW-ORS}

To derive an upper bound on the regret under SW-ORS, we first show that the latter can be expressed using the regret achieved under SW-KL-R-UCB.

\begin{theorem}\label{th:ORSchanging}
Under Assumptions \ref{as:lip} and \ref{assum:uniclosed}, for all $\Delta > 0$ and $\tau$ such that $\Delta > 4 r_K \tau \sigma$, the regret of SW-ORS per unit of time satisfies:
\als{	
R^{\text{SW-ORS}}(T)  &\leq \frac{2 R^{\pi}(T)}{K}  +  r_K H(\Delta,T) +   \frac{C_1 K T \log(\tau)}{\tau (\Delta - 4 r_K \tau \sigma)^2},
}
where $C_1>0$ is a constant, and $\pi$=SW-KL-R-UCB.
\end{theorem}

Choosing $\Delta = r_K \sigma^{1/4} \log(1/\sigma)$ in Theorem \ref{th:ORSchanging}, we deduce the following asymptotic regret upper bound for SW-ORS.
 
\begin{corollary}\label{prop:klUucbregret2}
Under Assumptions \ref{as:lip}-\ref{assum:uniclosed}-\ref{assum:separation2}, if $\tau = \sigma^{-3/4} \log(1/\sigma) / 8 $, then
\eqs{
\lim \sup_{T\to\infty} \frac{R^{\pi}(T)}{T}  \leq C \Phi(K) \sigma^{\frac{1}{4}} \log \Lp \frac{1}{\sigma} \Rp ( 1 + K o(1) ) \;,\; \sigma \to 0^{+},
}
for some constant $C>0$.
\end{corollary}

Note that $\sigma^{1/4}\log(1/\sigma)$ tends to 0 as $\sigma\to 0$, which indicates that the regret per unit time vanishes when we slow down the evolution of $\theta(t)$, i.e., SW-ORS tracks the best transmission rate if $\theta(t)$ evolves slowly. Also observe that the performance guarantee on SW-ORS depends on the size $K$ of the decision space only through $\Phi(K)$.

\section{Extension to MIMO Systems} \label{sec:mimo}

This section deals with MIMO systems, and extends all previous results and algorithms to these systems. In MIMO, the transmitter has the possibility of using several antennas to transmit several data streams in parallel. The maximal number of streams that can be transmitted in parallel is equal to the minimum between the number of transmit and receive antennas. We use the term ``mode'' to denote the number of streams to be transmitted i.e single steam mode, dual stream mode etc.

	We leverage structural properties of the achieved throughput as a function of the selected (mode, rate) pair, to propose an optimal sequential (mode, rate) selection scheme. Subsections \ref{sec:mimomodels} to \ref{sec:gors} deal with stationary radio environments. The extension to non-stationary environments is presented in subsection~\ref{nonstat2}. All results presented in this section are straightforward extensions of results derived in Sections \ref{section4}-\ref{sec:nonstat}, and their proofs are left to the reader.

\subsection{Model}\label{sec:mimomodels}  

In MIMO systems, the transmitter has to select, for each transmission, both a mode and a rate. As before the set of available rates is denoted by ${\cal R}$. The set of modes is ${\cal M}=\{1,\ldots,M\}$. Hence the set of possible decisions is ${\cal D}={\cal M}\times {\cal R}$. Let $D=|{\cal D}|$. With a slight abuse of terminology, we say that the (mode, rate) pair $(m,k)$ is selected  if the mode $m$ and rate $r_k$ are used. For $d\in {\cal D}$, we define the corresponding mode and rate as $m(d)$ and $r(d)$, respectively, i.e., $d=(m(d),r(d))$. Again, after each transmission, the transmitter is informed on whether the corresponding packet has been successfully received. Based on the observed past transmission successes and failures at the various (mode, rate) pairs, the transmitter has to select a (mode, rate) pair for the next packet transmission. We denote by $\Pi$ the set of all possible sequential (mode, rate) selection schemes.

For the $i$-th packet transmitted using (mode, pair) $d=(m,k)$, a binary random variable $X_d(i)$ represents the success or failure of the transmission. For any decision $d$, the r.v.s $(X_d(i), i=1, 2, \ldots)$ are i.i.d. Bernoulli with mean $\theta_d$. The expected reward of decision $d$ is denoted by $\mu_d=r(d)\theta_d$. To simplify the presentation, we assume that the best decision $d^\star$ is unique: $d^\star = \arg\max_{d\in {\cal D}} \mu_d$, and we define $\mu^\star=\mu_{d^\star}$.

\subsection{Correlations and Graphical Unimodality}

To design efficient (mode, rate) selection algorithms, we exploit structural properties of the throughputs obtained selecting different (mode, rate) pairs: (i) The successes and failures of transmissions at various rates are correlated; (ii) the throughput satisfies a property referred to as {\it graphical unimodality}. 
 
\subsubsection{Correlations} 

If a transmission is successful at a high rate at a given mode, it has to be successful at a lower rate using the same mode, and similarly, if a low-rate transmission fails, then a transmitting at a higher rate would also fail. Formally assume that at a given time, (mode, rate) pair $(m,k)$ (resp. rate $(m,l)$) has already been selected $(i-1)$ (resp. $(j-1)$) times. Then:
\begin{equation}\label{eq:dep1}
\left( k<l \hbox{ and } X_{(m,k)}(i)=0\right) \Longrightarrow (X_{(m,l)}(j)=0),
\end{equation}
\begin{equation}\label{eq:dep2}
\left( k>l \hbox{ and } X_{(m,k)}(i)=1\right) \Longrightarrow (X_{(m,l)}(j)=1).
\end{equation}
Again using simple coupling arguments, we can readily show that an equivalent way of expressing the correlations is to state that the following assumption holds.

\begin{assumption}\label{as:corr}
$\theta=(\theta_d,d\in {\cal D})\in {\cal T}'$, where ${\cal T}'=\{\eta\in [0,1]^D: \eta_{(m,k)}\ge \eta_{(m,l)},\forall m, \forall k<l\}$.
\end{assumption}

\subsubsection{Graphical Unimodality} 

Graphical unimodality extends the notion of unimodality to the case where decisions are the vertices of an undirected graph $G=({\cal D},E)$. When $(d,d')\in E$, we say that the two decisions $d$ and $d'$ are neighbors. Define ${\cal N}(d)=\{d'\in {\cal D}: (d,d')\in E\}$ as the set of neighbors of $m$. Graphical unimodality expresses the fact that when the optimal decision is $d^\star$, then for any $d\in {\cal D}$, there exists a path in $G$ from $d$ to $d^\star$ along which the expected reward is increased. In other words there is no {\it local} maximum in terms of expected reward except at $d^\star$, where the notion of locality is defined through that of neighborhood ${\cal N}(d), d\in {\cal D}$. Formally: 

\begin{assumption}\label{as:graph}
$\theta\in {\cal U}_G$, where ${\cal U}_G$ is the set of parameters $\theta\in [0,1]^D$ such that, if $d^\star=\arg\max_d r(d)\theta_d$, for any $d\in {\cal D}$, there exists a path $(d_0=d,d_1,\ldots,d_p=d^\star)$ in $G$ such that for any $i=1,\ldots,p$, $\mu_{d_i} > \mu_{d_{i-1}}$.
\end{assumption}

\begin{figure}[htb]
\begin{center}
\includegraphics[width=10.2cm]{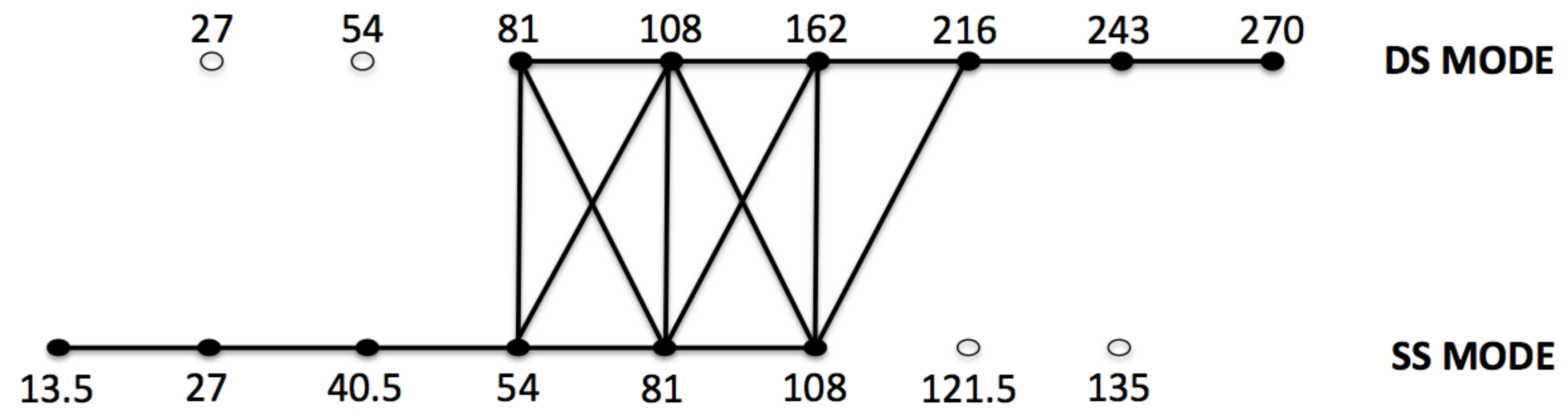} 
\end{center}
\vspace{-3mm}
\caption{Graph providing unimodality in MIMO 802.11n systems. Rates are in Mbit/s, two MIMO modes are considered, single-stream (SS) and double-stream (DS) modes.}
\label{fig2}

\end{figure}

Note that as considered earlier, when modes do not exist, unimodality of the mean rewards is a particular case of graphical unimodality where $G=({\cal R},E)$ and $E=\{(r_1,r_2),\ldots,(r_{K-1},r_K) \}$. In practice for 802.11n MIMO systems, as discussed in more details in Section \ref{sec:num}, we can find a graph $G$ such that the throughput or average reward obtained at various (mode, rate) pairs is graphically unimodal with respect to $G$. Such a graph is presented in Figure \ref{fig2}. It has been constructed exploiting various observations and empirical results from the literature. First, for a given mode (SS or DS), the throughput is unimodal in the rate. Then, when the SNR is relatively low, it has been observed that using SS mode is always better than using DS mode; this explains why for example, the (mode, rate) pair (SS,13.5) has no neighbor in the DS mode. Similarly, when the SNR is very high, then it is always optimal to use DS mode. Finally when the SNR is neither low nor high, there is no clear choice between SS and DS mode, which explains why we need links between the two modes in the graph. 
 
\medskip

\subsection{Graphical Unimodal MAB}

As earlier, we consider an alternative system, where the transmission of a packet is assumed to last for exactly one slot, see Section \ref{section4}. The regret of an algorithm $\pi$ up to slot $T$ is then defined as:
$$
R^\pi(T) = \mu^\star T - \sum_{d\in {\cal D}} \mu_d\mathbb{E}[t_d^\pi(T)],
$$
where $t_d^\pi(T)$ denotes the number of times decision $d$ has been taken up to slot $T$. Applying the same arguments as those used in Section \ref{section4}, we know that an asymptotically optimal sequential decision algorithm is also asymptotically optimal in the original system (where the time it takes to transmit a packet depends on the selected rate). 

\medskip\noindent
{\bf $(P_G)$ Graphically Unimodal MAB.} We have a set ${\cal D}$ of possible decisions. If decision $d$ is taken for the $i$-th time, we received a reward $r(d)X_d(i)$. $(X_d(i),i=1,2,...)$ are i.i.d. with Bernoulli distribution with mean $\theta_d$. The structure of rewards across decisions are expressed through Assumptions \ref{as:corr} and \ref{as:graph}, or equivalently $\theta\in {\cal T}'\cap {\cal U}_G$ for some graph $G$. The objective is to design a decision scheme minimizing the regret $R^\pi(T)$ over all possible algorithms $\pi\in \Pi$.

\subsection{Lower bound on regret}

	To state the lower bound on regret, we define the following sets: for any $d\in {\cal D}$, $N(d)=\{d'\in {\cal N}(d): r(d)\theta_{d}\le r(d')\}$.

\medskip
\begin{theorem}\label{th:lowerGraph}
Let $\pi\in \Pi$ a uniformly good sequential decision algorithm for the MAB problem $(P_G)$. We have:
$$
\lim\sup_{T\to\infty}{R^\pi(T)\over \log(T)}\ge c_G(\theta),
$$
where
$$
c_G(\theta) = \sum_{d\in N(d^\star)} {r(d^\star)\theta_{d^\star}- r(d)\theta_d\over I(\theta_d,{r(d^\star)\theta_{d^\star}\over r(d)}) }.
$$
\end{theorem}

Observe that the lower bound on the regret depends on the graph $G$, and is proportional to the number of neighbors in $G$ of the optimal decision $d^\star$. To achieve a regret as small as possible, it is therefore important to design a graph $G$ as sparse as possible while preserving the graphical unimodality property. In practice, we can build a graph $G$ for 802.11n MIMO systems with small degree, irrespective of the number of available (mode, rate) pairs.  

\subsection{G-ORS Algorithm}\label{sec:gors}

Next we devise a (mode, rate) pair selection algorithm, referred to as G-ORS ("G" stand for "Graphical") whose regret matches the regret lower bound derived in Theorem \ref{th:lowerGraph} in the case of stationary environments. This algorithm is an extension of the ORS algorithm. As before, $\hat\mu_d(n)$ denotes the empirical average reward obtained using decision $d$ up to slot $n$. The leader $L(n)$ at slot $n$ is the decision with maximum empirical average reward. Further define $l_d(n)$ as the number of times up to slot $n$ that decision $d$ has been the leader. Finally, let $\gamma$ the maximum degree of a vertex in $G$. 

Under G-ORS algorithm, the index $b_d(n)$ of decision $d$ in slot $n$ is given by:
\begin{align*}
b_d(n)= & \max \Big\{ q \in [0,r(d)] : t_d(n)I\big(\frac{\hat{\mu}_d(n)}{r(d)}, \frac{q}{r(d)} \big) \\
&\leq \log (l_{L(n)}(n))+c\log(\log (l_{L(n)}(n))) \Big\},
\end{align*}
where $c$ is a positive constant. For the $n$-th slot, G-ORS selects the decision in ${\cal N}(L(n))$ with maximum index. Ties are broken arbitrarily.

\begin{separation}
\vspace{-0.1cm}
    {\bf Algorithm 5} G-ORS
\vspace{-0.5cm}\separator
\vspace{-0.2cm}

Select all decisions $d\in {\cal D}$ once.

For $n = D+1,\dots$, select decision $d(n)$ where:
\vspace{0.2cm}
 $$ d(n) = \begin{cases}  L(n)  & \text{if } (l_{L(n)}(n) - 1)/\gamma \in \NN, \\
	  \displaystyle \arg\max_{d \in {\cal N}(L(n))} b_d(n) & \text{otherwise.}
	  \end{cases} $$
\vspace{0.3cm}
\end{separation}

The next theorem states that the regret achieved under the ORS algorithm matches the lower bound derived in Theorem \ref{th:lowerGraph}. 

\medskip
\begin{theorem}\label{th:GORS2}
Fix $\theta\in {\cal T}'\cap{\cal U}_G$. For all $\epsilon > 0$, under algorithm $\pi=$ G-ORS, the regret at time $T$ is bounded by:
$$
R^{\pi}(T) \leq  (1 + \epsilon) c_G(\theta) \log (T) + O(\log(\log(T))).
$$
As a consequence:
$$
\lim\sup_{T\to\infty}{R^\pi(T)\over \log(T)}\le c_G(\theta).
$$
\end{theorem}

\subsection{Non-stationary environments: SW-G-ORS Algorithm}\label{nonstat2}

	We consider a non stationary environment where parameter evolve in a Lipschitz manner as described in section~\ref{sec:nonstat}. We introduce SW-G-ORS which is a straightforward extension of G-ORS to non stationary environments using a finite window of size $\tau$. The definition of  $t_d^\tau(n)$ $\hat\mu_d^\tau(n)$ and $l_d^\tau(n)$ and $b_d^\tau(n)$ remain unchanged.

\begin{separation}
\vspace{-0.1cm}
    {\bf Algorithm 5} SW-G-ORS
\vspace{-0.5cm}\separator
\vspace{-0.2cm}

For $n = 1,\dots,D$, select decision $d(n)=n$.

For $n = D+1,\dots$, select decision $d(n)$ where:
\vspace{0.2cm}
 $$ d(n) = \begin{cases}  L^{\tau}(n)  & \text{if } (l_{L^{\tau}(n)}^{\tau}(n) - 1)/\gamma \in \NN, \\
	  \displaystyle \arg\max_{d \in {\cal N}(L^{\tau}(n))} b^{\tau}_d(n) & \text{otherwise.}
	  \end{cases} $$
\vspace{0.3cm}
\end{separation}

The regret per unit of time of SW-G-ORS is given by proposition~\ref{prop:SWGORSregret}.
\begin{proposition}\label{prop:SWGORSregret}
	Consider assumption~\ref{assum:separation2}, and parameters $\Delta = r_K \sigma^{1/4} \log(1/\sigma)$ and \newline $\tau = \sigma^{-3/4} \log(1/\sigma) / 8 $. Then there exists a constant $C > 0$ independent of $D$ such that the regret per unit of time of SW-ORS is upper bounded by:	
	\eqs{
	 \lim \sup_{T} \frac{R^{\pi}(T)}{T}  \leq C \Phi(D) \sigma^{\frac{1}{4}} \log \Lp \frac{1}{\sigma} \Rp ( 1 + D o(1) ) \;,\; \sigma \to 0^{+}. 
	}
\end{proposition}

\section{Numerical Experiments}\label{sec:num}

In this section, we illustrate the efficiency of our algorithms using traces that are either artificially generated or extracted from test-beds. Artificial traces allow us to build a performance benchmark including various kinds of radio channel scenarios as those used in \cite{bicket2005bit}. They also provide the opportunity to create non-stationary radio environments (in the literature, RA mechanisms are mostly evaluated in stationary environments).

\subsection{802.11g systems}

\subsubsection{Artificial traces} 
\begin{figure}[t]
  \centering
  \subfigure[Steep]{
   \label{fig:steep}
  \includegraphics*[width=0.48\columnwidth ]{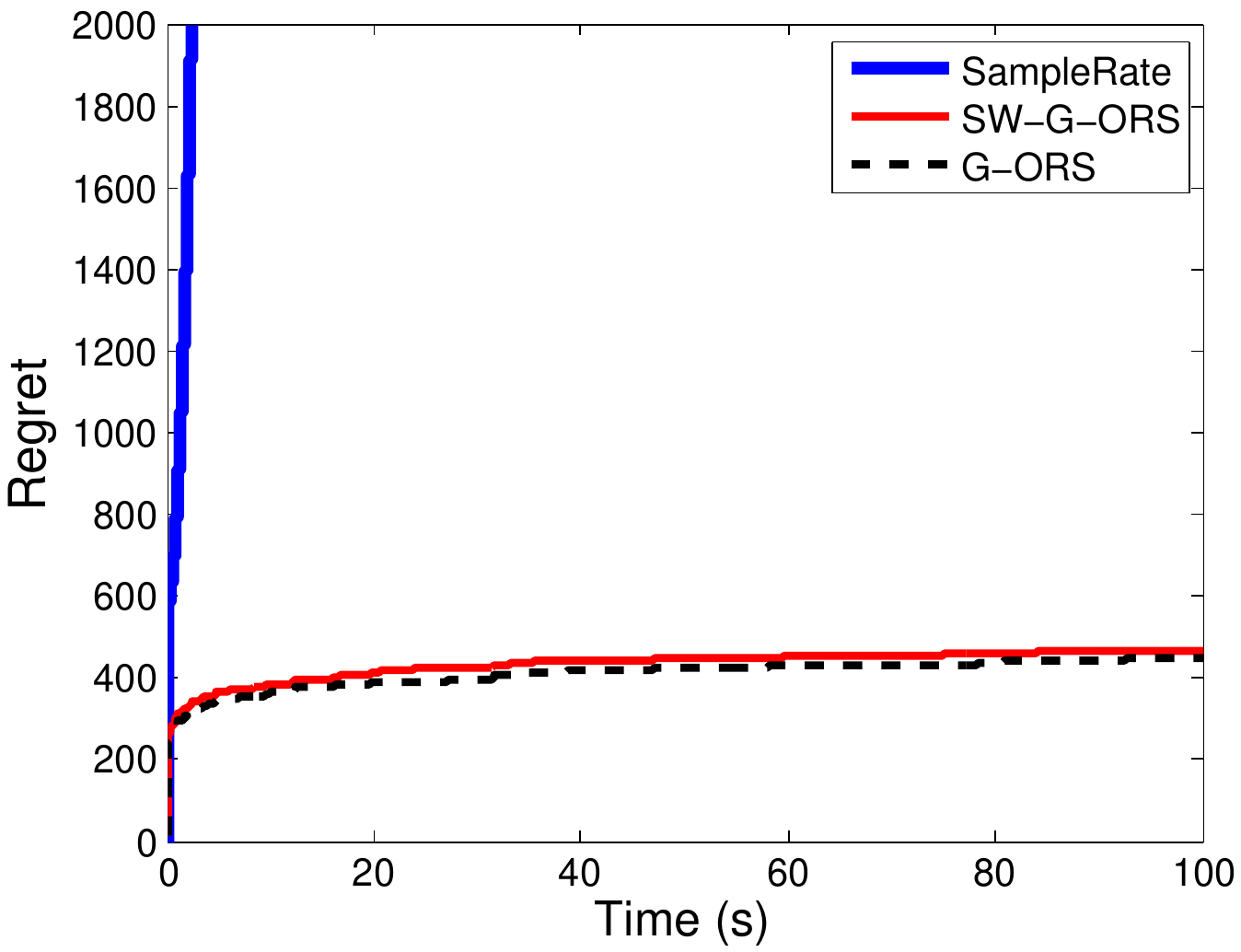}} \hspace{-4mm}
  \subfigure[Steep (zoom)]{
  \label{fig:steep_all}
  \includegraphics*[width=0.48\columnwidth ]{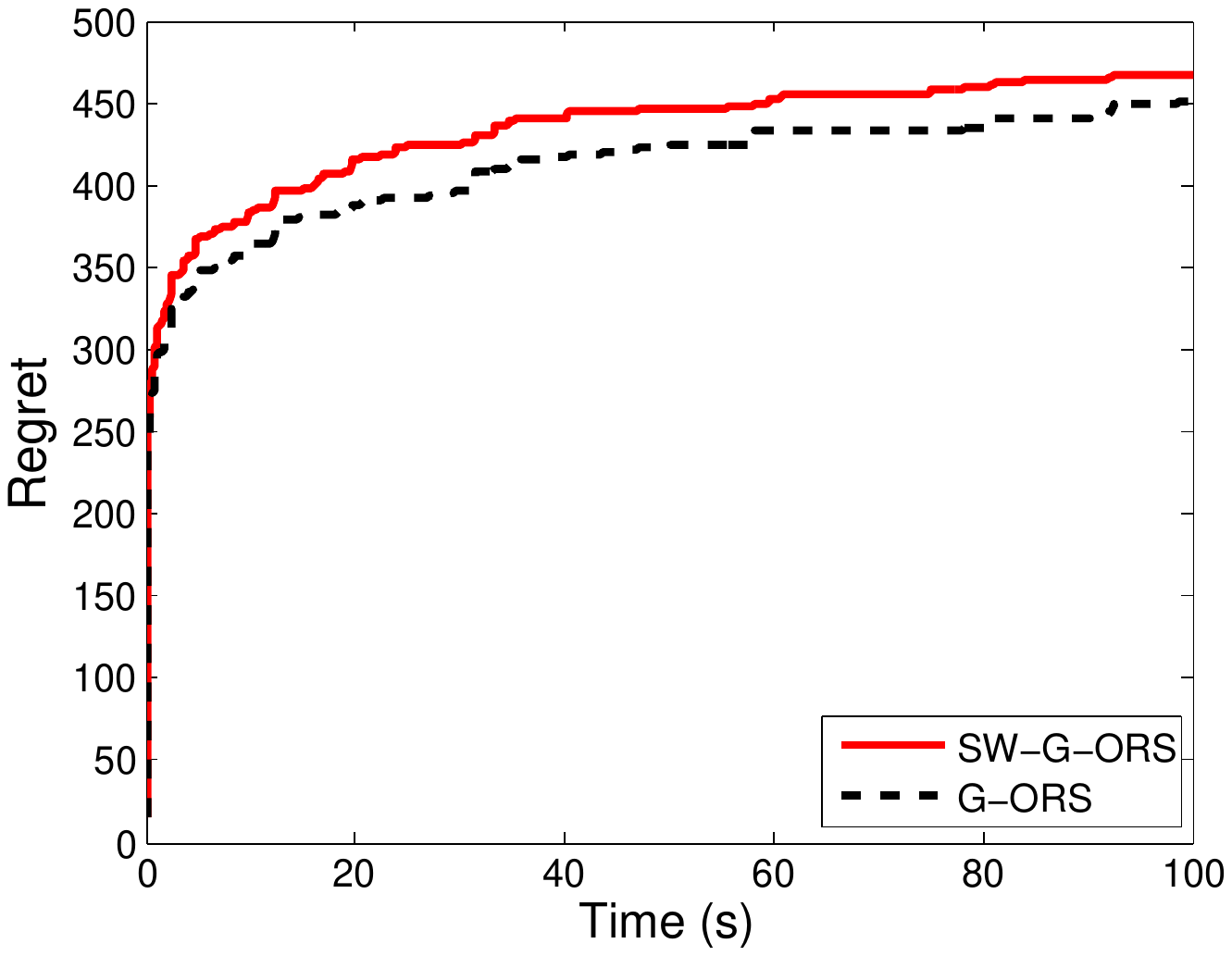}} \hspace{-4mm}
 \subfigure[Gradual]{
  \label{fig:gradual}
  \includegraphics*[width=0.48\columnwidth ]{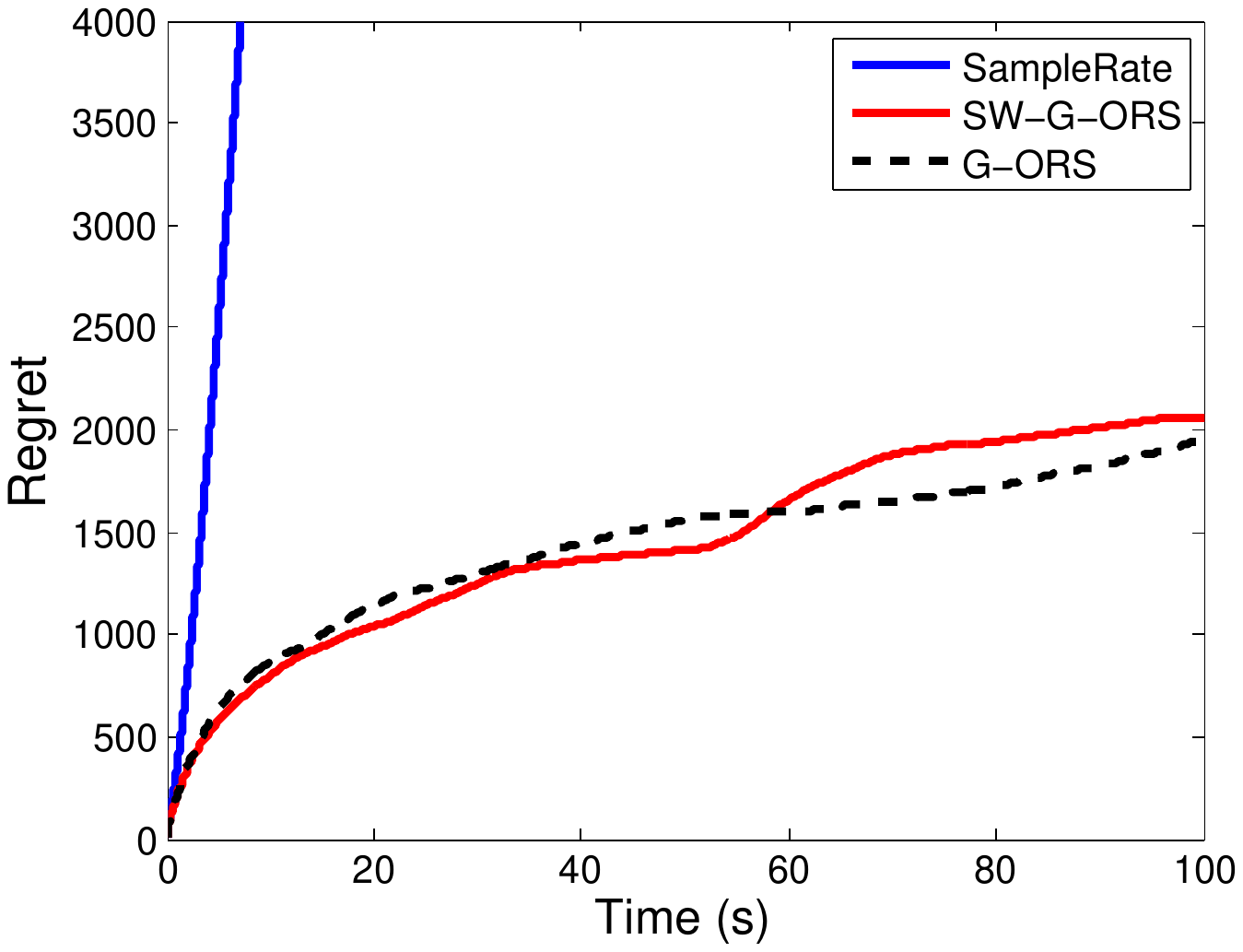}} \hspace{-4mm}
  \subfigure[Lossy]{
  \label{fig:lossy}
  \includegraphics*[width=0.48\columnwidth ]{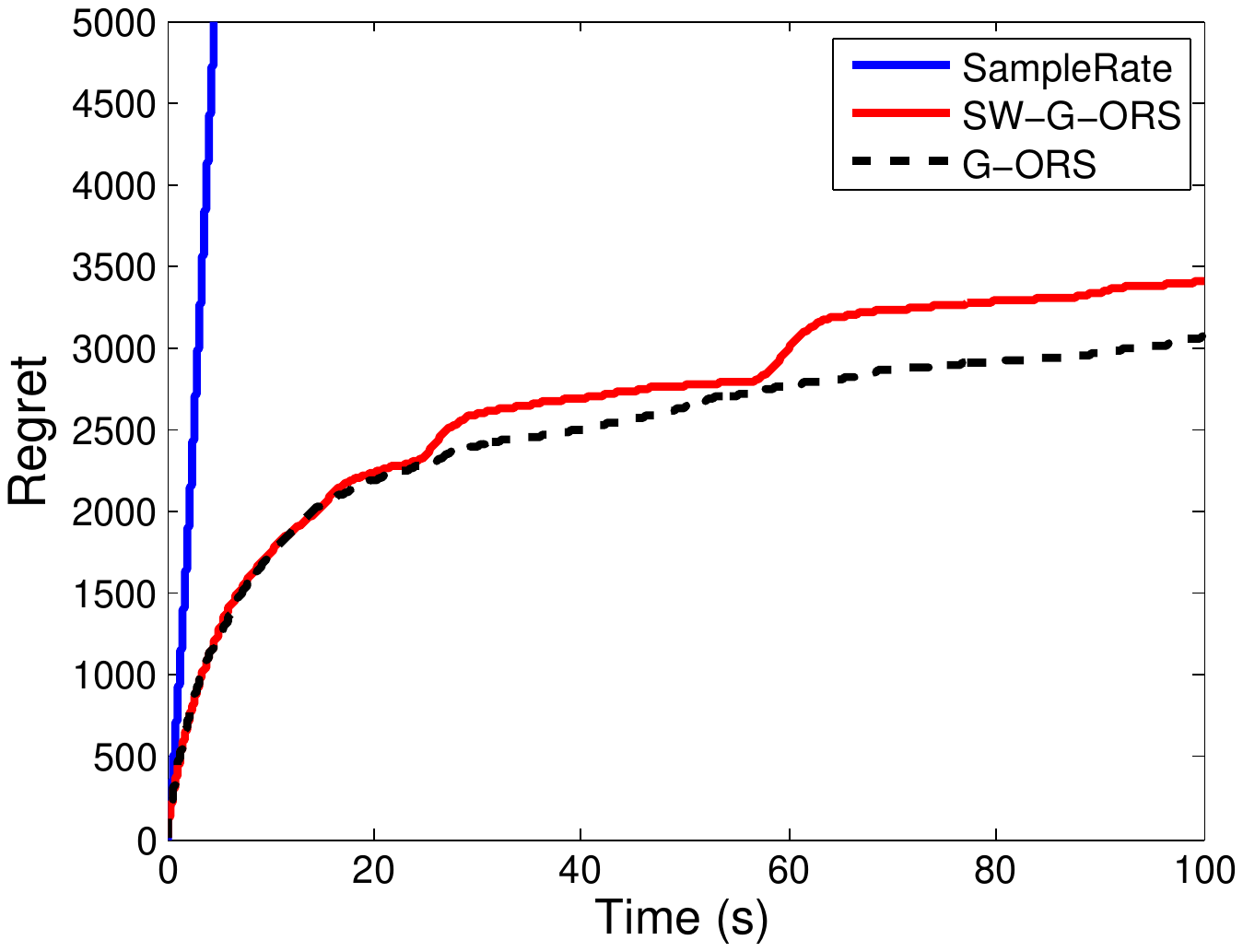}}
\vspace{-3mm}
\caption{Regret vs. time in stationary environments under SampleRate, G-ORS, and SW-G-ORS.}  
  \label{fig:stationary}
  \vspace{-0.5cm}
\end{figure}

We first consider 802.11g with 8 available rates from 6 to 54 Mbit/s. Algorithms are tested in three different scenarios as in \cite{bicket2005bit}: {\it steep}, {\it gradual}, and {\it lossy}. In steep scenarios, the successful transmission probability is either very high or very low. In gradual scenarios, the best rate is the highest rate with success probability higher than 0.5. Finally in lossy scenarios, the best rate has a low success probability, i.e., less than 0.5. In stationary environments, the success transmission probabilities at the various rates are (steep) $\theta=(.99, .98, .96, .93, 0.9, .1, .06, .04)$, (gradual) $\theta=(.95, .9, .8, .65, .45, .25, .15, .1)$, and (lossy) $\theta=(.9, .8, .7, .55, .45, .35, .2, .1)$. Observe that in all cases, $\theta\in {\cal T}\cap{\cal U}_G$ (unimodality holds). We compare G-ORS and SW-G-ORS to SampleRate, where the size of the sliding window is taken equal to 10s. SampleRate explores new rates every ten packet transmissions, and hence has a regret linearly increasing with time. G-ORS and SW-G-ORS explore rates in an optimal manner, and significantly outperform SampleRate as demonstrated see Fig. \ref{fig:stationary}.  

\begin{figure}[thb]
  \centering
 \includegraphics*[width=0.48\columnwidth]{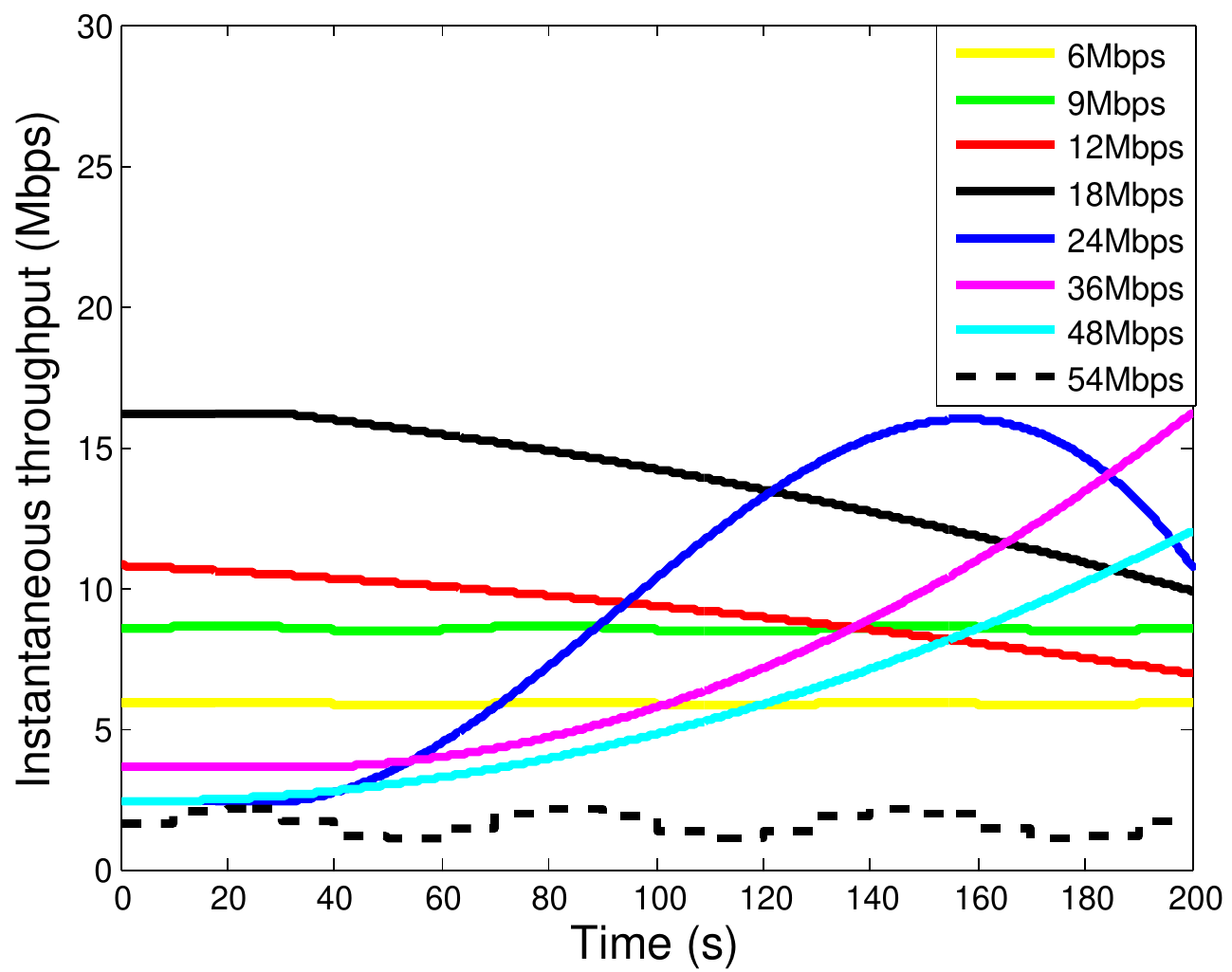} \hspace{-3mm}
  \includegraphics*[width=0.48\columnwidth]{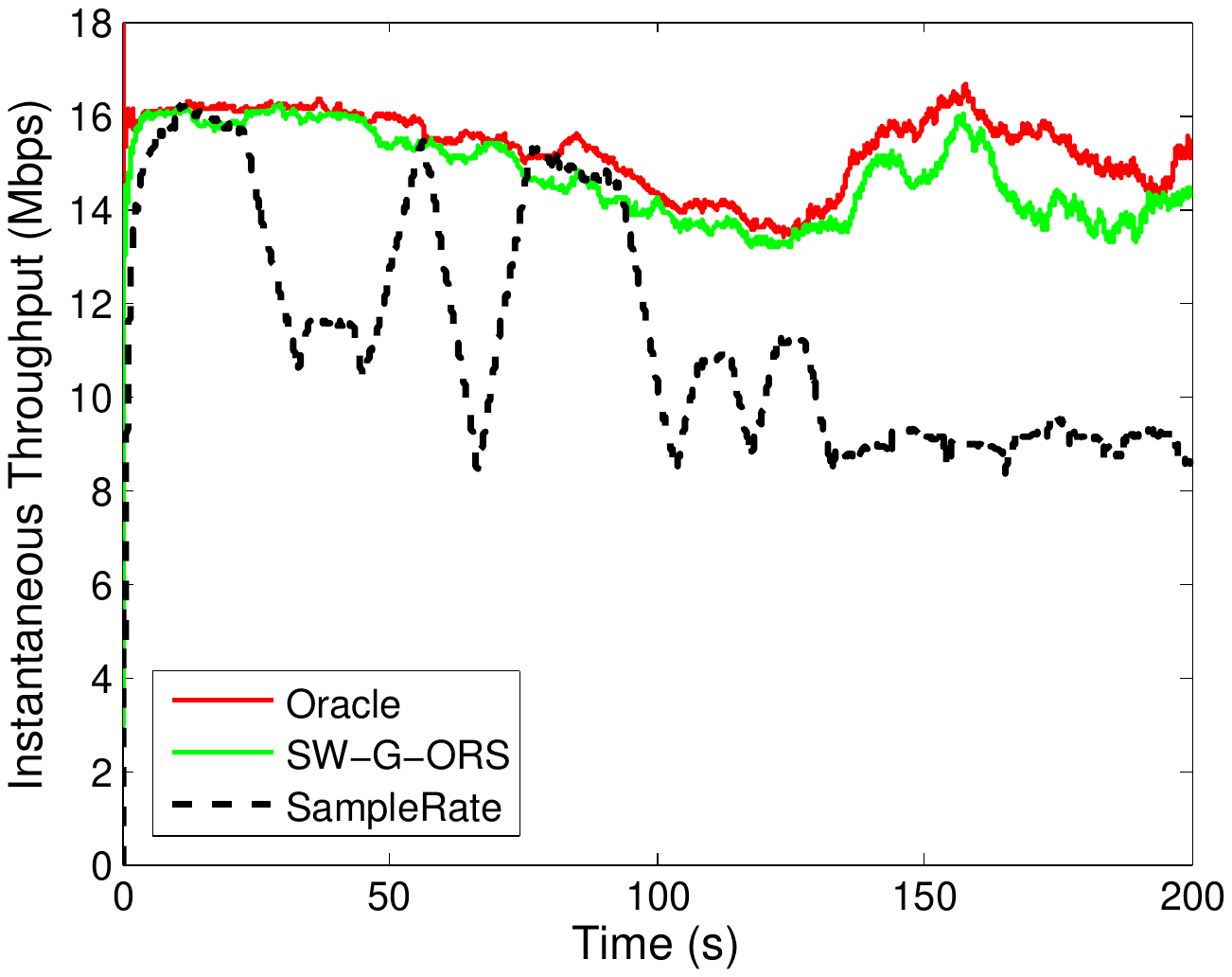}
  \vspace{-3mm}
\caption{Artificially generated non-stationary environment: (left) throughput at different rates; (right) throughput (averaged over 10s) under SW-G-ORS, SampleRate, and the Oracle algorithm.}
  \label{fig:nonst}
  \vspace{-0.5cm}
\end{figure}

For non-stationary environments, we artificially generate varying success probabilities $\theta(t)$ as depicted in Fig. \ref{fig:nonst} (left). At the beginning, the value of $\theta$ corresponds to a steep scenario. It then evolves to a gradual and finally lossy scenario. Fig. \ref{fig:nonst} (right) compares the performance of SW-G-ORS to that of SampleRate and of an oracle algorithm (that always knows the best rate for transmission). SW-G-ORS again outperforms SampleRate, and its performance is close to that of the Oracle algorithm (aware of the success probabilities at various rates).  

\subsubsection{Test-bed traces}

\begin{figure}[htp]
  \centering
  \subfigure[Stationary environment]{
  \label{fig:st_trace}
  \includegraphics*[width=0.48\columnwidth]{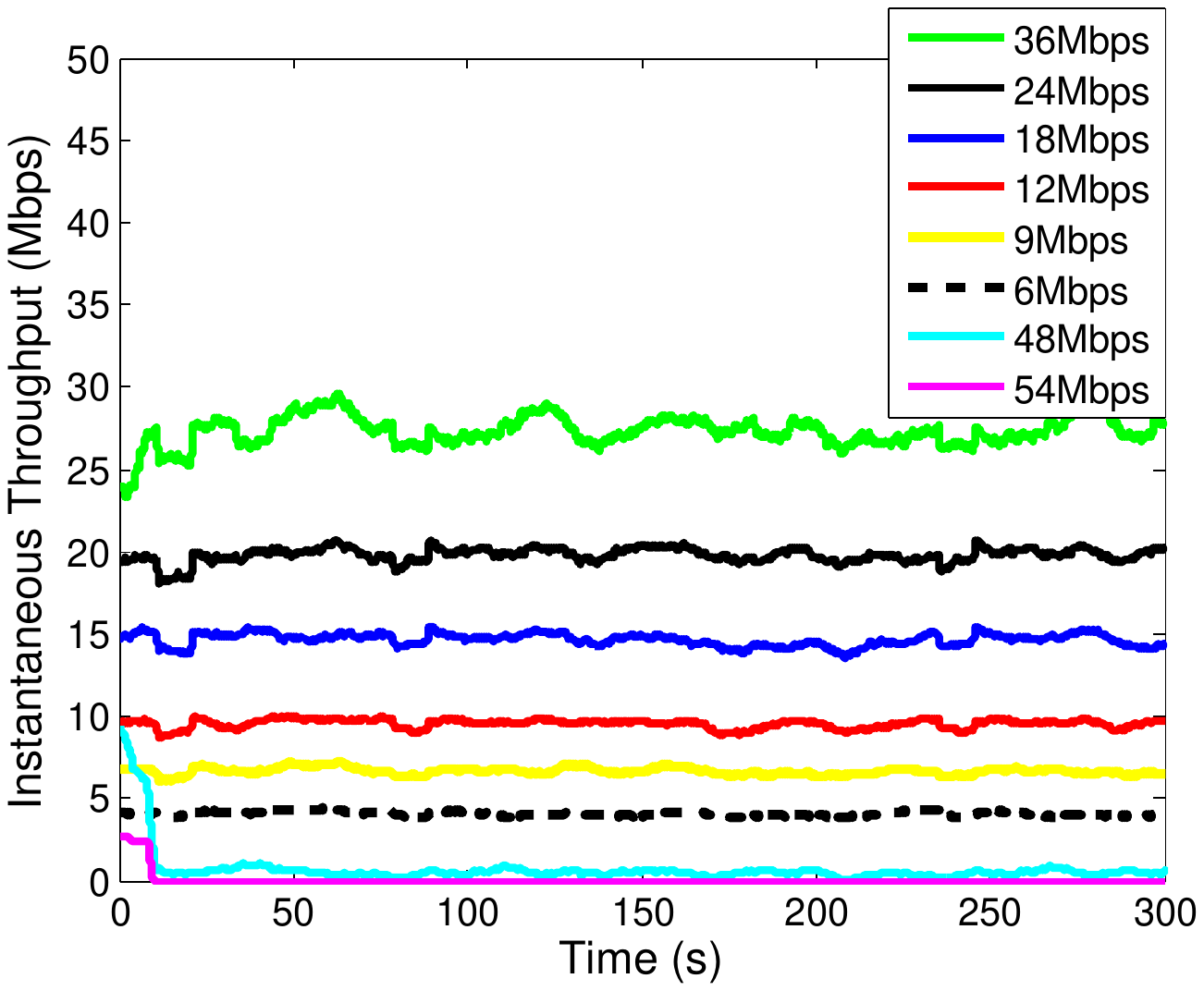}
  \includegraphics*[width=0.48\columnwidth]{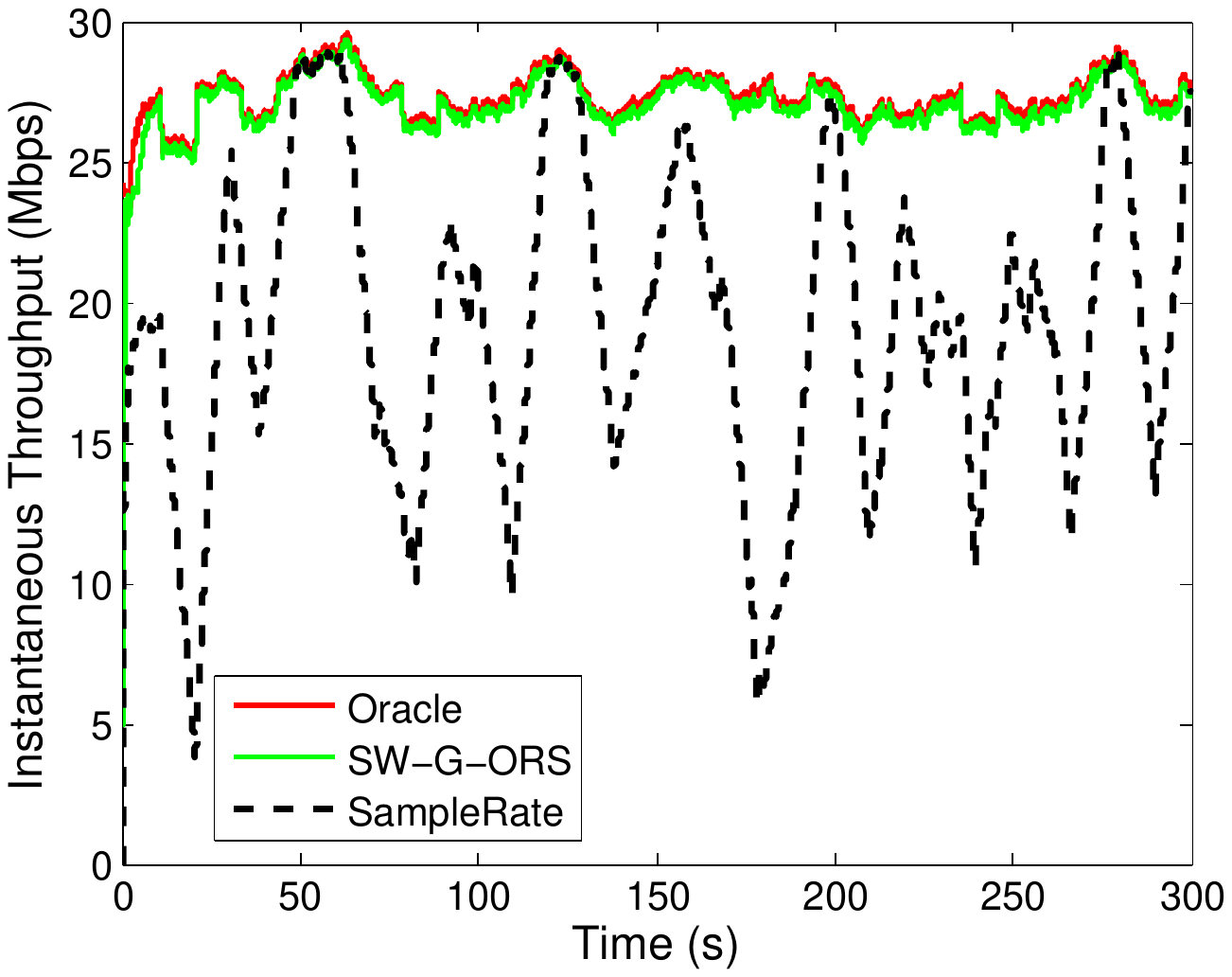}}
 \hspace{-3mm}
 \subfigure[Non-Stationary environment]{
  \label{fig:st_result}
  \includegraphics*[width=0.48\columnwidth]{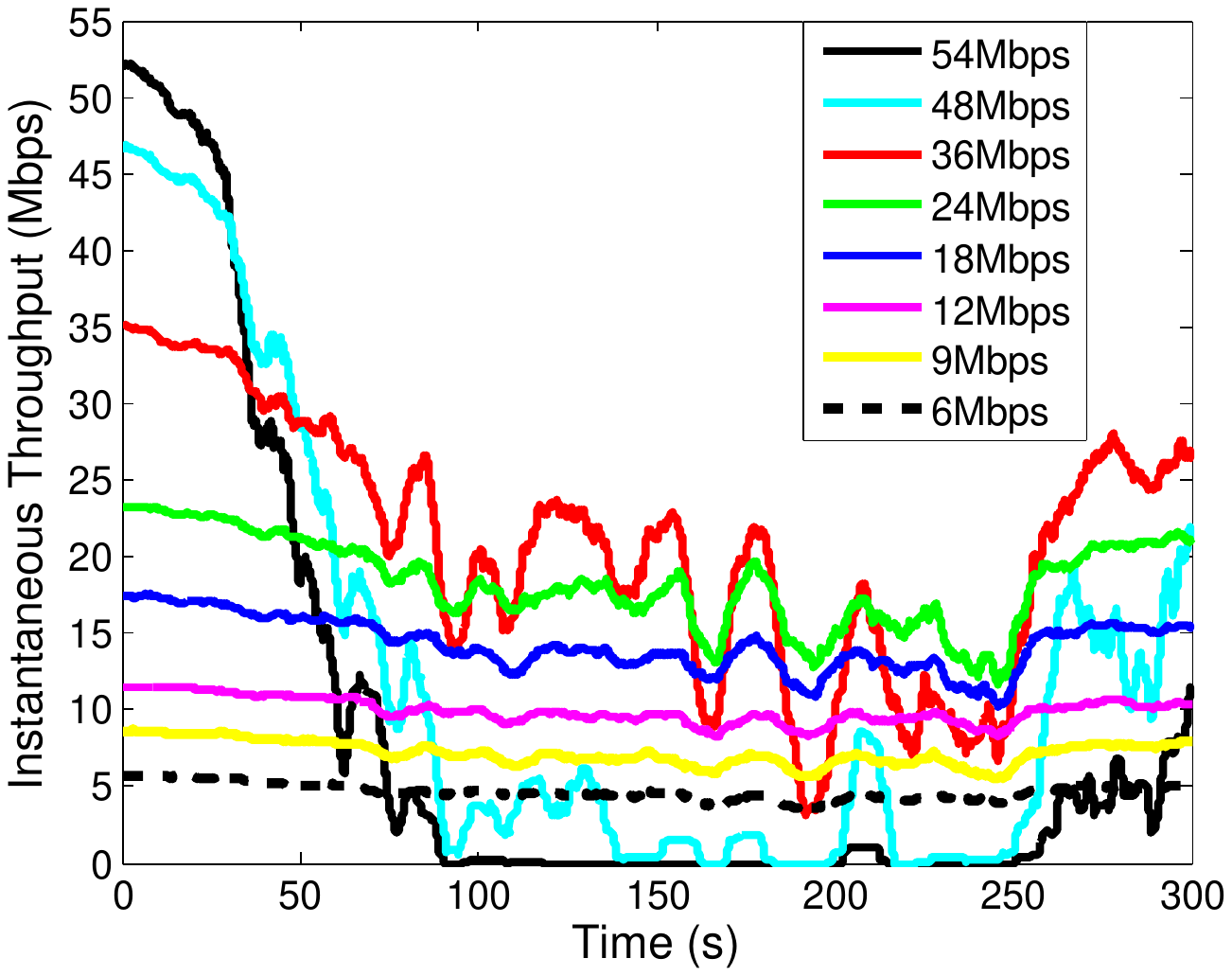}
  \includegraphics*[width=0.48\columnwidth]{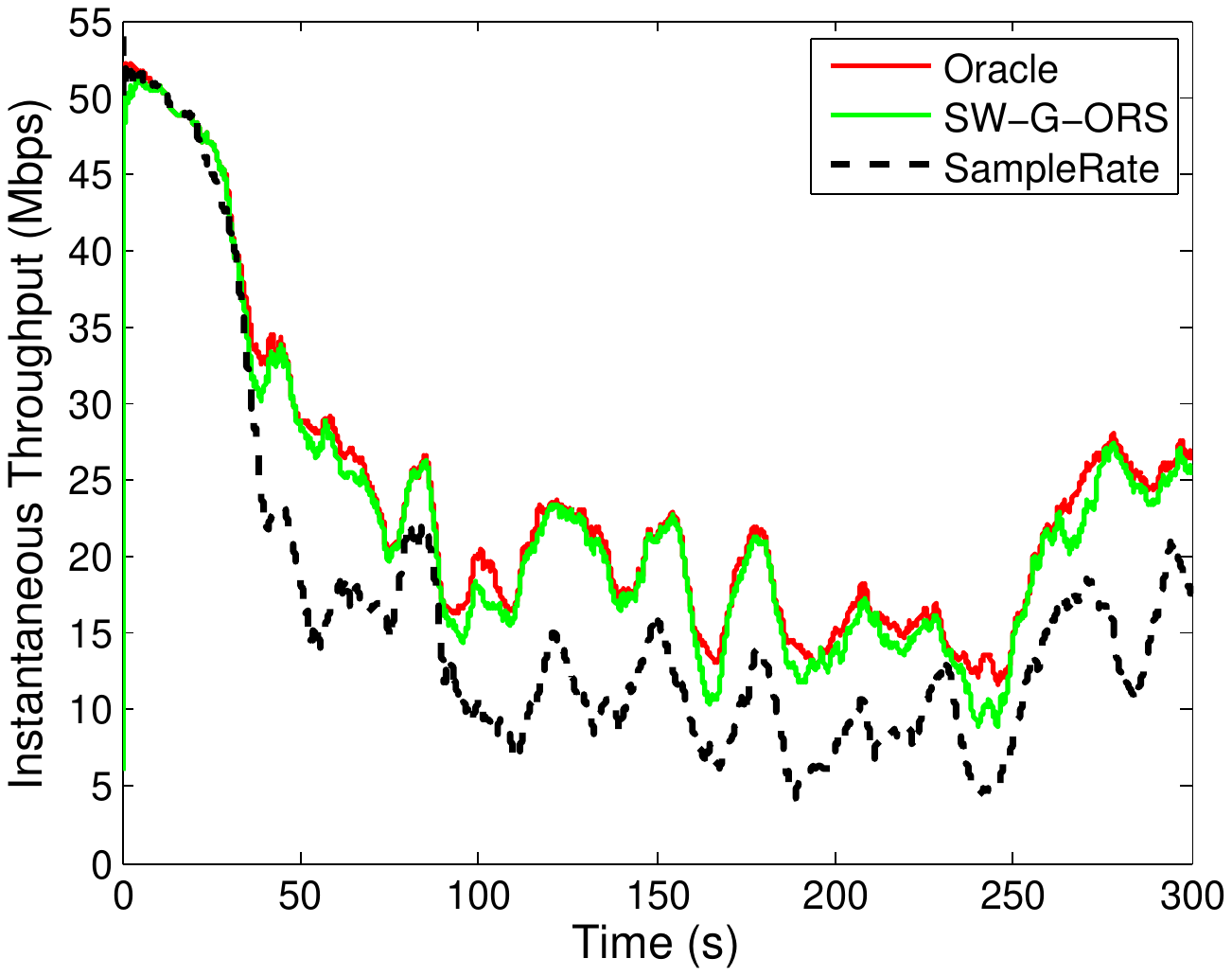}} 
 \hspace{-3mm}
\caption{802.11g test-bed traces. Throughput evolution at different rates (left), and throughput under SW-G-ORS, SampleRate, and the Oracle algorithm (right) in stationary (top) and non-stationary (bottom) environment.}
  \label{fig:st_channel}
  \vspace{-0.5cm}
\end{figure}

We now present results obtained on our 802.11g test-bed, consisting of two 802.11g nodes connected in ad-hoc mode. We collect traces recording the throughputs at the 8 available rates, and then use these traces to test SampleRate, and SW-G-ORS algorithm. Packets are of size 1500 bytes. We generate two kinds of traces: (a) when the two nodes have fixed positions, the successful packet transmission probabilities are roughly constant -- we have a stationary environment; (b) the receiver is then moved (at the speed of a pedestrian), generating a non-stationary environment. The results are presented in Fig. \ref{fig:st_channel}. Again in both scenarios, SW-G-ORS clearly outperforms SampleRate, and exhibits a performance close that of the Oracle algorithm.

\subsection{802.11n MIMO systems}

Next we investigate the performance of SW-G-ORS in 802.11n MIMO systems with two modes, SS and DS, as in \cite{Pefkianakis:2010, deek2013}. We use frame aggregation (30 packets per frame) which is essential in 802.11n. SW-G-ORS is compared to MiRA \cite{Pefkianakis:2010} and SampleRate. To define SW-G-ORS, we use the graph $G$ depicted in Fig. \ref{fig2}. The sliding window for SW-G-ORS and SampleRate  is taken equal to 1s. MiRA is a RA algorithm specifically designed for MIMO systems. It zigzags between MIMO modes to find the best (mode, rate) pair. In its implementation, we use, as suggested in \cite{Pefkianakis:2010}, the following parameters: $\alpha = 1/8$, $\beta=1/4$, and $T_0=0.2ms$.

{\it Artificial traces.} To generate artificial traces, we use results from \cite{deek2013}, and more specifically, the mapping between channel measurements (the SNR and diffSNR\footnote{The maximal gap between the SNRs measured at the various antennas.}, see \cite{deek2013}) and the packet transmission success probabilities. For non-stationary scenarios, we artificially vary (smoothly) the SNR and diffSNR, and then deduce the corresponding evolution of the PER at various (mode, rate) pairs.

{\it Test-bed traces.} We also exploit real test-bed traces extracted from \cite{deek2013}. These traces correspond to stationary environments, and to generate real non-stationary traces, we let the system evolve between 5 stationary scenarios. 

\begin{figure}[htp]
  \centering
  \subfigure[Artificial traces]{
  \label{fig:st_trace}
  \includegraphics*[width=0.48\columnwidth]{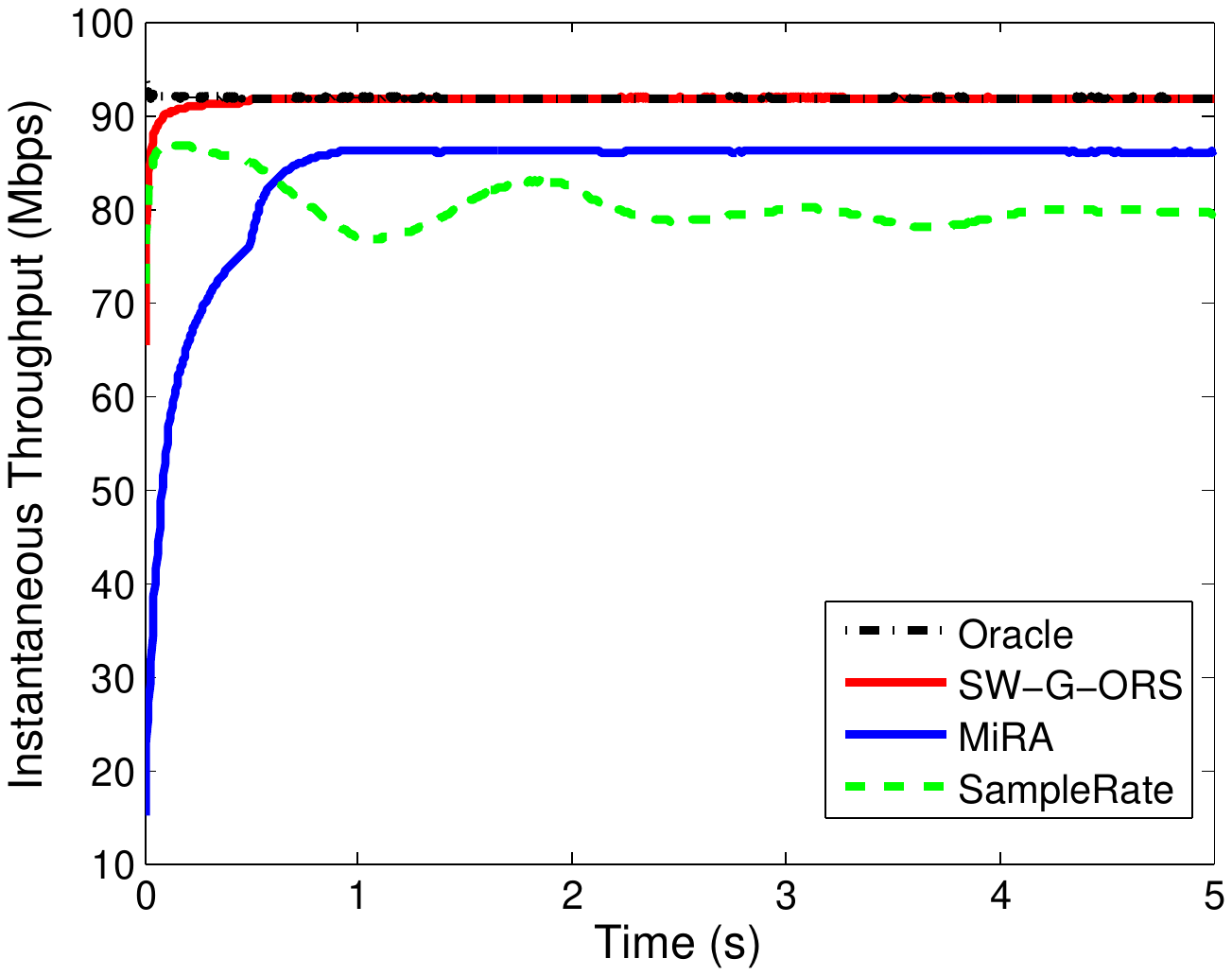}
  \includegraphics*[width=0.48\columnwidth]{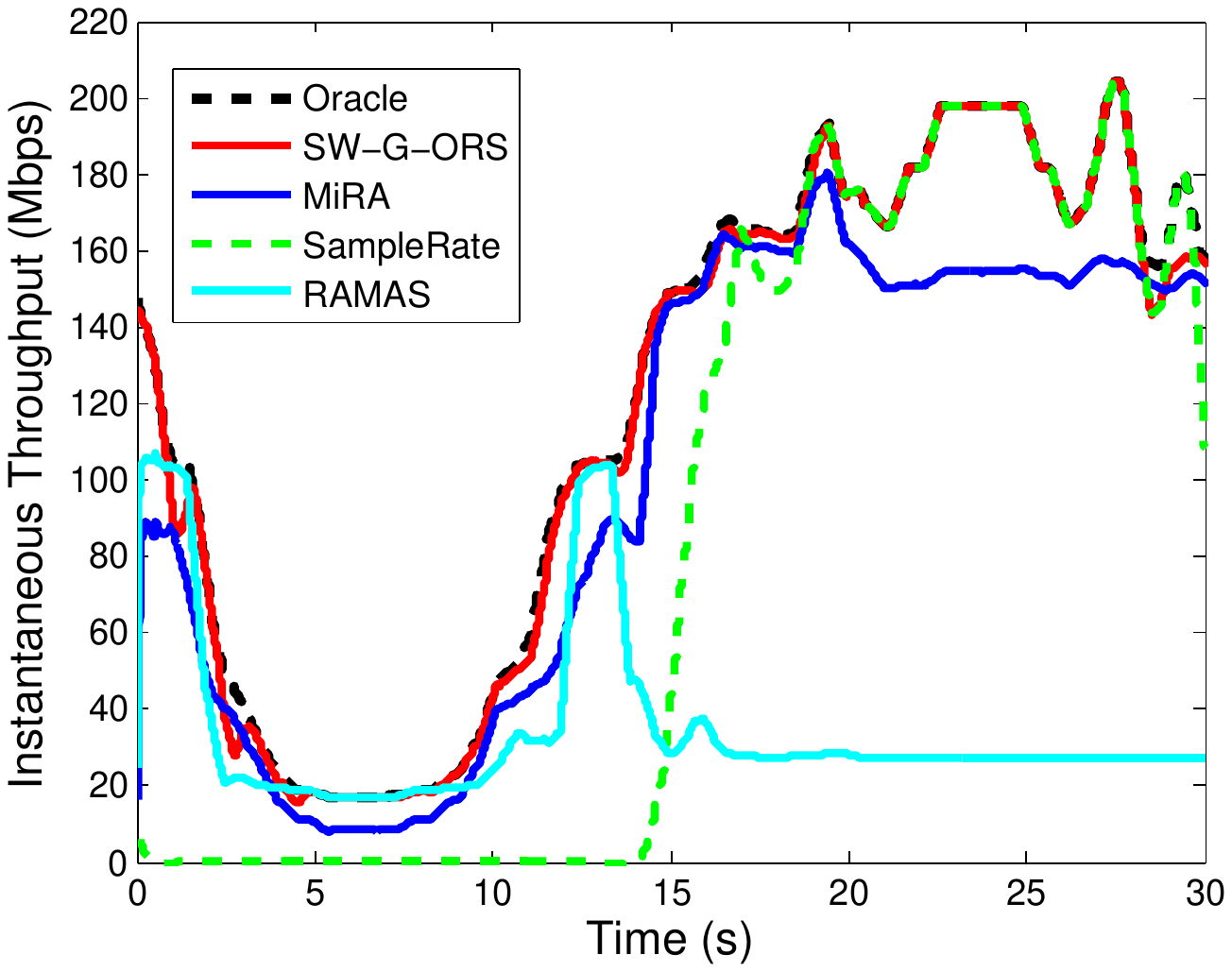}
  }
 \hspace{-3mm}
 \subfigure[Test-bed traces]{
  \label{fig:st_result}
  \includegraphics*[width=0.48\columnwidth]{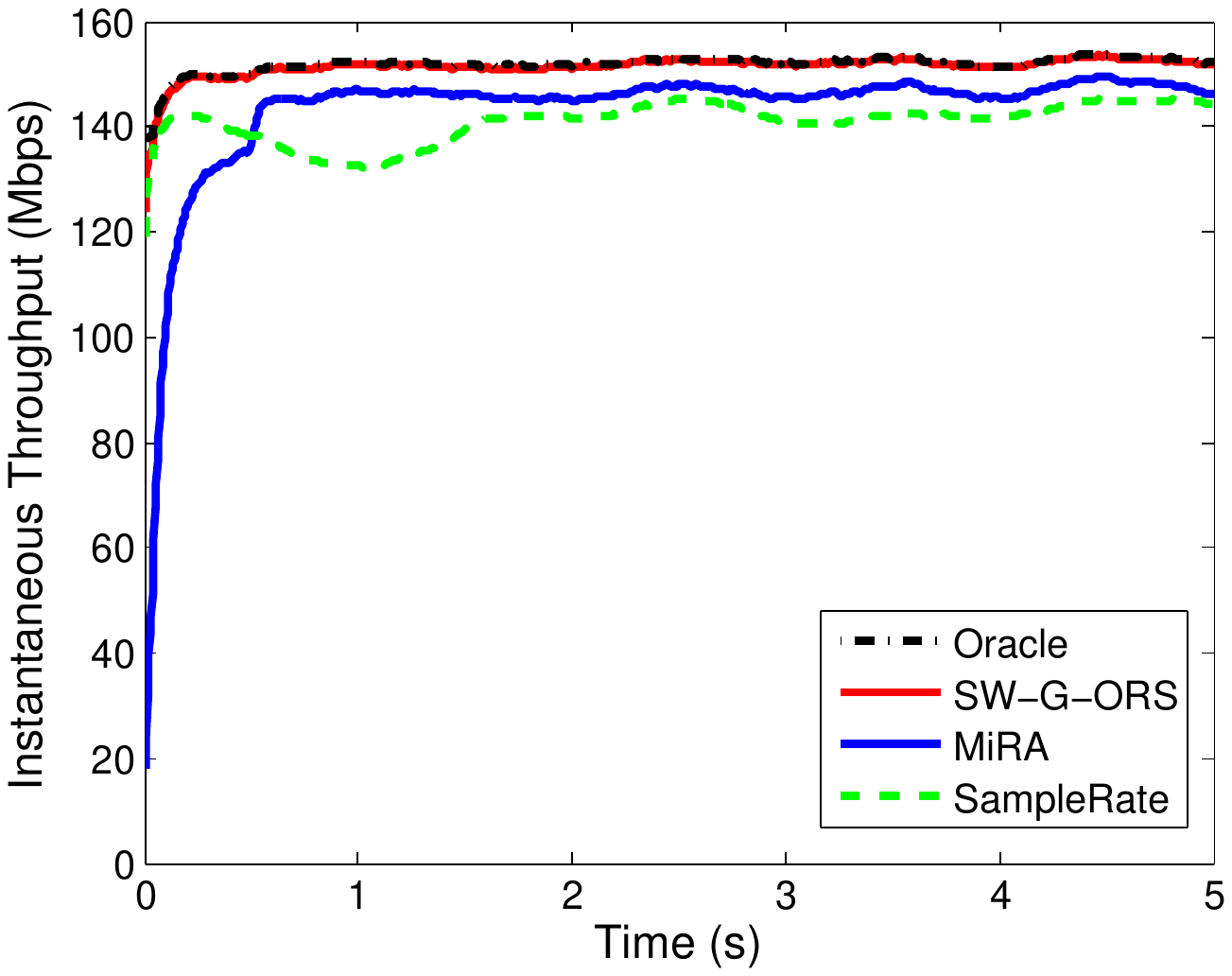}
  \includegraphics*[width=0.48\columnwidth]{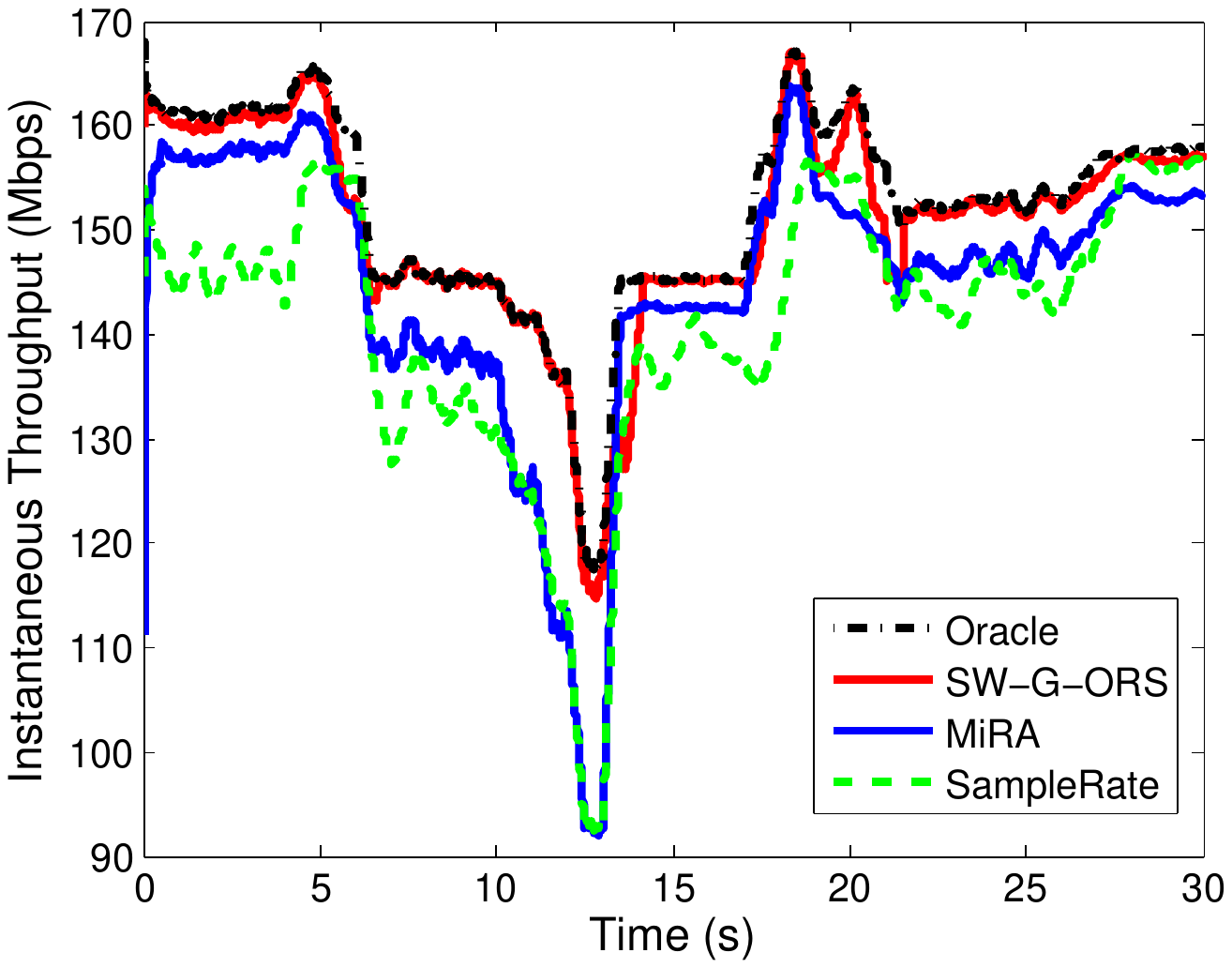}} 
 \hspace{-3mm}
\caption{Throughput under SW-G-ORS, SampleRate, MiRA, and the Oracle algorithm in 802.11n systems: (left) stationary environments, (right) non-stationary environments; (top) artificial traces, (bottom) test-bed traces.}
  \label{fig:st_channel2}
\end{figure}

Results are presented in Fig. \ref{fig:st_channel2}. Instantaneous throughputs are computed on a window of size 0.5s. In stationary environments, we observe, as expected, that SW-G-ORS is able to learn the best (mode, rate) pair very rapidly, faster than any other algorithm. In the tested scenarios, we find that both MiRA and SampleRate were also able to find the best pair (there are scenarios where SampleRate is not able to do it \cite{Pefkianakis:2010}). Note however that SW-G-ORS provides a better throughput than MiRA and SampleRate (these algorithms do not explore (mode, rate) pairs in an optimal way). In non-stationary scenarios, the throughput of SW-G-ORS is really close to that of the Oracle algorithm. MiRA and SampleRate do seem to be able track the best (mode, rate) pair, but the performance loss compared to SW-G-ORS can be quite significant.

\section{Conclusion}

In this paper, we investigated the fundamental limits of sampling approaches for the design of RA adaptation algorithms in 802.11 systems. We developed G-ORS, an algorithm that provably learns as fast as it is possible the best MIMO mode and rate for transmission. The proposed design methodology is based on online stochastic optimisation techniques: it is versatile, and can be easily adapted to evolving 802.11 standards. Our numerical experiments showed that G-ORS outperforms state-of-the-art sampling-based RA algorithms. This is not surprising as G-ORS is by design optimal. This performance superiority is due to the fact that under G-ORS, the way sub-optimal mode and rate pairs are explored is carefully and optimally controlled. 


\appendix

\section{Proof of Theorem \ref{th:lower_dep}}

We derive here the regret lower bounds for the MAB problem $(P_U)$. To this aim, we apply the techniques used by Graves and Lai \cite{graves1997} to investigate efficient adaptive decision rules in controlled Markov chains. We recall here their general framework. Consider a controlled Markov chain $(X_t)_{t\ge 0}$ on a finite state space ${\cal S}$ with a control set $U$. The transition probabilities given control $u\in U$ are parametrized by $\theta$ taking values in a compact metric space $\Theta$: the probability to move from state $x$ to state $y$ given the control $u$ and the parameter $\theta$ is $p(x,y;u,\theta)$. The parameter $\theta$ is not known. The decision maker is provided with a finite set of stationary control laws $G=\{g_1,\ldots,g_K\}$ where each control law $g_j$ is a mapping from ${\cal S}$ to $U$: when control law $g_j$ is applied in state $x$, the applied control is $u=g_j(x)$. It is assumed that if the decision maker always selects the same control law $g$ the Markov chain is then irreducible with stationary distribution $\pi_\theta^g$. Now the reward obtained when applying control $u$ in state $x$ is denoted by $r(x,u)$, so that the expected reward achieved under control law $g$ is: $\mu_\theta(g)=\sum_xr(x,g(x))\pi_\theta^g(x)$. There is an optimal control law given $\theta$ whose expected reward is denoted $\mu_\theta^{\star}\in \arg\max_{g\in G} \mu_\theta(g)$. Now the objective of the decision maker is to sequentially control laws so as to maximize the expected reward up to a given time horizon $T$. As for MAB problems, the performance of a decision scheme can be quantified through the notion of regret which compares the expected reward to that obtained by always applying the optimal control law.

We now apply the above framework to our MAB problem. For $(P_U)$, the parameter $\theta$ takes values in ${\cal T}\cap {\cal U}$. The Markov chain has values in ${\cal S}=\{0,r_1,\ldots,r_K\}$. The set of control laws is $G=\{1,\ldots,K\}$. These laws are constant, in the sense that the control applied by control law $k$ does not depend on the state of the Markov chain, and corresponds to selecting rate $r_k$. The transition probabilities are given as follows: for all $x,y\in {\cal S}$, 
$$
p(x,y;k,\theta)=p(y;k,\theta)=\left\{
\begin{array}{ll}
\theta_k, & \hbox{ if }y=r_k,\\
1-\theta_k, & \hbox{ if }y=0.
\end{array}\right.
$$
Finally, the reward $r(x,k)$ does not depend on the state and is equal to $r_k\theta_k$, which is also the expected reward obtained by always using control law $k$. 

We now fix $\theta\in {\cal T}\cap {\cal U}$. Define $I^k(\theta,\lambda)=I(\theta_k,\lambda_k)$ for any $k$. Further define the set $B(\theta)$ consisting of all {\it bad} parameters $\lambda\in {\cal T}\cap{\cal U}$ such that $k^{\star}$ is not optimal under parameter $\lambda$, but which are statistically {\it indistinguishable} from $\theta$:
$$
B(\theta)=\{ \lambda \in {\cal T}\cap {\cal U} : \lambda_{k^{\star}} = \theta_{k^{\star}} \textrm{ and }  \max_k  r_k\lambda_k > r_{k^{\star}}\lambda_{k^{\star}} \},
$$
$B(\theta)$ can be written as the union of sets $B_k(\theta)$, $k=1,\ldots,K$ defined as:
$$
B_k(\theta)=\{ \lambda \in B(\theta) :  r_k\lambda_k > r_{k^{\star}}\lambda_{k^{\star}} \}.
$$
Note that $B_k(\theta)=\emptyset$ if $r_k<r_{k^\star}\theta_{k^\star}$. Define $P=\{k: r_k\ge r_{k^\star}\theta_{k^\star}\}$. Observe that there exist $k_0$ and $k_1$ such that $k_0\le k^\star\le k_1$ and $P=\{k_0,\ldots,k_1\}$. Further define $P'=P\setminus\{k^\star\}$.

By applying Theorem 1 in \cite{graves1997}, we know that $c(\theta)$ is the minimal value of the following LP:
\begin {eqnarray}
\textrm{min }  & \sum_k c_k(r_{k^{\star}}\theta_{k^{\star}} - r_k\theta_k) \\
\textrm{s.t. } & \inf_{\lambda \in B_k(\theta)} \sum_{l \neq k^{\star}} c_lI^{l}(\theta,\lambda) \geq 1, \quad \forall k \in P' \label{eq:con1}\\
& c_k \geq 0, \quad \forall k .
\end {eqnarray}

Next we show that the constraints (\ref{eq:con1}) on the $c_k$'s are equivalent to:
\begin{equation}\label{eq:con2}
\min_{k\in N(k^\star)} c_{k}I(\theta_k,{r_{k^\star}\theta_{k^\star}\over r_k}) \ge 1.
\end{equation}
To this aim, let $k\in P'$. Without loss of generality assume that $k>k^\star$. We prove that:
\begin{equation}\label{eq:inf}
\inf_{\lambda \in B_k(\theta)} \sum_{l \neq k^{\star}} c_lI^{l}(\theta,\lambda) = \sum_{l=k^\star+1}^k c_lI(\theta_l,{r_{k^\star}\theta_{k^\star}\over r_l}).
\end{equation}
This is simply due to the following two observations:
\begin{itemize}
\item for all $\lambda\in B_k(\theta)$, we have $\lambda_{k^\star}r_{k^\star}= \theta_{k^\star}r_{k^\star}$ and $\lambda_{k}r_{k}> \lambda_{k^\star}r_{k^\star}$, which using the unimodality of $\lambda$, implies that for any $l\in \{k^\star,\ldots,k\}$, $\lambda_{l}r_l\ge \theta_{k^\star}r_{k^\star}$. Hence:
$$
\sum_{l \neq k^{\star}} c_lI^{l}(\theta,\lambda) \ge \sum_{l=k^\star+1}^k c_lI(\theta_l,{r_{k^\star}\theta_{k^\star}\over r_l}).
$$
\item For $\epsilon >0$, define $\lambda_\epsilon$ as follows: for all $l\in \{k^\star,\ldots,k\}$, $\lambda_l = (1+(l-k^\star)\epsilon) {r_{k^\star}\theta_{k^\star}\over r_l}$, and for all $l\notin \{k^\star,\ldots,k\}$, $\lambda_l=\theta_l$. By construction, $\lambda_\epsilon\in B_k(\theta)$, and 
$$
\lim_{\epsilon\to 0} \sum_{l \neq k^{\star}} c_lI^{l}(\theta,\lambda_\epsilon) = \sum_{l=k^\star+1}^k c_lI(\theta_l,{r_{k^\star}\theta_{k^\star}\over r_l}).
$$
\end{itemize}

From (\ref{eq:inf}), we deduce that constraints (\ref{eq:con1}) are equivalent to (\ref{eq:con2}) (indeed, only the constraints related to $k\in N(k^\star)$ are really active, and for $k\in N(k^\star)$, (\ref{eq:con1}) is equivalent to $c_{k}I(\theta_k,{r_{k^\star}\theta_{k^\star}\over r_k}) \ge 1$). With the constraints (\ref{eq:con2}), the optimization problem becomes straightforward to solve, and its solution yields: 
$$
c(\theta) = \sum_{k\in N(k^\star)} {r_{k^{\star}}\theta_{k^{\star}} - r_k\theta_k) \over I(\theta_k,{r_{k^\star}\theta_{k^\star}\over r_k})}.
$$
\ep

\section{Proof of Theorem \ref{th:lower_indep}}

The proof of Theorem \ref{th:lower_indep} is similar to that of Theorem \ref{th:lower_dep}. The only difference is that in absence of correlations and unimodality, we can only assume that the parameter $\theta$ takes values in $[0,1]^K$. 

In what follows, we fix $\theta\in [0,1]^K$, and denote by $k^{\star}$ the index of the optimal rate. The sets $B(\theta)$ and $B_k(\theta)$ are now defined as:
$$
B(\theta)=\{ \lambda \in [0,1]^K : \lambda_{k^{\star}} = \theta_{k^{\star}} \textrm{ and }  \max_k  r_k\lambda_k > r_{k^{\star}}\lambda_{k^{\star}} \}.
$$
$$
B_k(\theta)=\{ \lambda \in B(\theta) :  r_k\lambda_k > r_{k^{\star}}\lambda_{k^{\star}} \}.
$$
By applying Theorem 1 in \cite{graves1997}, we know that $c'(\theta)$ is the minimal value of the following optimization problem:
\begin {eqnarray*}
\begin {array}{ll}
\textrm{min }  & \sum_k c_k(r_{k^{\star}}\theta_{k^{\star}} - r_k\theta_k) \\
\textrm{s.t. } & \inf_{\lambda \in B_k(\theta)} \sum_{l \neq k^{\star}} c_lI^{l}(\theta,\lambda) \geq 1, \quad \forall k \neq k^{\star} \\
& c_k \geq 0, \quad \forall k .
\end{array}
\end {eqnarray*}
We now solve :
$
\inf_{\lambda \in B_k(\theta)} \sum_{l \neq k^{\star}} c_lI^{l}(\theta,\lambda).
$  
Remark that for $k<k_0$, $B_k(\theta)=\emptyset$, because $r_{k^{\star}}\theta_{k^{\star}}>r_k$ for $k<k_0$. Let $k\ge k_0$. For any fixed $\theta$ and all $l$,
$$
I^l(\theta,\lambda) = \theta_l \log \frac{\theta_l}{\lambda_l} + (1-\theta_l) \log \frac{1-\theta_l}{1-\lambda_l}
$$
is a convex in $\lambda_l$ and it achieves its minimum (equal to $0$) at $\lambda_l = \theta_l$.
Hence, we can choose $\lambda_l = \theta_l$, for all $l \neq k$. Since $\lambda \in B_k(\theta)$, we must have $\lambda_k > \frac{r_{k^{\star}}\theta_{k^{\star}}}{r_k} $. Hence from the convexity of $I(\theta_k,\lambda_k)$ in $\lambda_k$, we get:
$$
\inf_{\lambda_k:\lambda_k > \frac{r_{k^{\star}}\theta_{k^{\star}}}{r_k}}I(\theta_k,\lambda_k)=I(\theta_k,\frac{r_{k^{\star}}\theta_{k^{\star}}}{r_k}).
$$
We deduce that
$
\inf_{\lambda \in B_k(\theta)} \sum_{l \neq k^{\star}} c_lI^{l}(\theta,\lambda)  =  c_kI(\theta_k, \frac{r_{k^{\star}}\theta_{k^{\star}}}{r_k}),
$
where the infimum is reached for $\lambda=(\theta_1,\dots,\theta_{k-1}, \frac{r_{k^{\star}}\theta_{k^{\star}}}{r_k}, \theta_{k+1}, \dots, \theta_K)$. Finally we obtain that:
$$
c'(\theta) =\sum_{k=k_0, k\neq k^{\star}}^K {r_{k^{\star}}\theta_{k^{\star}}-r_k\theta_k\over I(\theta_k, \frac{r_{k^{\star}}\theta_{k^{\star}}}{r_k})}.
$$

\section{Proof of Theorem \ref{th:KL-UCB-R2}} 

{\bf Notations.} Throughout the proof, by a slight abuse of notation, we omit the floor/ceiling functions when it does not create ambiguity.	Consider a suboptimal rate $k \neq k^\star$. If $k$ has only one neighbor, we denote it by ${k}_2$ and we must have $\theta_k r_k < \theta_{{k}_2} r_{{k}_2}$ since $k$ is suboptimal. Otherwise we denote by ${k}_1$ and ${k}_2$ the neighbors of $k$ with $\theta_{{k}_1} r_{{k}_1} <  \theta_k r_k < \theta_{{k}_2} r_{{k}_2}$.

Define the difference between the average reward of $k$ and $k^\prime$ : $\Delta_{k,k^\prime} = | \theta_{k^\prime} r_{k^\prime} - \theta_{k} r_{k}|  > 0$. We use the notation: 
\eqs{
t_{k,k^\prime}(n) = \sum_{n^\prime=1}^n \indic \{ L(n) = k, k(n) = k^\prime \}.
}
$t_{k,k^\prime}(n)$ is the number of times up time $n$ that $k^\prime$ has been selected given that $k$ was the leader.

\medskip
\noindent
{\bf Proof.} Let $T>0$. The regret $R^{ORS}(T)$ of ORS algorithm up to time $T$ is:
$$
R^{ORS}(T) = \sum_{k\neq k^\star} (r_{k^\star} \theta_{k^\star} - r_k \theta_k  ) \EE[ \sum_{n=1}^T   \indic\{ k(n)=k\} ].
$$
We use the following decomposition:
$$
\indic\{ k(n)=k\}=\indic\{L(n)=k^\star,k(n)=k\} +\indic\{L(n)\neq k^\star,k(n)=k\}.
$$
Now 
\begin{align*}
\sum_{k\neq k^\star}   (r_{k^\star} \theta_{k^\star} - r_k \theta_k  ) & \EE[ \sum_{n=1}^T   \indic\{ L(n)\neq k^\star, k(n)=k\} ]\\
& \le r_{k^\star}\sum_{k\neq k^\star}\EE[ \sum_{n=1}^T   \indic\{ L(n)\neq k^\star,k(n)=k\} ] \\
&\le r_{k^\star}\sum_{k\neq k^\star}\EE[ l_k(T)].
\end{align*}
Observing that when $L(n)=k^\star$, the algorithm selects a decision $k \in {\cal N}(k^\star)$, we deduce that: 
\begin{align*}
R^{ORS}(T) \leq  r_{k^\star} \sum_{k \neq k^\star } \EE[  l_k(T) ]   +  \sum_{k \in N(k^\star)} (r_{k^\star} \theta_{k^\star} - r_k \theta_k  ) \EE[ \sum_{n=1}^T   \indic \{ L(n) = k^\star , k(n) = k \} ] 
\end{align*}

Then we analyze the two terms in the r.h.s. in the above inequality. The first term corresponds to the average number of times where $k^\star$ is not the leader, while the second term represents the accumulated regret when the leader is $k^\star$. The following result states that the first term is $O(\log(\log(T)))$:

\begin{theorem} \label{lemma:l1bound}
For $k \neq k^\star$, $ \EE[ l_k(T) ] = O(\log(\log(T)))$.
\end{theorem}

From the above theorem, we conclude that the leader is $k^\star$ except for a negligible number of instants (in expectation). When $k^\star$ is the leader, ORS behaves as KL-R-UCB restricted to the set $N(k^\star)$ of possible decisions. Following the same analysis as in \cite{garivier2011} (the analysis of KL-UCB), we can show that for all $\epsilon > 0$ there are constants $C_1 \leq 7$ , $C_2(\epsilon)$ and $\beta(\epsilon) > 0$ such that:
\begin{align}
\EE[ \sum_{n=1}^T  \indic \{L(n) = k^\star , k(n) = k \} ]  & \leq  \EE[ \sum_{n=1}^T  \indic \{  b_{k}(n) \geq   b_{k^\star}(n) \} ]    \sk 
& \leq (1 + \epsilon) \frac{\log (T)}{I(\theta_k, \frac{r_{k^\star} \theta_{k^\star}}{r_k})}   + C_1 \log(\log(T)) + \frac{C_2(\epsilon)}{T^{\beta(\epsilon)}}.
\end{align}
Combining the above bound with Theorem \ref{lemma:l1bound}, we get:
\begin{align}\label{ref:regretlead2}
R^{ORS}(T) & \leq   (1 + \epsilon) c(\theta) \log (T) +  O( \log(\log(T)) ),
\end{align}
which concludes the proof of Theorem \ref{th:KL-UCB-R2}.
\ep

It remains to show that Theorem \ref{lemma:l1bound} holds, which is done in the next section.

\section{Proof of Theorem \ref{lemma:l1bound}}

The proof of Theorem \ref{lemma:l1bound} is technical, and requires a few preliminary results presented in \ref{sec:conc}, \ref{sec:dev}, and \ref{sec:kl}. The theorem itself is proved in \ref{sec:pro}.

\subsection{Concentration inequalities}\label{sec:conc}
 
 We recall the Hoeffding's inequality which is used throughout the proofs.

\begin{lemma}\label{lem:hoeffding}[Hoeffding's inequality]
Let $\{ Z_t \}_{1 \leq t \leq n}$ be a sequence of independent random variables with, for any $t$, $ Z_t \in [a_t,b_t]$ almost surely. We have, for any $\delta >0$:
$$ 
\PP\bigg[   \big| \sum_{t=1}^n (Z_t - \EE[Z_t]) \big| \geq \delta  \bigg] \leq 2 \exp\left( -\frac{2 \delta^2}{\sum_{t=1}^n (b_t - a_t)^2} \right).
$$
\end{lemma} 
 
We also prove a concentration inequality of independent interest for sums of bounded independent variables with a random number of summands. Lemma~\ref{lem:concentr} is not a simple consequence of Hoeffding's inequality. 

\begin{lemma}\label{lem:concentr}
Let $\{ Z_t \}_{t \in \mathbb{Z}}$ be a sequence of independent random variables with values in $[0,B]$. Define ${\cal F}_n$ the $\sigma$-algebra generated by $\{ Z_t \}_{t \leq n}$ and the filtration ${\cal F} = ( {\cal F}_n )_{n \in \mathbb{Z}} $. Consider $s \in \NN$, $n_0 \in \mathbb{Z}$ and $T \geq n_0$. We define $S_n = \sum_{t=n_0}^n B_t (Z_t - \EE[Z_t])$, where $B_t \in \{0,1\}$ is a ${\cal F}_{t-1}$-measurable random variable. Further define $t_n = \sum_{t=n_0}^n B_t$. Define $\phi \in \{n_0,\dots,T+1\}$ a ${\cal F}$-stopping time such that either $t_{\phi} \geq s$ or $\phi = T+1$. 

 Then we have that:
 \eqs{
 \PP[ S_{\phi}  \geq t_{\phi} \delta \;,\;  \phi \leq T  ] \leq \exp \Lp - \frac{2 s \delta^2 }{B^2} \Rp.
 }
 As a consequence:
 \eqs{
 \PP[ | S_{\phi} | \geq t_{\phi} \delta \;,\;  \phi \leq T  ] \leq 2 \exp \Lp - \frac{2 s \delta^2 }{B^2} \Rp.
 }
\end{lemma}

\bp
Let $\lambda > 0$, and define $G_n = \exp( \lambda(S_n - \delta t_n)  ) \indic \{n \leq T \}$. We have that:
	\als{
	\PP[ S_{\phi}  \geq t_{\phi} \delta \;,\;  \phi \leq T  ] &= \PP[ \exp( \lambda(S_{\phi}  - \delta t_{\phi} ) ) \indic \{\phi \leq T \} \geq 1] \sk 
		&= \PP[ G_{\phi}  \geq 1] \leq \EE[ G_{\phi} ].
	}
Next we provide an upper bound $\mathbb{E}[G_{\phi}]$. We define the following quantities:
	\als{ 
	Y_t &=  B_t [  \lambda (Z_t - \EE[Z_t])  - \lambda^2 B^2/8  ] \sk
	\tilde{G}_n &=  \exp \Lp \sum_{t=n_0}^n Y_t \Rp \indic \{n \leq T \}.
	}
	So that $G$ can be written:
	\eqs{ 
	G_n = \tilde{G}_n \exp (- t_n ( \lambda \delta - \lambda^2 B^2/8 ) ).
	}
	Setting $\lambda = 4 \delta/B^2$:
	\eqs{ 
	G_n  = \tilde{G}_n \exp(-2 t_n \delta^2 / B^2).
	}
	Using the fact that $t_{\phi} \geq s$ if $\phi \leq T$,  we can upper bound $G_{\phi}$ by:
	\eqs{
	G_{\phi} =  \tilde{G}_{\phi} \exp(-2 t_{\phi} \delta^2 / B^2) \leq  \tilde{G}_{\phi}  \exp(-2 s \delta^2 / B^2).
	}
	It is noted that the above inequality holds even when $\phi = T+1$, since $G_{T+1} = \tilde{G}_{T+1} = 0$. Hence:
		\eqs{
	\EE[ G_{\phi} ] \leq \EE[ \tilde{G}_{\phi} ] \exp(-2 s \delta^2 / B^2).
	}
	We prove that $(\tilde{G}_n)_n$ is a super-martingale. We have that $\EE[ \tilde{G}_{T+1} | {\cal F}_T] = 0 \leq \tilde{G}_{T}$. For $n \leq T-1$, since $B_{n+1}$ is ${\cal F}_n$ measurable:
	\eqs{
	\EE[ \tilde{G}_{n+1} | {\cal F}_n] = \tilde{G}_n (  (1 - B_{n+1}) +   B_{n+1} \EE[ \exp(Y_{n+1}) ] )  . 
	}
	As proven by Hoeffding (\cite{Hoeffding1963}[eq. 4.16]) since $Z_{n+1} \in [0, B]$:
	\eqs{
	\EE[ \exp(\lambda (Z_{n+1} - \EE[ Z_{n+1}] ) ) ] \leq \exp(\lambda^2 B^2/8 ),
	}
 so $\EE[ \exp(Y_{n+1}) ] \leq 1$ and $(\tilde{G}_n)_n$ is indeed a supermartingale: $\EE[ \tilde{G}_{n+1} | {\cal F}_n] \leq \tilde{G}_n$. Since $\phi \leq T+1$ almost surely, and $(\tilde{G}_n)_n$ is a supermartingale, Doob's optional stopping theorem yields: $\EE[ \tilde{G}_{\phi}] \leq \EE[\tilde{G}_{n_0-1}] = 1$, and so
	\eqs{
		\PP[ S_{\phi}  \geq t_{\phi} \delta  ,  \phi \leq T  ] \leq \mathbb{E}[G_\phi]\le \EE[ \tilde{G}_{\phi} ] \exp(-2 s \delta^2 / B^2) \leq \exp(-2 s \delta^2 / B^2).
	}
	which concludes the proof. The second inequality is obtained by symmetry.

\ep

\subsection{Deviation bounds}\label{sec:dev} 

The following two lemmas are used repeatedly in the proof of Theorem~\ref{lemma:l1bound}. They are corollaries of Lemma~\ref{lem:concentr} and allow us to show that certain families of events happen only rarely (in expectation). 
 
Lemma~\ref{lem:deviation} states that if a set of instants $\Lambda$ can be decomposed into a family of subsets $(\Lambda(s))_{s\ge 1}$ of instants (each subset has at most one instant) where $k$ is tried sufficiently many times ($t_k(n) \geq \epsilon s$, for $n \in \Lambda(s)$), then the expected number of instants in $\Lambda$ at which the average reward of $k$ is badly estimated is finite.
 
\begin{lemma}\label{lem:deviation}
Let $k\in\{ 1,\ldots,K\}$, and $\epsilon > 0$. Define ${\cal F}_n$ the $\sigma$-algebra generated by $( X_k(t) )_{1 \leq t \leq n, 1 \leq k \leq K}$. Let $\Lambda \subset \NN$ be a (random) set of instants. Assume that there exists a sequence of (random) sets $(\Lambda(s))_{s\ge 1}$ such that (i) $\Lambda \subset \cup_{s \geq 1} \Lambda(s)$, (ii) for all $s\ge 1$ and all $n\in \Lambda(s)$, $t_k(n) \ge \epsilon s$, (iii) $|\Lambda(s)| \leq 1$, and (iv) the event $n \in \Lambda(s)$ is ${\cal F}_n$-measurable. Then for all $\delta > 0$:
\eq{\label{eq:ineq1}
\EE[ \sum_{n \geq 1} \indic\{ n \in \Lambda , | \hat\mu_k(n) - \EE[\hat\mu_k(n)] | > \delta \} ]  \leq  \frac{r_k^2}{\epsilon \delta^2}.
}
\end{lemma}
\bp
Let $T \geq 1$. For all $s \geq 1$, since $\Lambda(s)$ has at most one element, define $\phi_s = T+1$ if $\Lambda(s) \cap \{1 , \dots, T \}$ is empty and $\{ \phi_s \} = \Lambda(s)$ otherwise. Since $\Lambda \subset \cup_{s \geq 1} \Lambda(s)$, we have:
\eqs{
\sum_{n = 1}^{T} \indic\{ n \in \Lambda , | \hat\mu_k(n) - \EE[\hat\mu_k(n)]  | > \delta \} \leq \sum_{s \geq 1} \indic \{ | \hat\mu_k(\phi_s) - \EE[\hat\mu_k(\phi_s)]  | > \delta , \phi_s \leq T \}.
}
Taking expectations:
\eqs{
\EE[ \sum_{n = 1}^{T} \indic\{ n \in \Lambda , | \hat\mu_k(n) - \EE[\hat\mu_k(n)]  | > \delta \} ] \leq \sum_{s \geq 1} \PP [ | \hat\mu_k(\phi_s) - \EE[\hat\mu_k(\phi_s)]  | > \delta , \phi_s \leq T  ].
} 
 Since $\phi_s$ is a stopping time upper bounded by $T+1$, and that $t_k(\phi_s) \geq \epsilon s$ we can apply Lemma~\ref{lem:concentr} to obtain:
 \eqs{
\EE[ \sum_{n = 1}^{T} \indic\{ n \in \Lambda , | \hat\mu_k(n) - \EE[\hat\mu_k(n)]  | > \delta \} ] \leq  \sum_{s \geq 1} 2 \exp \Lp - \frac{ 2 s \epsilon \delta^2}{r_k^2} \Rp  \leq \frac{r_k^2}{\epsilon \delta^2}.
} 
  We have used the inequality: $\sum_{s \geq 1} e^{- s w} \leq \int_{0}^{+\infty} e^{- u w} du = 1/w$. Since the above reasoning is valid for all $T$, we obtain the claim~\eqref{eq:ineq1}.
\ep

A useful corollary of Lemma~\ref{lem:deviation} is obtained by choosing $\delta = \Delta_{k,k^\prime}/2$, when arms $k$ and $k^\prime$ are separated by at least $\Delta_{k,k^\prime}$.
 
\begin{lemma}\label{cor:deviation}
Let $k, k'\in\{ 1,\ldots,K\}$ with $k\neq k'$ and $\epsilon >0$. Define ${\cal F}_n$ the $\sigma$-algebra generated by $( X_k(i) )_{1 \leq i \leq n, 1 \leq k \leq K}$. Let $\Lambda \subset \NN$ be a (random) set of instants. Assume that there exists a sequence of (random) sets $(\Lambda(s))_{s\ge 1}$ such that (i) $\Lambda \subset \cup_{s \geq 1} \Lambda(s)$, (ii) for all $s\ge 1$ and all $n\in \Lambda(s)$, $t_k(n)\ge \epsilon s$ and $t_{k'}(n)\ge \epsilon s$, (iii) for all $s$ we have $|\Lambda(s)| \leq 1$ almost surely and (iv) for all $n \in \Lambda$, we have $\EE[\hat\mu_k(n)] \leq  \EE[\hat\mu_{k^\prime}(n)] - \Delta_{k,k^\prime}$ (v) the event $n \in \Lambda(s)$ is ${\cal F}_n$-measurable. Then:
\begin{equation}\label{eq:ineq2}
\EE[ \sum_{n \geq 1} \indic\{ n \in \Lambda , \hat\mu_k(n) > \hat\mu_{k^\prime}(n)  \} ] \leq \frac{ 4( r_k^2 + r_{k^\prime}^2) }{\epsilon \Delta_{k,k^\prime}^2}.
\end{equation}

\end{lemma}

\subsection{KL divergence }\label{sec:kl} 
 
We present results related to the KL divergence that will be instrumental when manipulating indexes $b_k(n)$. Lemma~\ref{lem:pinsker} gives an upper and a lower bound for the KL divergence. The lower bound is Pinsker's inequality. The upper bound is due to the fact that $I(p,q)$ is convex in its second argument.
\begin{lemma}\label{lem:pinsker}
For all $p,q \in [0,1]^2$, $p \leq q$:
\eq{\label{eq:pinkser_ineq}
 2 (p - q)^2 \leq I(p,q) \leq \frac{ (p-q)^2 }{q(1-q)}.
 }
 and 
 \eq{\label{eq:pinkser_equiv}
  I(p,q) \sim \frac{ (p-q)^2 }{q(1-q)} \;,\; q \to p^+
 }
 
\end{lemma}
\bp
	The lower bound is Pinsker's inequality. 
	For the upper bound, we have: 
	\eqs{
	\frac{\partial I}{\partial q}\Lp p,q \Rp = \frac{q-p}{q(1-q)}.
	}
	Since $q \mapsto \frac{\partial I}{\partial q}(p,q)$ is increasing, the fundamental theorem of calculus gives the announced result:
	\eqs{
	I(p,q) \leq \int_{p}^{q} \frac{\partial I}{\partial u} \Lp p,u \Rp du \leq \frac{(p-q)^2}{q(1-q)}. 
	}
	The equivalence comes from a Taylor development of $q \to I(p,q)$ at $p$, since: 
	\als{
	\frac{\partial I}{\partial q}(p,q) |_{q=p} = 0, \sk
	\frac{\partial^2 I}{\partial q^2}(p,q) |_{q=p} = \frac{1}{q(1-q)}.
	}
\ep	

\medskip
Lemma~\ref{lem:deviation_result} is straightforward from \cite{garivier2011}[Theorem 10]. It should be observed that this result is not a direct application of Sanov's theorem; Lemma \ref{lem:deviation_result} provides sharper bounds in certain cases.	

\begin{lemma}\label{lem:deviation_result}
For $1 \leq t_k(n)  \leq \tau$ and $\delta > 0$, if $\{ X_k(i) \}_{1 \leq i \leq \tau}$ are i.i.d Bernoulli random variables with parameter $\theta_k$, we have that:
$$
\PP \Lb t_k(n) I \Lp  \frac{1}{t_k(n)} \sum_{i=1}^{t_k(n)} X_k(i) , \theta_k \Rp \geq \delta  \Rb \leq 2 e \ceil{\delta \log(\tau) }\exp(-\delta).	
$$
\end{lemma}
	
\subsection{Proof of Theorem~\ref{lemma:l1bound} }\label{sec:pro} 

Let $k$ be the index of a suboptimal rate under. Let  $\delta > 0$, $\epsilon > 0$ small enough (we provide a more precise definition later on). To derive an upper bound of $\mathbb{E}[l_k(T)]$, we decompose the set of times where $k$ is the leader into the following sets:
$$
\{n \leq T: L(n)=k\}\subset A_\epsilon\cup B_{\epsilon}^T,
$$
where
\begin{align*}
A_{\epsilon} & =  \{ n:	L(n) = k ,  t_{k_2}(n) \geq \epsilon l_k(n) \}\\
B_{\epsilon}^T &= \{  n \leq T: L(n) = k , t_{k_2}(n) \leq \epsilon l_k(n)\}.
\end{align*}
Hence we have:
\begin{equation*}
 \mathbb{E}[l_k(T)] \leq \mathbb{E}\big[ |A_{\epsilon}| + |B_{\epsilon}^T| \big],
\end{equation*}	
Next we provide upper bounds of $\mathbb{E}[ |A_{\epsilon}|]$ and $\mathbb{E}[ |B_{\epsilon}^T |]$. 

\medskip
\noindent
\underline{Bound on $\mathbb{E} |A_\epsilon|$.} Let $n \in A_{\epsilon}$ and assume that $l_k(n)=s$. By design of the algorithm, $t_{k}(n) \geq s/3$. Also  $t_{k_2}(n) \geq \epsilon l_k(n) = \epsilon s$. We apply Lemma~\ref{cor:deviation} with $\Lambda(s) = \{ n  \in A_{\epsilon} , l_k(n) = s  \}$, $\Lambda =\cup_{s\ge 1}\Lambda(s)$. Of course, for any $s$, $|\Lambda(s)| \leq 1$. We have: $A_\epsilon =\{n\in \Lambda:\hat{\mu}_k(n)\ge \hat{\mu}_{k_2}(n)\}$, since when $n\in A_\epsilon$, $k$ is the leader. Lemma~\ref{cor:deviation} can be applied with $k'=k_2$. We get: $\EE| A_{\epsilon}|  < \infty$.

\medskip
\noindent
\underline{Bound on $\mathbb{E} |B_{\epsilon}^T|$.} We introduce the following sets:
\begin{itemize}
\item $C_\delta$ is the set of instants at which the average reward of the leader $k$ is badly estimated:
\eqs{
C_{\delta} = \{  n:  L(n) = k ,  | \hat{\mu}_k(n) - \theta_k r_k | > \delta  \}.
}
\item $D_{\delta} = D_{\delta,k}\cup D_{\delta,k_1}$ where $D_{\delta,k'}=\{ n: L(n) = k , k(n) = k^\prime ,  | \hat{\mu}_{k^\prime}(n) - \theta_{k^\prime} r_{k^\prime} | > \delta \}$  is the set of instants at which $k$ is the leader, $k^\prime$ is selected and the average reward of $k^\prime$ is badly estimated.
\item $E^T = \{ n \leq T: L(n) = k , b_{k_2}(n) \leq \theta_{k_2} r_{k_2} \}$, is the set of instants at which $k$ is the leader, and the upper confidence index $b_{k_2}(n)$ underestimates the average reward $\theta_{k_2} r_{k_2}$.
\end{itemize}
We first prove that $|B_{\epsilon}^T| \leq 12(|C_\delta| + |D_{\delta}| + |E^T|) + O(1)$ as $T$ grows large, and then provide upper bounds on $\mathbb{E}|C_\delta|$, $\mathbb{E}|D_\delta|$, and $\mathbb{E}|E^T|$. 
Let $n \in B_{\epsilon}^T$. When $k$ is the leader, the selected decision is either $k_1$, or $k$, or $k_2$, and hence:
$$
l_k(n) = t_{k,k_1}(n) + t_{k,k}(n) + t_{k,k_2}(n),
$$
where recall that $t_{k,k'}(n)$ denotes the number of times up to time $n$ when $k$ is the leader and $k'$ is selected. Since $n \in B_{\epsilon}^T$, $t_{k,k_2}(n)\le \epsilon l_k(n)$, from which we deduce that:
$$
(1 - \epsilon) l_k(n) \le t_{k,k_1}(n) + t_{k,k}(n).
$$ 
Choose $\epsilon < 1/6$. With this choice, from the previous inequality, we must have that either (a) $t_{k,k_1}(n) \geq l_k(n)/3$ or (b) $t_{k,k}(n) \geq l_k(n)/2 + 1$.

\medskip
(a) Assume that $t_{k,k_1}(n) \geq l_k(n)/3$.  Since $t_{k,k_1}(n)$ is only incremented when $k_1$ is selected and $k$ is the leader, and since $n \mapsto l_k(n)$ is increasing, there exists a unique $\phi(n) < n$ such that $L(\phi(n)) = k$, $k(\phi(n)) = k_1$, $t_{k,k_1}(\phi(n)) = \floor{l_k(n)/6}$. $\phi(n)$ is indeed unique because $t_{k,k_1}(\phi(n))$ is incremented at time $\phi(n)$.
	
Next we prove by contradiction that for $l_k(n) \geq l_0$ large enough and $\delta$ small enough, we must have $\phi(n) \in C_\delta \cup D_{\delta} \cup E^T$. Assume that $\phi(n) \notin C_\delta \cup D_{\delta} \cup E^T$. Then $b_{k_2}(\phi(n)) \geq \theta_{k_2} r_{k_2}$, $\hat\mu_{k_1}(\phi(n)) \leq \theta_{k_1} r_{k_1} + \delta$. Using Pinsker's inequality and the fact that $t_{k_1}(\phi(n)) \geq t_{k,k_1}(\phi(n))$:
\begin{align*}
b_{k_1}(\phi(n)) & \leq \hat\mu_{k_1}(\phi(n)) + \sqrt{ \frac{ \log(l_k(\phi(n))) + c \log(\log(l_k(\phi(n))))   }{ 2 t_{k_1}(\phi(n)) }   }\\
& \leq   \theta_{k_1} r_{k_1} + \delta + \sqrt{\frac{ \log(l_k(n)) + c \log(\log(l_k(n)))}{ 2 \floor{ l_k(n)/6}}}.
\end{align*} 
Now select $\delta < (\theta_{k_2} r_{k_2} - \theta_{k} r_{k})/2$ and $l_0$ such that  $\sqrt{ (\log(l_0) + c \log(\log(l_0)))/ 2 \floor{l_0/6}}  \leq \delta$.
If $l_k(n) \geq l_0$: 
$$
b_{k_1}(\phi(n))\leq   \theta_{k_1} r_{k_1} + 2 \delta < \theta_{k_2} r_{k_2} \leq b_{k_2}(\phi(n)),
$$
which implies that $k_1$ cannot be selected at time $\phi(n)$ (because $b_{k_1}(\phi(n)) < b_{k_2}(\phi(n))$), a contradiction.

\medskip	
(b) Assume that $t_{k,k}(n) \geq l_k(n)/2 + 1 = l_k(n)/3 + l_k(n)/6 + 1$. There are at least $l_k(n)/6 + 1$ instants $\tilde{n}$ such that $l_{k}(\tilde{n}) - 1$ is not a multiple of $3$, $L(\tilde{n}) = k$ and $k(\tilde{n}) = k$. By the same reasoning as in (a) there exists a unique $\phi(n) < n$ such that $L(\phi(n)) = k$, $k(\phi(n)) = k$ , $t_{k,k}(\phi(n)) = \floor{l_k(n)/6}$ and $(l_k(\phi(n)) -1)$ is not a multiple of $3$. So $b_k(\phi(n)) \geq b_{k_2}(\phi(n))$. The same reasoning as that applied in (a) (replacing $k_1$ by $k$) yields $\phi(n) \in C_\delta \cup D_{\delta} \cup E^T$.

\medskip
 We define $B_{\epsilon,l_0}^T = \{ n: n \in B_{\epsilon}^T , l_k(n) \geq l_0 \}$, and we have that $|B_{\epsilon}^T| \leq l_0 +  |B_{\epsilon,l_0}^T|$. We have defined a mapping $\phi$ from $B_{\epsilon,l_0}^T$ to $C_{\delta} \cup D_{\delta} \cup E^T$. To bound the size of $B_{\epsilon,l_0}^T$, we use the following decomposition:
\eqs{
\{ n: n \in B_{\epsilon,l_0}^T , l_k(n) \geq l_0 \} \subset \cup_{n^\prime \in  C_{\delta} \cup D_{\delta} \cup E^T} \{ n: n \in B_{\epsilon,l_0}^T , \phi(n) = n^\prime\}.
}
Let us fix $n^\prime$. If $n \in B_{\epsilon,l_0}^T$ and $\phi(n) = n^\prime$, then $\floor{l_k(n)/6} \in \{ t_{k,k_1}(n^\prime), t_{k,k}(n^\prime) \}$ and $l_k(n)$ is incremented at time $n$ because $L(n) = k$. Therefore:
\eqs{
 | \{ n: n \in B_{\epsilon,l_0}^T , \phi(n) = n^\prime\} | \leq 12.
 }
Using union bound, we obtain the desired result:
\eqs{
|B_{\epsilon}^T| \leq  l_0 +  |B_{\epsilon,l_0}^T| \leq   O(1) + 12  ( |C_{\delta}| +  |D_{\delta}| + |E^T|).
}

\medskip
\noindent	
\underline{Bound on $\mathbb{E} |C_\delta|$.} We apply Lemma~\ref{lem:deviation} with $\Lambda(s) = \{ n: L(n)=k, l_k(n) = s \}$, and $\Lambda = \cup_{s\ge 1}\Lambda(s)$. Then of course, $|\Lambda(s)| \leq 1$ for all $s$. Moreover by design, $t_{k}(n) \geq s/3$ when $n \in \Lambda(s)$, so we can choose any $\epsilon < 1/3$ in Lemma~\ref{lem:deviation}. Now $C_{\delta}=\{n\in \Lambda: | \hat{\mu}_k(n) - \theta_k r_k | > \delta  \}$. From (\ref{eq:ineq1}), we get $\EE|C_{\delta}| < \infty$.

\medskip
\noindent
\underline{Bound on $\mathbb{E} |D_{\delta} |$.}
Let $k^\prime\in \{k_1,k\}$. Define for any $s$, $\Lambda(s) = \{ n : L(n)=k, k(n) = k^\prime , t_{k^\prime}(n) = s \}$, and $\Lambda=\cup_{s\ge 1}\Lambda(s)$. We have $|\Lambda(s)| \leq 1$, and for any $n\in \Lambda(s)$, $t_{k^\prime}(n)=s \ge \epsilon s$ for any $\epsilon <1$. We can now apply Lemma~\ref{lem:deviation} (where $k$ is replaced by $k^\prime$). Note that $D_{\delta} = \{n\in \Lambda: | \hat{\mu}_{k^\prime}(n) - \theta_{k^\prime} r_{k^\prime} | > \delta\}$, and hence (\ref{eq:ineq1}) leads to $\EE| D_{\delta,k^\prime}| <\infty$, and thus $\EE| D_{\delta}| <\infty$.

\medskip
\noindent
\underline{Bound on $\mathbb{E} |E^T |$.} 
We can show as in~\cite{garivier2011} (the analysis of KL-UCB) that $\mathbb{E}|E^T| = O(\log(\log(T)))$ (more precisely, this result is a simple application of Theorem 10 in ~\cite{garivier2011}). 

We have shown that $\mathbb{E} |B_{\epsilon}^T| = O(\log(\log(T)))$, and hence $\mathbb{E}[l_k(T)]=O(\log(\log(T)))$, which concludes the proof of Theorem \ref{lemma:l1bound}.
\ep

\section{Proofs for non-stationary environments}
To simplify the notation, we remove the superscript $^{\tau}$ throughout the proofs, e.g $t_k^\tau(n)$ and $l_k^\tau(n)$ are denoted by $t_k(n)$ and $l_k(n)$.
\subsection{A lemma for sums over a sliding window} 
We will use Lemma~\ref{lem:window_sum} repeatedly to bound the number of times some events occur over a sliding window of size $\tau$. 
\begin{lemma}\label{lem:window_sum}
	Let $A \subset \NN$, and $\tau \in \NN$ fixed. Define $a(n) = \sum_{n^\prime=n-\tau}^{n-1} \indic \{ n^\prime \in A \}$. Then for all $T \in \NN$ and $s \in \NN$ we have the inequality:
	\eq{ \label{eq:wind}
	\sum_{n=1}^{T} \indic \{ n \in A ,  a(n) \leq s  \} \leq s \ceil{ T/\tau }.
	}

As a consequence, for all $k \in \{1 , \dots , K \}$, we have:
	\al{
	\sum_{n=1}^{T} \indic \{ k(n) = k ,  t_k(n) \leq s  \} & \leq s \ceil{ T/\tau }, \label{eq:window_sum1} \\
	\sum_{n=1}^{T} \indic \{ L(n) = k ,  l_k(n) \leq s  \} & \leq s \ceil{ T/\tau }. \nonumber
	}
These inequalities are obtained by choosing $A = \{ n : k(n) = k \}$ and $A = \{ n : L(n) = k \}$ in (\ref{eq:wind}).
\end{lemma}
\bp
We decompose $\{1 , \dots , T \}$ into intervals of size $\tau$: $\{1 , \dots , \tau \}$ , $\{\tau + 1 , \dots , 2 \tau \}$ etc. We have:
	\eq{\label{eq:time_windows}
	\sum_{n=1}^{T} \indic\{   n \in A ,  a(n) \leq s  \} \leq \sum_{i=0}^{ \ceil{ T/\tau }-1 } \sum_{n=1}^{\tau} \indic\{ n + i \tau \in A , a(n + i \tau) \leq s  \}. 
	}
Fix $i$ and assume that $\sum_{n=1}^{\tau} \indic\{ n + i \tau \in A , a(n + i \tau) \leq s  \} > s$. Then there must exist $n^\prime < \tau$ such that $n^\prime \in A$ and $\sum_{n=1}^{n^\prime} \indic\{ n + i \tau \in A , a(n + i \tau) \leq s  \}  = s$. Since $a(n^\prime + i \tau) \geq \sum_{n=1}^{n^\prime} \indic\{ n + i \tau \in A , a(n + i \tau) \leq s  \}$, we have $a(n^\prime + i \tau) \geq s$. As $n^\prime \in A$, we must have $a(n^{\prime\prime} + i \tau) \geq (s+1)$ for all $n^{\prime\prime} > n^\prime$ such that $n^{\prime\prime} \in A$. So 
	\eqs{
	\sum_{n=1}^{\tau} \indic\{ n + i \tau  \in A , a(n + i \tau) \leq s  \} =  \sum_{n=1}^{n^\prime} \indic\{ n + i \tau  \in A , a(n + i \tau) \leq s  \} = s,
	}
	which is a contradiction. Hence, for all $i$:
	\eqs{
	\sum_{n=1}^{\tau} \indic\{ n + i \tau  \in A , a(n + i \tau) \leq s  \} \leq s,
	}
and substituting in~\eqref{eq:time_windows} gives the desired result:
\eqs{
	\sum_{n=1}^{T} \indic\{  n \in A , a(n) \leq s  \} \leq \sum_{i=0}^{\ceil{ T/\tau }-1 } s =  s \ceil{ T/\tau }.
}
\ep

\subsection{Deviation bound} 

We prove a deviation bound similar to that of Lemma~\ref{lem:deviation} for non-stationary environments.

\begin{lemma}\label{lem:deviation_ns}
Let $k\in\{ 1,\ldots,K\}$, $n_0 \in \NN$ and $\epsilon > 0$. Let $\Lambda \subset \NN$ be a (random) set of instants. Assume that there exists a sequence of (random) sets $(\Lambda(s))_{s\ge 1}$ such that (i) $\Lambda \subset \cup_{s \geq 1} \Lambda(s)$, (ii) for all $s\ge 1$ and all $n\in \Lambda(s)$, $t_k(n) \ge \epsilon s$, and (iii) for all $s \geq 1$ $|\Lambda(s) \cap [n_0,n_0+\tau] | \leq 1$. Then for all $\delta > 0$:
	\eqs{
\EE[ \sum_{n = n_0}^{n_0 + \tau} \indic\{ n \in \Lambda , | \hat\mu_k(n) - \EE[\hat\mu_k(n)] | > \delta \} ]  \leq  \frac{\log(\tau) r_k^2}{2 \epsilon \delta^2} + 2.
}\end{lemma}

\bp Fix $s_0 \geq 1$. We use the following decomposition, depending on the value of $s$ with respect to $s_0$:
	\als{
	\{ n \in \Lambda , | \hat\mu_k(n) - \EE[\hat\mu_k(n)] | > \delta \} \subset A \cup B,
	}
where
\als{
A &= \{n_0,\dots,n_0 + \tau\} \cap ( \cup_{ 1 \leq s \leq s_0} \Lambda(s) ) , \sk
B &=  \{n_0,\dots,n_0 + \tau\} \cap \{ n \in \cup_{ s \geq s_0} \Lambda(s) : | \hat\mu_k(n) - \EE[\hat\mu_k(n)] | > \delta  \}.
}
Since for all $s$, $|\Lambda(s) \cap \{n_0, \dots, n_0 + \tau \}|  \leq 1$, we have $|A| \leq  s_0$. The expected size of $B$ is upper bounded by:
	\als{
		E[ |B| ] &\leq \sum_{n = n_0}^{n_0 + \tau}  \PP[  n \in \cup_{s \geq s_0} \Lambda(s) , | \hat\mu_k(n) - \EE[\hat\mu_k(n)] | > \delta        ] \sk
						 &\leq \sum_{n = n_0}^{n_0 + \tau}  \PP[  | \hat\mu_k(n) - \EE[\hat\mu_k(n)] | > \delta   , t_k(n) \geq \epsilon s_0].
	}
For a given $n$, we apply Lemma~\ref{lem:concentr} with $n-\tau$ in place of $n_0$, and $\phi = n$ if $t_k(n) \geq \epsilon s_0$ and $\phi = T+1$ otherwise. It is noted that $\phi$ is indeed a stopping time. We get:
	\eqs{
	\PP[  | \hat\mu_k(n) - \EE[\hat\mu_k(n)] | > \delta   , t_k(n) \geq \epsilon s_0] \leq 2 \exp \Lp - \frac{2 s_0 \epsilon \delta^2 }{r_k^2} \Rp.
	}
	Therefore, setting $s_0 =  r_k^2 \log(\tau) / (2 \epsilon \delta^2)$,
		\eqs{
		E[ |B| ] \leq 2 \tau \exp \Lp - \frac{2 s_0 \epsilon \delta^2}{r_k^2} \Rp = 2.
	}
Finally we obtain the announced result:
	\eq{
\EE[ \sum_{n = n_0}^{n_0 + \tau} \indic\{ n \in \Lambda , | \hat\mu_k(n) - \EE[\hat\mu_k(n)] | > \delta \} ]  \leq   \frac{\log(\tau) r_k^2}{2 \epsilon \delta^2} + 2.
}
	\ep
	
\begin{lemma}\label{cor:deviation_ns}
	Consider $k,k^\prime \in\{ 1,\ldots,K\}$, $n_0 \in \NN$ and $\epsilon > 0$. Let $\Lambda \subset \NN$ be a (random) set of instants. Assume that there exists a sequence of (random) sets $(\Lambda(s))_{s\ge 1}$ such that (i) $\Lambda \subset \cup_{s \geq 1} \Lambda(s)$, and (ii) for all $s\ge 1$ and all $n\in \Lambda(s)$, $t_k(n) \ge \epsilon s$, $t_{k^\prime}(n) \ge \epsilon s$ and (iii) for all $s \geq 1$ $|\Lambda(s) \cap [n_0,n_0+\tau] | \leq 1$ and (iv) for all $n \in \Lambda$, we have $\EE[\hat\mu_k(n)] \leq  \EE[\hat\mu_{k^\prime}(n)] - \Delta_{k,k^\prime}$.

Then for all $\delta > 0$:
	\eqs{
\EE[ \sum_{n = n_0}^{n_0 + \tau} \indic\{ n \in \Lambda , \hat\mu_k(n) > \hat\mu_{k^\prime}(n) \} ]  \leq  \frac{2 \log(\tau) ( r_k^2 + r_{k^\prime}^2)}{ \epsilon \Delta_{k,k^\prime}^2} + 4.
}\end{lemma}	
	
\subsection{Proof of Theorem~\ref{th:KLUCBchanging}}

Recall that due to the changing environment and the use of a sliding window, the empirical reward is a biased estimator of the average reward, and that its bias is upper bounded by $\sigma \tau r_K$. 

To ease the regret analysis, we first provide simple bounds on the distribution of the empirical reward distribution. Unlike in the stationary case, the empirical reward $\hat\mu_{k}(n)$ is not a sum of $t_k(n)$ i.i.d. Bernoulli variables. In order to work with i.i.d. random variables only, we introduce $\underline{\hat\mu}_{k}(n)$ and $\overline{\hat\mu}_{k}(n)$ which are sums of $t_k(n)$ i.i.d. Bernoulli variables, and such that $\underline{\hat\mu}_{k}(n) \leq {\hat\mu}_{k}(n) \leq \overline{\hat\mu}_{k}(n)$ in distribution. This means that for all $\mu \geq 0$, $\PP[ \underline{\hat\mu}_{k}(n) \geq \mu  ] \leq \PP[ {\hat\mu}_{k}(n) \geq \mu  ]  \leq \PP[ \overline{\hat\mu}_{k}(n) \geq \mu  ]$. By definition:
\eqs{
\hat\mu_{k}(n)  = \frac{1}{t_k(n)} \sum_{ n^\prime = n-\tau }^{n-1}  r_k X_k(n^\prime) \indic \{k(n^\prime) = k \},
}
where for $n - \tau \leq n^\prime \leq n-1$, $X_k(n^\prime)$ is a Bernoulli random variable whose mean lies in  $[\theta_k(n) - \tau \sigma , \theta_k(n) + \tau \sigma]$. Hence if we define
\als{
\underline{\hat\mu}_{k}(n)  &= \frac{1}{t_k(n)} \sum_{ n^\prime = n-\tau }^{n}  r_k \underline{X}_k(n^\prime) \indic \{k(n^\prime) = k \},\sk
\overline{\hat\mu}_{k}(n)  &= \frac{1}{t_k(n)} \sum_{ n^\prime = n-\tau }^{n}  r_k \overline{X}_k(n^\prime) \indic \{k(n^\prime) = k \} ,
}
where $\underline{X}_k(n^\prime)$ and $\overline{X}_k(n^\prime)$ are Bernoulli random variables with means $\theta_k(n) - \tau \sigma$ and $\theta_k(n) + \tau \sigma$, respectively, then of course, $\underline{\hat\mu}_{k}(n) \leq {\hat\mu}_{k}(n) \leq \overline{\hat\mu}_{k}(n)$ in distribution.
	
\medskip	
Now the regret under $\pi$=SW-ORS is given by:
\eqs{
R^{\pi}(T) = \sum_{n=1}^T \sum_{k=1}^K ( r_{k^\star(n)} \theta_{k^\star}(n) - r_k \theta_k(n)  ) \PP [ k(n) = k ]. 
}
Let $\epsilon >0$ and ${\cal K}^\tau = (1+\epsilon) \frac{  \log(\tau) + c \log(\log(\tau))}{I_{\min }}$. We introduce the following sets of events: 

\medskip
\noindent
(i) $A  = \cup_{k=1}^{K} A_k = \cup_{k=1}^{K} (A_{k,1} \cup A_{k,2})$, where
\begin{align*}
A_{k,1} &= \{ 1 \leq n \leq T : k(n) = k,  | r_k \theta_k(n) -  r_{k^\star(n)} \theta_{k^\star}(n) |  <  2 r_K \tau \sigma \},\\
A_{k,2} &= \{ n \notin A_{k,1} : k(n) = k, I \Lp \theta_k(n) + \tau \sigma,  \frac{r_{k^\star(n)}}{r_k}( \theta_{k^\star}(n) - \tau \sigma ) \Rp < I_{\min} \}.
\end{align*}
$A_k$ is the set of times at which $k$ is chosen, and $k$ is "close" to the optimal decision. Two decisions are close if either the difference between their average rewards is smaller than the error caused by the changing environment $2 r_K \tau \sigma$, or their KL-divergence number is smaller than $I_{\min}$, taking into account the error caused by the changing environment. Note that, by definition, $|A| \leq G(T,I_{\min},\tau,\sigma)$.

\medskip
\noindent
(ii)  $B = \{ 1 \leq n \leq T : b_{k^\star}(n) \leq  r_{k^\star(n)}( \theta_{k^\star}(n) - \tau \sigma )  \}$. $B$ is the set of times at which the index $b_{k^\star}(n)$ underestimates the average reward of the optimal decision (with an error greater than the bias $r_{k^\star(n)} \tau \sigma$).

\medskip
\noindent
(iii) $C =  \cup_{k=1}^{K} C_k$ ,  $C_k = \{ 1 \leq n \leq T : k(n) = k , t_k(n) \leq {\cal K}^\tau \}$. $C_k$ is the set of times at which $k$ is selected and it has been tried less than ${\cal K}^\tau$ times.

\medskip
\noindent
(iv) $D =  \cup_{k=1}^{K} D_k$, $D_k = \{ 1 \leq n \leq T : k(n) = k, n \notin (A \cup B \cup C) \}$. $D_k$ is the set of times where (a) $k$ is chosen, (b) $k$ has been tried more than ${\cal K}^\tau$ times, (c) $k$ is not close to the optimal decision, and (d) the average reward of the optimal decision is not underestimated.

\medskip
We will show that:
\begin{equation}\label{eq:i}
n \in A_k \Rightarrow r_{k^\star(n)} \theta_{k^\star}(n) - r_k \theta_k(n) \leq r_K \Lp \sqrt{\frac{I_{\min}}{2}} + 2 \tau \sigma \Rp,
\end{equation}
and the following inequalities
$$
\EE[|B|] \leq O( T/\tau),\quad \EE[|C_k|] \leq {\cal K}^\tau \ceil{T/\tau},\quad \EE[|D_k]] \leq  \frac{T}{(\tau \log(\tau)^c)^{g_0 \epsilon^2}}.
$$

We deduce that:
\als{
R^{\pi}(T) & \leq  r_K \Lp \sqrt{\frac{I_{\min}}{2}} + 2 \tau \sigma \Rp G(T,I_{\min},\tau,\sigma) + O(T/\tau) + r_K K {\cal K}^\tau \floor{T/\tau} \sk
&+ \frac{K T}{(\tau \log(\tau)^c)^{g_0 \epsilon^2}} ,
}
which proves Theorem \ref{th:KLUCBchanging}.

\medskip
\noindent
\underline{Proof of (\ref{eq:i}).} Let $n \in A_k$. If $n \in A_{k,1}$, by definition we have $|r_{k^\star(n)} \theta_{k^\star}(n) - r_k \theta_k(n) |  <  2 r_K \tau \sigma$. If $n \in A_{k,2}$, then by definition $n \notin A_{k,1}$ so that: $\theta_k(n) + \tau \sigma < r_{k^\star(n)}( \theta_{k^\star(n)} - \tau \sigma )/r_k$. Furthermore:
\eqs{
I(\theta_k(n) + \tau \sigma,  r_{k^\star(n)}( \theta_{k^\star(n)} - \tau \sigma )/r_k )  < I_{\min}.
}
Using Pinsker's inequality (Lemma~\ref{lem:pinsker}), we get:
\eqs{ 
I_{min} \geq 2 \Lp \frac{r_{k^\star(n)}}{r_k}( \theta_{k^\star}(n) - \tau \sigma ) - \theta_k(n) + \tau \sigma \Rp^2,
}
so that:
\eqs{
| r_{k^\star(n)} \theta_{k^\star}(n) - r_k \theta_k(n) | \leq r_K \Lp \sqrt{\frac{I_{min}}{2}} + 2 \tau \sigma \Rp,
}
which completes the proof of (\ref{eq:i}).

\medskip
\noindent
\underline{Bound on $\mathbb{E}[|B|]$.} Let $n\in B$. Note that $\underline{\hat\mu}_{k^\star}(n) \leq {\hat\mu}_{k^\star}(n)\le b_{k^\star}(n)$ in distribution (the second inequality actually holds almost surely). Since $b_{k^\star}(n) \leq  r_{k^\star}( \theta_{k^\star}(n) - \sigma \tau)$, we deduce that: $\underline{\hat\mu}_{k^\star}(n) \leq r_{k^\star}( \theta_{k^\star}(n) - \sigma \tau)$. Now we have:
\als{
\PP [ n \in B] &=  \PP [ b_{k^\star}(n) \leq  r_{k^\star(n)}( \theta_{k^\star}(n) - \sigma \tau)  ] \sk
& =  \PP \Lb t_{k^\star}(n) I \Lp\frac{\hat\mu_{k^\star}(n)}{r_{k^\star}} , \theta_{k^\star}(n) - \sigma \tau \Rp \geq \log(\tau) + c \log(\log(\tau)) \Rb \sk
& \stackrel{(a)}{\leq}  \PP \Lb t_{k^\star}(n) I \Lp \frac{ \underline{\hat\mu}_{k^\star}(n)}{r_{k^\star}} , \theta_{k^\star}(n) - \sigma \tau \Rp \geq \log(\tau) + c \log(\log(\tau)) \Rb \sk
&\stackrel{(b)}{\leq} \frac{2 e}{\tau (\log(\tau))^{c-2}},  
}
where (a) is due to the fact that $\underline{\hat\mu}_{k^\star}(n) \leq {\hat\mu}_{k^\star}(n)$ in distribution, and (b) is obtained applying Lemma \ref{lem:deviation_result}. Hence: $\EE [|B|] \leq O(T/\tau)$.
	
\medskip	
\noindent
\underline{Bound on $\mathbb{E}[|C_k|]$.} Using Lemma~\ref{lem:window_sum}, we get $|C_k| \leq   {\cal K}^\tau \ceil{ T/\tau }$, and hence $|C| \leq  K {\cal K}^\tau \floor{ T/\tau }$.

\medskip
\noindent
\underline{Bound on $\mathbb{E}[|D_k|]$.} We will prove that $n \in D_k$ implies that $\overline{\hat\mu}_{k}(n)$ deviates from its expectation by at least $r_k f(\epsilon,I_{min}) > 0$ so that:
\eqs{ \PP[ n \in D_k ] \leq \PP \Lb \overline{\hat\mu}_{k}(n) - \EE[ \overline{\hat\mu}_{k}(n)] > r_k f(\epsilon,I_{\min}) \Rb. }

Let $n \in D_k$. Since $k(n) = k$ and $b_{k^\star}(n) \geq r_{k^\star(n)}(\theta_{k^\star}(n) - \sigma \tau)$, we have $b_k(n) \geq r_{k^\star(n)}(\theta_{k^\star}(n) - \sigma \tau)$. We decompose $D_k$ as follows:
\als{
	D_{k} &= D_{k,1} \cup D_{k,2} \sk
	D_{k,1} &= \{ n \in D_k:  \overline{\hat\mu}_{k}(n) \geq  r_{k^\star(n)}( \theta_{k^\star}(n) - \sigma \tau)   \} \sk
	D_{k,2} &= \{ n \in D_k:  \overline{\hat\mu}_{k}(n) \leq   r_{k^\star(n)}( \theta_{k^\star}(n) - \sigma \tau)   \}
} 
	If $n \in D_{k,1}$, $\overline{\hat\mu}_{k}(n) - \EE[ \overline{\hat\mu}_{k}(n)] \geq r_{k^\star(n)}( \theta_{k^\star}(n) - \sigma \tau) - r_k( \theta_{k}(n) + \sigma \tau) > 0$ so that $\overline{\hat\mu}_{k}(n)$ indeed deviates from its expectation. Now let $n \in D_{k,2}$. We have:
\als{
\PP[ n \in D_{k,2} ] & \leq  \PP[ b_k(n) \geq  r_{k^\star(n)}\theta_{k^\star}(n) - \sigma \tau , n \in D_{k,2}] \sk
& =  \PP\Lb t_k(n)I \Lp \frac{ {\hat\mu}_{k}(n)}{r_k} ,  \frac{r_{k^\star(n)}}{r_k}( \theta_{k^\star}(n) - \sigma \tau)   \Rp \leq  \log(\tau) + c \log(\log(\tau))      , n \in D_{k,2}\Rb \sk
& \stackrel{(a)}{\leq} \PP \Lb  {\cal K}^\tau I \Lp \frac{ \overline{\hat\mu}_{k}(n)}{r_k} ,  \frac{r_{k^\star(n)}}{r_k}( \theta_{k^\star}(n) - \sigma \tau)   \Rp \leq  \log(\tau) + c \log(\log(\tau)),  t_k(n) \geq {\cal K}^\tau \Rb \sk
&= \PP \Lb I \Lp \frac{ \overline{\hat\mu}_{k}(n)}{r_k} , \frac{r_{k^\star(n)}}{r_k}( \theta_{k^\star}(n) - \sigma \tau)  \Rp \leq \frac{I_{\min}}{1+\epsilon} , t_k(n) \geq {\cal K}^\tau\Rb ,
}
where in (a), we used the facts that: $\overline{\hat\mu}_{k}(n) \leq   r_{k^\star(n)}( \theta_{k^\star}(n) - \sigma \tau)$, $\overline{\hat\mu}_{k}(n) \geq {\hat\mu}_{k}(n)$ in distribution, and $t_k(n)\ge {\cal K}^\tau$ ($n\notin C$). By continuity and monotonicity of the KL divergence, there exists a unique positive function $f$ such that: 
	\als{
	I \Lp \theta_{k}(n) + \sigma \tau + f(\epsilon,I_{\min} )  , \frac{r_{k^\star(n)}}{r_k}( \theta_{k^\star}(n) - \sigma \tau)  \Rp &= \frac{I_{min}}{1+\epsilon},\sk
	\theta_{k}(n) + \sigma \tau + f(\epsilon,I_{\min} ) &\leq \frac{r_{k^\star(n)}}{r_k}( \theta_{k^\star}(n) - \sigma \tau).
	}
	
	We are interested in the asymptotic behavior of $f$ when $\epsilon$ , $I_{min}$ both tend to $0$ .  Define $\theta^\prime$ , $\theta^{\prime\prime}$ and $\theta_0$ such that
	\eqs{  \theta_{k}(n) + \sigma \tau  \leq \theta^\prime  \leq \theta^{\prime\prime} \leq \theta_0 = \frac{r_{k^\star(n)}}{r_k}( \theta_{k^\star}(n) - \sigma \tau).}
	and
	\eqs{
	I( \theta^\prime, \theta_0) = I_{\min} \;\;, \;\;
	I( \theta^{\prime\prime}, \theta_0) = \frac{I_{\min}}{1 + \epsilon}.
	}
	Using the equivalent \eqref{eq:pinkser_equiv} given in Lemma~\ref{lem:pinsker}, there exists a function $a$ such that:
	\als{
		\frac{ ( \theta_0 - \theta^\prime )^2  }{ \theta_0(1-\theta_0) }(1 + a( \theta_0 - \theta^\prime))  &= I_{\min}, \sk
		\frac{ ( \theta_0 - \theta^{\prime\prime} )^2  }{ \theta_0(1-\theta_0) }(1 + a( \theta_0 - \theta^{\prime\prime} ) )  &= \frac{I_{\min}}{1 + \epsilon}.
	}
 with $a(\delta) \to 0$ when $\delta \to 0^+$. It is noted that $0 \leq \theta_0 - \theta^{\prime\prime} \leq \theta_0 - \theta^{\prime} = o(1)$ when $I_{\min} \to 0^{+}$ by continuity of the KL divergence. Hence:
	\eqs{
	 \theta^{\prime\prime} - \theta^\prime =  \Lp \frac{\epsilon}{2} + o(1) \Rp \sqrt{ \theta_0(1-\theta_0) I_{\min} }.
	}
	Using the inequality
	\als{
	f(\epsilon,I_{min} ) =  \theta^{\prime\prime} - (\theta_{k}(n) + \sigma \tau)  \geq \theta^{\prime\prime} - \theta^{\prime} = \frac{\epsilon}{2} \sqrt{ \theta_0(1-\theta_0) I_{\min} },
	}
	we have proved that:	
	\als{
		2 f(\epsilon,I_{min} )^2 \geq \epsilon^2 g_0 I_{\min} + o(\epsilon^2)
		}
	with
	\als{
	g_0 &=  \frac{1}{2} \min_{1 \leq n \leq T}  \min_{ \stackrel{k \neq k^{\star}(n),}{r_{k^\star(n)}( \theta_{k^\star}(n) - \sigma \tau) < r_k} }  \frac{r_{k^\star(n)}}{r_k}( \theta_{k^\star}(n) - \sigma \tau) \Lb 1 - \frac{r_{k^\star(n)}}{r_k}( \theta_{k^\star}(n) - \sigma \tau)  \Rb. 
	}
	
	Therefore, since $\EE[ \overline{\hat\mu}_{k}(n)] \leq r_k(\theta_{k}(n) + \sigma \tau)$, as claimed, we have
	\eqs{
	\PP[ n \in D_{k}] \leq \PP \Lb \overline{\hat\mu}_{k}(n) - \EE[ \overline{\hat\mu}_{k}(n)]  \geq r_k f(\epsilon,I_{\min}) \;,\; t_k(n) \geq {\cal K}^\tau \Rb.
	}

	We now apply Lemma~\ref{lem:concentr} with $n-\tau$ in place of $n_0$, ${\cal K}^\tau$ in place of $s$ and $\phi = n$ if $t_k(n) \geq {\cal K}^\tau$ and $\phi = T+1$ otherwise. We obtain, for all $n$: 
\als{
	\PP[ n \in D_k] &\leq \PP \Lb \overline{\hat\mu}_{k}(n) - \EE[ \overline{\hat\mu}_{k}(n)]  \geq r_k f(\epsilon,I_{\min}) ,  t_k(n) \geq {\cal K}^\tau \Rb \sk
	&\leq \exp \Lp - 2 {\cal K}^\tau f(\epsilon,I_{\min})^2 \Rp \leq \frac{1}{(\tau \log(\tau)^c)^{g_0 \epsilon^2}},
	}	
	and we get the desired bound by summing over $n$:
	\eqs{
	\EE[|D_k|] = \sum_{n=1}^T \PP[ n \in D_k] \leq \frac{T}{(\tau \log(\tau)^c)^{g_0 \epsilon^2}}.
	}

\subsection{Proof of theorem~\ref{th:ORSchanging}} 	

We first introduce some notations. For any set $A$ of instants, we use the notation: $A[n_0,n] = A \cap \{ n_0,\dots,n_0+\tau \}$. Let $n_0 \leq n$. We define $t_k(n_0,n)$ the number of times $k$ has been chosen during interval $\{ n_0,\dots,n_0+\tau \}$, $l_k(n_0,n)$ the number of times $k$ has been the leader, and $t_{k,k^\prime}(n_0,n)$ the number of times $k ^\prime$ has been chosen while $k$ was the leader:
\als{
t_k(n_0,n) &= \sum_{n^\prime=n_0}^n \indic \{ k(n^\prime) = k \},\sk
l_k(n_0,n) &= \sum_{n^\prime=n_0}^n \indic \{ L(n^\prime) = k \}, \sk
t_{k,k^\prime}(n_0,n) &= \sum_{n^\prime=n_0}^n \indic \{ L(n^\prime) = k, k(n^\prime) = k^\prime \}.
}
Note that $l_k(n-\tau,n) = l_k(n)$, $t_k(n-\tau,n) = t_k(n)$ and $t_{k,k^\prime}(n-\tau,n) = t_{k,k^\prime}(n)$. Given $\Delta > 0$, we define the set of instants at which the average reward of $k$ is separated from the average reward of its neighbours by at least $\Delta$:
\eqs{
	{\cal N}_k(\Delta) = \cap_{k^\prime \in N(k)} \{ n: |r_k \theta_k(n) - r_{k^\prime} \theta_{k^\prime}(n)  | > \Delta  \}.
}
We further define the amount of time that $k$ is suboptimal, $k$ is the leader, and it is well separated from its neighbors:
\eqs{
{\cal L}_k(\Delta) = \{ n: L(n) = k \neq k^\star(n), n \in {\cal N}_k(\Delta) \}.
}

By definition of the regret under $\pi=$SW-ORS:
	\eqs{
		R^{\pi}(T) = \sum_{n=1}^T \sum_{k \neq k^{\star}(n)} (\theta_{k^\star}(n) r_{k^\star} - \theta_k(n) r_k) \PP [ k(n) = k ].
	}
To bound the regret, as in the stationary case, we split the regret into two components: the regret accumulated when the leader is the optimal arm, and the regret generated when the leader is not the optimal arm. The regret when the leader is suboptimal satisfies: 
\als{
\sum_{n=1}^T \sum_{k \neq k^{\star}(n)} &(\theta_{k^\star}(n) r_{k^\star} - \theta_k r_k) \indic\{ k(n) = k  , L(n) \neq k^{\star}(n)\} \sk 
&\leq r_K \sum_{n=1}^T \indic\{ L(n) \neq k^{\star}(n)\} \leq r_K \sum_{n=1}^T \sum_{k \neq k^{\star}(n)} \indic\{ L(n)=k \neq k^{\star}(n)\} \sk
&\leq r_K \sum_{n=1}^T \sum_{k \neq k^{\star}(n)}  \indic\{ n \in {\cal L}_k(\Delta) \} + \indic\{ \exists k^{\prime} \in N(k): | \theta_{k}(n) r_k -  \theta_{k^\prime}(n) r_{k^\prime} | \leq \Delta  \} \sk
&\leq r_K \Lp \sum_{k=1}^K |{\cal L}_k(\Delta)[0,T]| + H(\Delta,T)  \Rp.
}
		
Therefore the regret satisfies:
\al{\label{eq:reget_non_stat}
R^{\pi}(T) &\leq  r_K \Lp H(\Delta,T) + \sum_{k=1}^K \EE[|{\cal L}_k(\Delta)[0,T]|] \Rp \sk
&+ \sum_{n=1}^T \sum_{k \in N(k^{\star}(n))}  (\theta_{k^\star}(n) r_{k^\star} - \theta_{k}(n) r_k) \PP[ k(n) = k ].
}
The second term of the r.h.s in \eqref{eq:reget_non_stat} is the regret of SW-ORS when $k^{\star}(n)$ is the leader. This term can be analyzed using the same techniques as those used for the analysis of SW-KL-R-UCB and is upper bounded by the regret of SW-KL-R-UCB. It remains to bound the first term of the r.h.s in \eqref{eq:reget_non_stat}.

\begin{theorem}\label{th:kluucb_chang_leader}
	Consider $\Delta > 4 r_K \tau \sigma$. Then for all $k$: 
\begin{equation}\label{eq:reget_non_stat2}
\EE[|{\cal L}_k(\Delta)[0,T]|] \leq C_1\times \frac{T \log(\tau)}{\tau (\Delta - 4 r_K \tau \sigma)^2},
\end{equation}
where $C_1>0$ does not depend on $T$, $\tau$, $\sigma$ and $\Delta$.
\end{theorem}

Substituting~\eqref{eq:reget_non_stat2} in~\eqref{eq:reget_non_stat}, we obtain the announced result.

\ep

\subsection{Proof of theorem~\ref{th:kluucb_chang_leader}} 

It remains to prove theorem~\ref{th:kluucb_chang_leader}. Define $\delta = (\Delta - 4 r_K \tau \sigma)/2$. We can decompose $\{1,\dots,T\}$ into at most $\ceil{T/\tau}$ intervals of size $\tau$. Therefore, to prove the theorem, it is sufficient to prove that for all $n_0 \in {\cal L}_k(\Delta)$ we have:
\eqs{
\EE[|{\cal L}_k(\Delta)[n_0,n_0+\tau] |] \leq O \Lp \frac{\log(\tau)}{\delta^2} \Rp.
} 
	
In the remaining of the proof, we consider an interval $\{ n_0,\dots,n_0+\tau \}$, with $n_0 \in {\cal L}_k(\Delta)$ fixed. It is noted that the best and worst neighbour of $k$ change with time. We define $k_1(n)$ and $k_2(n)$ to be the worst and the best neighbor of $k$ respectively at time $n$. From the Lipshitz assumption and the fact that $\Delta > 4 r_K \tau \sigma$, we have that for all $n \in \{ n_0,\dots,n_0+\tau \}$, $k_1(n) = k_1(n_0)$ and $k_1(n) = k_1(n_0)$. Indeed for all $n \in \{ n_0,\dots,n_0+\tau \}$:

\als{
	 \theta_{k_2(n_0)}(n) r_{k_2(n_0)} - \theta_{k}(n)  r_{k}  &\geq \theta_{k_2(n_0)}(n_0) r_{k_2(n_0)} - \theta_{k}(n_0) r_{k} - 2 (n-n_0) \sigma r_K \sk
	 &\geq \Delta - 2 r_K \tau \sigma \geq  2 r_K \tau \sigma > 0.
}
We denote $k_1=k_1(n_0) = k_1(n)$ and $k_2=k_2(n_0) = k_2(n)$ when this does not create ambiguity. We will use the fact that, for all $n \in \{ n_0,\dots,n_0+\tau \}$:
\als{
\EE[\hat\mu_{k_2}(n)] - \EE[\hat\mu_k(n)]  &\geq  r_{k_2} \theta_{k_2}(n) - r_{k} \theta_{k}(n)  - 2 r_K \tau \sigma, \sk
&\geq 	r_{k_2} \theta_{k_2}(n_0) - r_{k} \theta_{k}(n_0)  - 4 r_K \tau \sigma, \sk
&\geq \Delta  - 4 r_K \tau \sigma  = 2 \delta > 0.
}

We decompose ${\cal L}_k(\Delta)[n_0,n_0+\tau] = A_{\epsilon}^{n_0} \cup B_{\epsilon}^{n_0}$, with:
\begin{itemize}
\item[] $A_{\epsilon}^{n_0} = \{  n \in {\cal L}_k(\Delta)[n_0,n_0+\tau] , t_{k_2}(n) \geq \epsilon l_k(n_0,n) \}$ the set of times where $k$ is the leader, $k$ is not the optimal arm, and its best neighbor $k_2$ has been tried sufficiently many times during interval $\{ n_0,\dots,n_0+\tau \}$,
\item[] $B_{\epsilon}^{n_0} = \{ n \in {\cal L}_k(\Delta)[n_0,n_0+\tau] , t_{k_2}(n) \leq \epsilon l_k(n_0,n) \}$ the set of times where $k$ is the leader, $k$ is not the optimal arm, and its best neighbor $k_2$ has been little tried during interval $\{ n_0,\dots,n_0+\tau \}$.
\end{itemize}
	
\medskip
\noindent
\underline{Bound on $\mathbb{E}[A_{\epsilon}^{n_0}]$.} Let $n \in A_{\epsilon}^{n_0}$. We recall that $\EE[\hat\mu_{k_2}(n)] - \EE[\hat\mu_k(n)]  \geq 2 \delta$, so that the reward of $k$ or $k_2$ must be badly estimated at time $n$:
	\eqs{
	\PP[n \in A_{\epsilon}^{n_0}] \leq \PP[ |\hat\mu_{k}(n) - \EE[\hat\mu_{k}(n)]| > \delta] + \PP[ |\hat\mu_{k_2}(n) - \EE[\hat\mu_{k_2}(n)]| > \delta].
	}
We apply lemma~\ref{cor:deviation_ns}, with $k^\prime = k_2$,  $\Delta_{k,k^\prime} = 2 \delta$, $\Lambda(s) = \{ n \in A_{\epsilon}^{n_0}, l_k(n_0,n) =  s\}$, $t_{k_2}(n) \geq \epsilon l_k(n_0,n)  = \epsilon s$. By design of SW-ORS : $t_k(n) \geq l_k(n_0,n)/3 = s/3$. Using the fact that $|\Lambda(s)| \leq 1$ for all $s$, we have that:
\eqs{
\EE[ A_{\epsilon}^{n_0} ] \leq O \Lp \frac{\log(\tau)}{\epsilon \delta^2} \Rp.
}
	
\medskip
\noindent
\underline{Bound on $\mathbb{E}[B_{\epsilon}^{n_0}]$.} Define $l_0$ such that 
\eqs{
\sqrt{ \frac{ \log(l_0) + c \log(\log(l_0)) }{2 \floor{l_0/6}} } \leq \delta.
} 
In particular we can choose $l_0 = 6 (\log(1/\delta)/\delta^2) $. Indeed, with such a choice we have that 
\eqs{
\sqrt{ \frac{ \log(l_0) + c \log(\log(l_0)) }{2 \floor{l_0/6}} } \sim \delta/2 \;,\; \delta \to 0^+.
}
Let $\epsilon < 1/6$, and define the following sets:
\begin{itemize}
\item[] $C_\delta^{n_0}$ is the set of instants at which the average reward of the leader $k$ is badly estimated:
\als{
C_{\delta}^{n_0} &= \{ n \in \{ n_0,\dots,n_0+\tau \}:  L(n) = k \neq k^\star(n),  |\hat{\mu}_k(n) - \EE[\hat{\mu}_k(n)]| > \delta  \};
}
\item[] $D_{\delta}^{n_0} = D_{\delta,k}^{n_0}\cup D_{\delta,k_1}^{n_0}$ where $D_{\delta,k'}^{n_0}=\{ n: L(n) = k \neq k^\star(n), k(n) = k, | \hat{\mu}_{k^\prime}(n) - \EE[\hat{\mu}_{k^\prime}(n)] | > \delta \}$. $D_{\delta}^{n_0}$ is the set of instants at which $k$ is the leader, $k^\prime$ is selected and the average reward of $k^\prime$ is badly estimated.
\item[] $E^{n_0} = \{ n \leq T: L(n) = k \neq k^\star(n), b_{k_2}(n) \leq \EE[\hat{\mu}_{k_2}(n)] \}$ is the set of instants at which $k$ is the leader, and the upper confidence index $b_{k_2}(n)$ underestimates the average reward $\EE[\hat{\mu}_{k_2}(n)]$.
\end{itemize}
	
Let $n \in B_{\epsilon}^{n_0}$. Write $s = l_k(n_0,n)$, and we assume that $s \geq l_0$. Since $t_{k_2}(n_0,n) \leq \epsilon l_k(n_0,n)$ and the fact that $l_{k}(n_0,n) = t_{k_1}(n_0,n) + t_{k}(n_0,n) + t_{k_2}(n_0,n)$, we must have (a) $t_{k_1}(n_0,n) \geq s/3$ or  (b) $t_{k_1}(n_0,n) \geq s/2 + 1$. Since $t_{k,k}(n)$ and $t_{k,k_2}(n)$ are incremented only at times when $k(n) = k$ and $k(n) = k_2$ respectively, there must exist a unique index $\phi(n) \in \{ n_0,\dots,n_0+\tau \}$ such that either: (a) $t_{k,k_1}(\phi(n)) = \floor{s/6}$  and $k(\phi(n)) = k_1$; or (b) $t_{k,k_2}(\phi(n)) = \floor{s/2}$ and  $k(n) = k$ and $l_k(\phi(n))$ is not a multiple of $3$. In both cases, as in the proof of theorem~\ref{lemma:l1bound}, we must have that $\phi(n) \in C_\delta^{n_0} \cup D_{\delta}^{n_0} \cup E^{n_0}$.
	
We now upper bound the number of instants $n$ which are associated to the same $\phi(n)$. Let $n,n^\prime \in B_{\epsilon}^{n_0}$ and $s = l_k(n_0,n)$. We see that $\phi(n^\prime) = \phi(n)$ implies either $\floor{l_{k}(n_0,n^\prime)/6} = \floor{l_{k}(n_0,n)/6}$ or $\floor{l_{k}(n_0,n^\prime)/2} = \floor{l_{k}(n_0,n)/2}$. Furthermore, $n^\prime \mapsto l_k(n_0,n^\prime)$ is incremented at time $n^\prime$. Hence for all $n \in B_{\epsilon}^{n_0}$:
	\eqs{
	| n^\prime \in B_{\epsilon}^{n_0} , \phi(n^\prime) = \phi(n)| \leq  12.
	}
We have established that:
	\als{
	|B_{\epsilon}^{n_0}| &\leq l_0 + 12 (|C_\delta^{n_0}| + |D_\delta^{n_0}|  + |E^{n_0}| ) \sk
	 &= 6 \log(1/\delta)/\delta^2 + 12 (|C_\delta^{n_0}| + |D_\delta^{n_0}|  + |E^{n_0}| ).
	}
We complete the proof by providing bounds of the expected sizes of sets $C_\delta^{n_0}$, $D_\delta^{n_0}$ and $E^{n_0}$.

\medskip
\noindent
\underline{Bound of $\mathbb{E}[C_\delta^{n_0}]$}: Using lemma~\ref{lem:deviation_ns} with $\Lambda(s) = \{ n \in C_\delta^{n_0}, l_k(n_0,n) =  s\}$, and by design of SW-ORS: $t_k(n) \geq l_k(n_0,n)/3 = s/3$. Since $|\Lambda(s)| \leq 1$ for all $s$, we have that:
	 \eqs{
	 \EE[ |C_\delta^{n_0}| ] \leq  O \Lp \frac{\log(\tau)}{\delta^2} \Rp.
	 }
\medskip
\noindent
\underline{Bound of $\mathbb{E}[D_{\delta}^{n_0}]$}: Using lemma~\ref{lem:deviation_ns} with $\Lambda(s) = \{ n \in D_\delta^{n_0}, t_{k,k^\prime}(n_0,n) =  s\}$, and $|\Lambda(s)| \leq 1$ for all $s$, we have that:
	\eqs{
	 \EE[ |D_{\delta,k^\prime}^{n_0}| ] \leq  O \Lp \frac{\log(\tau)}{\delta^2} \Rp .
	 }

\medskip
\noindent
\underline{Bound of $\mathbb{E}[E^{n_0}]$}: By lemma~\ref{lem:deviation_result} since $l_k(n) \leq \tau$:
\als{
	\PP[  n \in E^{n_0} ] &\leq 2 e \ceil{ \log(\tau) (\log(\tau) + c \log(\log(\tau)) ) } \exp(- \log(\tau) + c \log(\log(\tau))) \sk
	 																	& \leq \frac{4 e}{\tau \log(\tau)^{c-2}}.
}
Thus
	\eqs{
	\EE[ |E^{n_0}| ] \leq \frac{4 e}{(\log \tau)^{c-2}}.
	}

Putting the various bounds all together, we have: 
	\eqs{
	\EE[|{\cal L}_k(\Delta)[n_0,n_0+\tau] |] \leq O \Lp \frac{\log(\tau)}{\delta^2} \Rp,
	} 
for all $n_0 \in {\cal L}_k(\Delta)$, uniformly in $\delta$, which concludes the proof. \ep

\newpage
\bibliographystyle{abbrv}
\bibliography{RA}

\begin{thebibliography}{10}

\bibitem{3gpp}
3GPP TR 25.848 V 4.0.0.

\bibitem{aguayo2004}
D.~Aguayo, J.~Bicket, S.~Biswas, G.~Judd, and R.~Morris.
\newblock Link-level measurements from an 802.11b mesh network.
\newblock In {\em Proceedings of the 2004 conference on Applications,
  technologies, architectures, and protocols for computer communications},
  SIGCOMM '04, pages 121--132. ACM, 2004.

\bibitem{auer2002}
P.~Auer, N.~Cesa-Bianchi, and P.~Fischer.
\newblock Finite time analysis of the multiarmed bandit problem.
\newblock {\em Machine Learning}, 47(2-3):235--256, 2002.

\bibitem{Auer2003}
P.~Auer, N.~Cesa-Bianchi, Y.~Freund, and R.~E. Schapire.
\newblock The nonstochastic multiarmed bandit problem.
\newblock {\em SIAM J. Comput.}, 32(1):48--77, Jan. 2003.

\bibitem{bicket2005bit}
J.~Bicket.
\newblock {\em Bit-rate selection in wireless networks}.
\newblock PhD thesis, Massachusetts Institute of Technology, 2005.

\bibitem{bubeck2012}
S.~Bubeck and N.~Cesa-Bianchi.
\newblock Regret analysis of stochastic and nonstochastic multi-armed bandit
  problems.
\newblock {\em Foundations and Trends in Machine Learning}, 5(1):1--122, 2012.

\bibitem{camp2008}
J.~Camp and E.~Knightly.
\newblock Modulation rate adaptation in urban and vehicular environments:
  cross-layer implementation and experimental evaluation.
\newblock In {\em Proceedings of the 14th ACM international conference on
  Mobile computing and networking}, MobiCom '08, pages 315--326. ACM, 2008.

\bibitem{crepaldi2012}
R.~Crepaldi, J.~Lee, R.~Etkin, S.-J. Lee, and R.~Kravets.
\newblock Csi-sf: Estimating wireless channel state using csi sampling amp;
  fusion.
\newblock In {\em INFOCOM, 2012 Proceedings IEEE}, pages 154--162, 2012.

\bibitem{deek2013}
L.~Deek, E.~Garcia-Villegas, E.~Belding, S.-J. Lee, and K.~Almeroth.
\newblock Joint rate and channel width adaptation in 802.11 mimo wireless
  networks.
\newblock In {\em Proceedings of IEEE Secon}, 2013.

\bibitem{freudenthaler2007}
K.~Freudenthaler, A.~Springer, and J.~Wehinger.
\newblock Novel sinr-to-cqi mapping maximizing the throughput in hsdpa.
\newblock In {\em Proceedings of IEEE WCNC}, 2007.

\bibitem{garivier2011}
A.~Garivier and O.~Capp\'e.
\newblock The {KL-UCB} algorithm for bounded stochastic bandits and beyond.
\newblock In {\em Proceedings of Conference On Learning Theory (COLT)}, 2011.

\bibitem{garivier08}
A.~Garivier and E.~Moulines.
\newblock On upper-confidence bound policies for non-stationary bandit
  problems, 2008.
\newblock ArXiv e-print. http://arxiv.org/abs/0805.3415.

\bibitem{garivier2011alt}
A.~Garivier and E.~Moulines.
\newblock On upper-confidence bound policies for switching bandit problems.
\newblock In {\em Proceedings of the 22nd international conference on
  Algorithmic learning theory}, ALT'11, pages 174--188, 2011.

\bibitem{graves1997}
T.~L. Graves and T.~L. Lai.
\newblock Asymptotically efficient adaptive choice of control laws in
  controlled markov chains.
\newblock {\em SIAM J. Control and Optimization}, 35(3):715--743, 1997.

\bibitem{halperin2010}
D.~Halperin, W.~Hu, A.~Sheth, and D.~Wetherall.
\newblock Predictable 802.11 packet delivery from wireless channel
  measurements.
\newblock {\em SIGCOMM Comput. Commun. Rev.}, 40(4):159--170, Aug. 2010.

\bibitem{haratcherev2004}
I.~Haratcherev, K.~Langendoen, R.~Lagendijk, and H.~Sips.
\newblock Hybrid rate control for ieee 802.11.
\newblock In {\em Proceedings of the ACM MobiWac}, 2004.

\bibitem{Hoeffding1963}
W.~Hoeffding.
\newblock Probability inequalities for sums of bounded random variables.
\newblock {\em Journal of the American Statistical Association}, 58(301):pp.
  13--30, 1963.

\bibitem{holland2001rate}
G.~Holland, N.~Vaidya, and P.~Bahl.
\newblock A rate-adaptive mac protocol for multi-hop wireless networks.
\newblock In {\em Proceedings of ACM Mobicom}, 2001.

\bibitem{judd2008}
G.~Judd, X.~Wang, and P.~Steenkiste.
\newblock Efficient channel-aware rate adaptation in dynamic environments.
\newblock In {\em Proceedings of ACM MobiSys}, 2008.

\bibitem{kamerman1997wavelan}
A.~Kamerman and L.~Monteban.
\newblock Wavelan-ii: a high-performance wireless lan for the unlicensed band.
\newblock {\em Bell Labs technical journal}, 2(3):118--133, 1997.

\bibitem{kim2008}
D.~Kim, B.~Jung, H.~Lee, D.~Sung, and H.~Yoon.
\newblock Optimal modulation and coding scheme selection in cellular networks
  with hybrid-arq error control.
\newblock {\em Wireless Communications, IEEE Transactions on},
  7(12):5195--5201, 2008.

\bibitem{kim2006cara}
J.~Kim, S.~Kim, S.~Choi, and D.~Qiao.
\newblock {CARA}: Collision-aware rate adaptation for {IEEE} 802.11 {WLANs}.
\newblock In {\em Proceedings of IEEE Infocom}, 2006.

\bibitem{kocsis2006}
L.~Kocsis and C.~Szepesv{\'a}ri.
\newblock Discounted ucb.
\newblock In {\em Proceedings of the 2dn {PASCAL} Challenges Workshop}, 2006.

\bibitem{lacage2004ieee}
M.~Lacage, M.~Manshaei, and T.~Turletti.
\newblock Ieee 802.11 rate adaptation: a practical approach.
\newblock In {\em Proceedings of MSWiM}, pages 126--134. ACM, 2004.

\bibitem{lai1985}
T.~Lai and H.~Robbins.
\newblock Asymptotically efficient adaptive allocation rules.
\newblock {\em Advances in Applied Mathematics}, 6(1):4--2, 1985.

\bibitem{nguyen2011}
D.~Nguyen and J.~Garcia-Luna-Aceves.
\newblock A practical approach to rate adaptation for multi-antenna systems.
\newblock In {\em Proceedings of IEEE ICNP}, 2011.

\bibitem{pang2005rate}
Q.~Pang, V.~Leung, and S.~Liew.
\newblock A rate adaptation algorithm for ieee 802.11 wlans based on mac-layer
  loss differentiation.
\newblock In {\em Proceedings of IEEE BroadNets}, 2005.

\bibitem{Pefkianakis:2010}
I.~Pefkianakis, Y.~Hu, S.~H. Wong, H.~Yang, and S.~Lu.
\newblock Mimo rate adaptation in 802.11n wireless networks.
\newblock In {\em Proceedings of ACM Mobicom}, 2010.

\bibitem{radunovic2011}
B.~Radunovic, A.~Proutiere, D.~Gunawardena, and P.~Key.
\newblock Dynamic channel, rate selection and scheduling for white spaces.
\newblock In {\em Proceedings of ACM CoNEXT}, 2011.

\bibitem{reis2006}
C.~Reis, R.~Mahajan, M.~Rodrig, D.~Wetherall, and J.~Zahorjan.
\newblock Measurement-based models of delivery and interference in static
  wireless networks.
\newblock {\em SIGCOMM Comput. Commun. Rev.}, 36(4):51--62, Aug. 2006.

\bibitem{robbins1952}
H.~Robbins.
\newblock Some aspects of the sequential design of experiments.
\newblock {\em Bulletin of the American Mathematical Society}, 58(5):527--535,
  1952.

\bibitem{sadeghi2002opportunistic}
B.~Sagdehi, V.~Kanodia, A.~Sabharwal, and E.~Knightly.
\newblock Opportunistic media access for multirate ad hoc networks.
\newblock In {\em Proceedings of ACM Mobicom}, 2002.

\bibitem{slivkins2011}
A.~Slivkins.
\newblock Contextual bandits with similarity information.
\newblock {\em Journal of Machine Learning Research - Proceedings Track},
  19:679--702, 2011.

\bibitem{slivkins2008}
A.~Slivkins and E.~Upfal.
\newblock Adapting to a changing environment: the brownian restless bandits.
\newblock In {\em COLT}, pages 343--354, 2008.

\bibitem{thompson1933}
W.~R. Thompson.
\newblock On the likelihood that one unknown probability exceeds another in
  view of the evidence of two samples.
\newblock {\em Biometrika}, 25(3/4):pp. 285--294, 1933.

\bibitem{vutukuru2009}
M.~Vutukuru, H.~Balakrishnan, and K.~Jamieson.
\newblock Cross-layer wireless bit rate adaptation.
\newblock In {\em Proceedings of ACM SIGCOMM}, 2009.

\bibitem{wong2006robust}
S.~Wong, H.~Yang, S.~Lu, and V.~Bharghavan.
\newblock Robust rate adaptation for 802.11 wireless networks.
\newblock In {\em Proceedings of ACM Mobicom}, 2006.

\bibitem{yu2009}
J.~Y. Yu and S.~Mannor.
\newblock Piecewise-stationary bandit problems with side observations.
\newblock In {\em ICML}, page 148, 2009.

\bibitem{yu2011}
J.~Y. Yu and S.~Mannor.
\newblock Unimodal bandits.
\newblock In {\em Proceedings of the 28th International Conference on Machine
  Learning (ICML-11)}, pages 41--48, New York, NY, USA, 2011. ACM.

\bibitem{zhang2008}
J.~Zhang, K.~Tan, J.~Zhao, H.~Wu, and Y.~Zhang.
\newblock A practical snr-guided rate adaptation.
\newblock In {\em INFOCOM 2008. The 27th Conference on Computer Communications.
  IEEE}, pages 2083--2091, 2008.

\end{thebibliography}

\end{document}